\documentclass[a4paper,twocolumn]{article}
\usepackage[ansinew]{inputenc}
\usepackage[english]{babel}
\usepackage{amsmath}
\usepackage{amssymb}
\usepackage[amssymb]{SIunits}
\usepackage{graphicx}
\usepackage[usenames,dvipsnames]{xcolor}		
\usepackage{epic}
\usepackage{booktabs}
\usepackage{hyperref}
\usepackage{color}
\usepackage{lscape}
\usepackage{fancyhdr}

\hypersetup{
   pdffitwindow=true,
   colorlinks=true,  
   breaklinks=true,  
   urlcolor= NavyBlue,   
   linkcolor=WildStrawberry,    
   citecolor=blue,   
   filecolor=magenta,
   menucolor=black,  
   pdftitle={A Physically-Consistent Chemical Dataset for the Simulation of N2-CH4 Shocked Flows Up to T=100,000K},     
   pdfauthor={Mario Lino da Silva},
   pdfcreator={Mario Lino da Silva}
}

\newcounter{daggerfootnote}
\newcommand*{\daggerfootnote}[1]{%
    \setcounter{daggerfootnote}{\value{footnote}}%
    \renewcommand*{\thefootnote}{\fnsymbol{footnote}}%
    \footnote[2]{#1}%
    \setcounter{footnote}{\value{daggerfootnote}}%
    \renewcommand*{\thefootnote}{\arabic{footnote}}%
    }

\title{\phantom{o}\\\vspace{-4cm}A Physically-Consistent Chemical Dataset for the Simulation of N$_{2}$--CH$_{4}$ Shocked Flows Up to T=100,000\kelvin}

\author{M. Lino da Silva and J. Vargas$^\dag$%
\thanks{Instituto de Plasmas e Fus\~{a}o Nuclear, Laborat\'{o}rio Associado, %
Instituto Superior T\'{e}cnico, Universidade T\'{e}cnica de Lisboa, Av. Rovisco Pais, %
1049--001, Lisboa, Portugal}
        }%

\date{December 24, 2017}

\begin{document}

\onecolumn


\maketitle%

\begin{abstract}
In the previous work carried out in the scope of the \emph{Validation of Aerothermochemistry Models for Re-Entry Applications}, it was verified that the G\"{o}k\c{c}en chemical dataset provided increasingly diverging results from experiments, as one considered shock speeds in excess of 5\kilo\metre\per\second. Namely, for shock velocities between 7 and 9\kilo\metre\per\second, more than one temporal peak in CN Violet radiation were predicted by models considering this kinetic dataset, in contradiction with experiments. This hinted at several of the rates from the dataset not being directly applicable in the temperature range of interest for such applications, often in excess of 10,000\kelvin.
Indeed, it has been found that several macroscopic rates from the G\"{o}k\c{c}en chemical dataset reached unphysical values at very high temperatures. Furthermore, many of the ionization rates have been found to be inadequate for the simulation of high-temperature N$_{2}$--CH$_{4}$ shocked flows. Here, we have carried an extensive update of the G\"{o}k\c{c}en chemical dataset, with the aim of at least reaching physically consistent rates for the whole T=100-100,000\kelvin\ temperature range. While it cannot really be claimed that such improved dataset is validated in such an extended temperature range (due to the scarcely available experimental data for such high temperature ranges), it is capable of providing more accurate simulations of high-speed shocked flows for this mixture, when compared to the G\"{o}k\c{c}en chemical dataset.\protect\daggerfootnote{This work as been carried out in the scope of ESA/ESTEC Contract 21029: Validation of Aerothermal Chemistry Models for Re-entry Applications}
\end{abstract}

\tableofcontents%

\section{Introduction}

In the times that preceded the entry of the Huygens spacecraft in Titan's atmosphere, there was the need to consolidate the corresponding aerothermodynamics database. G\"{o}k\c{c}en \cite{Gokcen:2005} carried out a much needed upgrade of the Nelson chemical model \cite{Nelson:1991}, taking into account a diverse array of chemical datasets for hydrocarbons, issued for the most from combustion applications. A validation of this dataset has been carried against the best available measurements of the time evolution of the species of interest, mainly obtained from shock-tube studies in the temperature range of 2,000--3,500\kelvin, followed by a sensitivity study that yielded a reduced dataset of 37 rates which was recommended for Titan atmospheric entry applications.\bigskip

Since then, several works have emphasized the need for considering more detailed approaches for the simulation of chemical processes in N$_2$--CH$_4$ flows. Besides the layers of more or less complex collisional-radiative models \cite{Magin:2006,Brandis:2009,Brandis:2010}, This has mainly consisted in replacing multitemperature approaches by state-specific approaches for the modeling of dissociation processes. This kind of improvements have been up to now restricted to the state-specific treatment of N$_2$ molecule dissociation processes  \cite{Magin:2006,Brandis:2009,Brandis:2010}, but still bring consistent improvements to the obtained results, as N$_2$ makes up to 98\% of Titan's atmosphere, and as CN production reactions are strongly dependent on the availability of atomic nitrogen atoms.\bigskip

In our previous report, we have applied recent multiquantum state-specific rates \cite{LinodaSilva:2007-1,LinodaSilva:2007-2} to the simulation of several shocked N$_2$--CH$_4$ flows for different shock velocities \cite{report1}. Although it has been found that the application of multiquantum state-specific datasets for N$_2$ dissociation indeed improved the quality of the numerical reproduction of experiments, it also has been found that for higher degrees of ionization typical of higher shock speeds, the time evolution of the species concentrations followed some unusual patterns, which are not apparent in the experimental data. More specifically, the concentration of the CN(B) radiative state was found to have two local maxima, which was not the case with experiments.\bigskip

This has led us to question the direct applicability of the G\"{o}k\c{c}en chemical dataset, for such high energy shocks. This dataset has been strictly validated for a temperature range of 2000--3500\kelvin\ \cite{Gokcen:2005}, yet different authors have been able to use it successfully, often for the reproduction of the 98\%N$_2$--2\%CH$_4$ experimental point at 5.15\kilo\metre\per\second\ and 13\pascal, obtained at the NASA Ames EAST facility \cite{Bose:2006}. This might be explained by the moderate post-shock translational temperature of T=12,600\kelvin, which is yet close enough to the top of the validity range of the G\"{o}k\c{c}en dataset. However, it is very well known that the post-shock temperature increases quadratically with shock velocity. Table \ref{tab:postshock} presents the postshock conditions for 3 different shocks reproduced in the University of Queensland X2 shock-tube.\bigskip

\begin{table*}[!htbp]
\caption{Post-shock Conditions, for a 98\%N$_2$--2\%CH$_4$ Mixture}%
\label{tab:postshock}%
\centering%
\begin{tabular}{c c l l}
\toprule%
\midrule%
v$_{\infty}$(\kilo\metre\per\second) & p$_{\infty}$(\pascal) & T$_0$(\kelvin) & p$_0$(\pascal)\\%
\midrule%
5.15  & 13 & 12,590 & 3200\notag\\%
7.40  & 13 & 25,680 & 6600\notag\\%
9.00  & 13 & 37,850 & 9770\notag\\%
\bottomrule%
\end{tabular}
\end{table*}

It is then clear that the high-temperature behavior of the G\"{o}k\c{c}en dataset should firstly be examined in order to check its consistency. For example, one may take advantage of the large number of cross-sections published in the literature for chemical reactions involving ions. Usually, published cross-sections extend at least to over 10\electronvolt, allowing for the production of rates valid up to very high electronic temperatures (assuming a Maxwellian distribution of the flow electrons). The update of the ionized rates dataset of G\"{o}k\c{c}en is a large task of this work, given that the ionized rates proposed by G\"{o}k\c{c}en are directly reported from the previous Nelson dataset \cite{Nelson:1991}, without any further updates. Additionally, simulations of high-speed shocked flows with a large degree of ionization have shown to produce results with some inconsistencies, further emphasizing the need for udates of these rates.\bigskip

The starting point of this work is then the reduced G\"{o}k\c{c}en chemical dataset, including 28 neutral and 9 ionized rates, which is reproduced in Table \ref{tab:Titan_set}.\bigskip

\begin{table*}[!htbp]
\caption{G\"{o}k\c{c}en Reduced Chemical Dataset for N$_2$--CH$_4$ flows}%
\label{tab:Titan_set}%
\centering%
\begin{tabular}{r r@{}l c}
\toprule%
\midrule%
No. & \multicolumn{2}{c}{Reaction} &   Rate (\centi\cubic\metre\per\mole\per\second)\\%
\midrule%
1  & N$_{2}$\ +\ Mol.&\ $\rightleftarrows$ N\ +\ N\ +\ Mol                    & $7.00\times 10^{21}T^{-1.60}\exp(-113,200/T)$\notag\\%
2  & N$_{2}$\ +\ Atom&\ $\rightleftarrows$ N\ +\ N\ +\ Atom                   & $3.00\times 10^{22}T^{-1.60}\exp(-113,200/T)$\notag\\%
3  & CH$_{4}$\ +\ M\phantom{$_{2}$}&\ $\rightleftarrows$ CH$_{3}$\ +\ H\ +\ M & $4.70\times 10^{47}T^{-8.20}\exp(-\phantom{0}59,200/T)$\notag\\%
4  & CH$_{3}$\ +\ M\phantom{$_{2}$}&\ $\rightleftarrows$ CH$_{2}$\ +\ H\ +\ M & $1.02\times 10^{16}\phantom{T^{-0.00}}\exp(-\phantom{0}45,600/T)$\notag\\%
5  & CH$_{3}$\ +\ M\phantom{$_{2}$}&\ $\rightleftarrows$ CH\ +\ H$_{2}$\ +\ M & $5.00\times 10^{15}\phantom{T^{-0.00}}\exp(-\phantom{0}42,800/T)$\notag\\%
6  & CH$_{2}$\ +\ M\phantom{$_{2}$}&\ $\rightleftarrows$ CH\ +\ H\ +\ M       & $4.00\times 10^{15}\phantom{T^{-0.00}}\exp(-\phantom{0}41,800/T)$\notag\\%
7  & CH$_{2}$\ +\ M\phantom{$_{2}$}&\ $\rightleftarrows$ C\ +\ H$_{2}$\ +\ M  & $1.30\times 10^{14}\phantom{T^{-0.00}}\exp(-\phantom{0}29,700/T)$\notag\\%
8  & CH\ +\ M\phantom{$_{2}$}&\ $\rightleftarrows$ C\ +\ H\ +\ M              & $1.90\times 10^{14}\phantom{T^{-0.00}}\exp(-\phantom{0}33,700/T)$\notag\\%
9  & C$_{2}$\ +\ M\phantom{$_{2}$}&\ $\rightleftarrows$ C\ +\ C\ +\ M         & $1.50\times 10^{16}\phantom{T^{-0.00}}\exp(-\phantom{0}71,600/T)$\notag\\%
10 & H$_{2}$\ +\ M\phantom{$_{2}$}&\ $\rightleftarrows$ H\ +\ H\ +\ M         & $2.23\times 10^{14}\phantom{T^{-0.00}}\exp(-\phantom{0}48,350/T)$\notag\\%
11 & CN\ +\ M\phantom{$_{2}$}&\ $\rightleftarrows$ C\ +\ N\ +\ M              & $2.53\times 10^{14}\phantom{T^{-0.00}}\exp(-\phantom{0}71,000/T)$\notag\\%
12 & NH\ +\ M\phantom{$_{2}$}&\ $\rightleftarrows$ N\ +\ H\ +\ M              & $1.80\times 10^{14}\phantom{T^{-0.00}}\exp(-\phantom{0}37,600/T)$\notag\\%
13 & HCN\ +\ M\phantom{$_{2}$}&\ $\rightleftarrows$ CN\ +\ H\ +\ M            & $3.57\times 10^{26}T^{-2.60}\exp(-\phantom{0}62,845/T)$\notag\\%
\midrule%
14 & CH$_{3}$\ +\ N\phantom{$_{2}$}&\ $\rightleftarrows$ HCN\ +\ H\ +\ H      & $7.00\times 10^{13}\phantom{T^{-0.00}\exp(-000,000/T)}$\notag\\%
15 & CH$_{3}$\ +\ H\phantom{$_{2}$}&\ $\rightleftarrows$ CH$_{2}$\ +\ H$_{2}$ & $6.03\times 10^{13}\phantom{T^{-0.00}}\exp(-\phantom{00}7,600/T)$\notag\\%
16 & CH$_{2}$\ +\ N$_{2}$&\ $\rightleftarrows$ HCN\ +\ NH                     & $4.82\times 10^{12}\phantom{T^{-0.00}}\exp(-\phantom{0}18,000/T)$\notag\\%
17 & CH$_{2}$\ +\ N\phantom{$_{2}$}&\ $\rightleftarrows$ HCN\ +\ H            & $5.00\times 10^{13}\phantom{T^{-0.00}\exp(-000,000/T)}$\notag\\%
18 & CH$_{2}$\ +\ H\phantom{$_{2}$}&\ $\rightleftarrows$ CH\ +\ H$_{2}$       & $6.03\times 10^{12}\phantom{T^{-0.00}}\exp(+\phantom{000,}900/T)$\notag\\%
19 & CH\ +\ N$_{2}$&\ $\rightleftarrows$ HCN\ +\ N                            & $4.40\times 10^{12}\phantom{T^{-0.00}}\exp(-\phantom{0}11,060/T)$\notag\\%
20 & CH\ +\ C\phantom{$_{2}$}&\ $\rightleftarrows$ C$_{2}$\ +\ H              & $2.00\times 10^{14}\phantom{T^{-0.00}\exp(-000,000/T)}$\notag\\%
21 & C$_{2}$\ +\ N$_{2}$&\ $\rightleftarrows$ CN\ +\ CN                       & $1.50\times 10^{13}\phantom{T^{-0.00}}\exp(-\phantom{0}21,000/T)$\notag\\%
22 & CN\ +\ H$_{2}$&\ $\rightleftarrows$ HCN\ +\ H                            & $2.95\times 10^{05}\phantom{T^{-0.00}}\exp(-\phantom{00}1,130/T)$\notag\\%
23 & CN\ +\ C\phantom{$_{2}$}&\ $\rightleftarrows$ C$_{2}$\ +\ N              & $5.00\times 10^{13}\phantom{T^{-0.00}}\exp(-\phantom{0}13,000/T)$\notag\\%
24 & N\ +\ H$_{2}$&\ $\rightleftarrows$ NH\ +\ H                              & $1.60\times 10^{14}\phantom{T^{-0.00}}\exp(-\phantom{0}12,650/T)$\notag\\%
25 & C\ +\ N$_{2}$&\ $\rightleftarrows$ CN\ +\ N                              & $5.24\times 10^{13}\phantom{T^{-0.00}}\exp(-\phantom{0}22,600/T)$\notag\\%
26 & C\ +\ H$_{2}$&\ $\rightleftarrows$ CH\ +\ H                              & $4.00\times 10^{14}\phantom{T^{-0.00}}\exp(-\phantom{0}11,700/T)$\notag\\%
27 & H\ +\ N$_{2}$&\ $\rightleftarrows$ NH\ +\ N                              & $3.00\times 10^{12}T^{-0.50}\exp(-\phantom{0}71,400/T)$\notag\\%
28 & H\ +\ CH$_{4}$&\ $\rightleftarrows$ CH$_{3}$\ +\ H$_{2}$                 & $1.32\times 10^{04}T^{-3.00}\exp(-\phantom{00}4,045/T)$\notag\\%
\midrule%
29 & N\ +\ N\phantom{$_{2}$}&\ $\rightleftarrows$ N$_{2}^{+}$\ +\ e$^{-}$     & $4.40\times 10^{07}T^{-1.50}\exp(-\phantom{0}67,500/T)$\notag\\%
30 & C\ +\ N\phantom{$_{2}$}&\ $\rightleftarrows$ CN$^{+}$\ +\ e$^{-}$        & $1.00\times 10^{15}T^{-1.50}\exp(-164,400/T)$\notag\\%
31 & N\ +\ e$^{-}$ &\ $\rightleftarrows$ N$^{+}$\ +\ e$^{-}$ +\ e$^{-}$       & $2.50\times 10^{34}T^{-3.82}\exp(-168,600/T)$\notag\\%
32 & C\ +\ e$^{-}$ &\ $\rightleftarrows$ C$^{+}$\ +\ e$^{-}$ +\ e$^{-}$       & $3.70\times 10^{31}T^{-3.00}\exp(-130,720/T)$\notag\\%
33 & H\ +\ e$^{-}$ &\ $\rightleftarrows$ H$^{+}$\ +\ e$^{-}$ +\ e$^{-}$       & $2.20\times 10^{30}T^{-2.80}\exp(-157,800/T)$\notag\\%
34 & Ar\ +\ e$^{-}$ &\ $\rightleftarrows$ Ar$^{+}$\ +\ e$^{-}$ +\ e$^{-}$     & $2.50\times 10^{34}T^{-3.82}\exp(-181,700/T)$\notag\\%
35 & CN$^{+}$\ +\ N\phantom{$_{2}$} &\ $\rightleftarrows$ CN\ +\ N$^{+}$      & $9.80\times 10^{12}\phantom{T^{-0.00}}\exp(-\phantom{0}40,700/T)$\notag\\%
36 & C$^{+}$\ +\ N$_{2}$ &\ $\rightleftarrows$ N$_{2}^{+}$\ +\ C              & $1.11\times 10^{14}T^{-0.11}\exp(-\phantom{0}50,000/T)$\notag\\%
\bottomrule%
\end{tabular}
\end{table*}

\clearpage

Three main objectives can be identified for this review of the G\"{o}k\c{c}en chemical dataset:

\begin{itemize}
\item Production of adequate backward rates
\item Verification of the consistency of the neutral species dataset, for a broadened temperature range
\item Update of the ionic species chemical dataset
\end{itemize}

The first point will be covered in Section \ref{sec:keq}, the second in Section \ref{sec:neutral}, and the third in Section \ref{sec:ions}, along with a sensitivity study to yield a reduced set of high-temperature N$_2$--CH$_4$ ionization reactions.

\section{Calculation of Thermodynamic Properties}
\label{sec:keq}

The calculation of N$_2$--CH$_4$ thermodynamic properties is essentially related to the problem of detailed balancing for the calculation of reverse chemical kinetic rates. Here we will describe the method used for the practical calculation of equilibrium constants K$_{eq}$(T) (Section \ref{sec:keqsub}), followed by a reminder on the proper calculation of reverse rates, for the case of nonequilibrium flow conditions (Section \ref{sec:detbal}).

\subsection{Derivation of Equilibrium Constants}
\label{sec:keqsub}

This work aims at providing a consistent set of direct K$_f$ and reverse K$_b$ reactions for a large temperature range (T=100--100,000\kelvin). As it is well known, equilibrium coefficients for specific reactions are directly related to the equilibrium concentrations of their constituents. For example, a reaction of the type $A + B \rightleftarrows C + D$ has for equilibrium constant:

\begin{equation}
\label{eq:Keq}
K_{eq}(T)=\frac{[A(T)]_{eq}[B(T)]_{eq}}{[C(T)]_{eq}[D(T)]_{eq}}
\end{equation}

It is very challenging to produce a set of equilibrium constants valid for a large temperature range. When developing or using a chemical equilibrium code, one is faced with two complementary difficulties.

\begin{enumerate}
\item The code has to have an accurate set of partition functions for all the relevant species in the selected temperature range.
\item The code has to be numerically capable of resolving very low concentrations.
\end{enumerate}

\begin{figure}[!htbp]
\centering
\includegraphics[width=.6\textwidth]{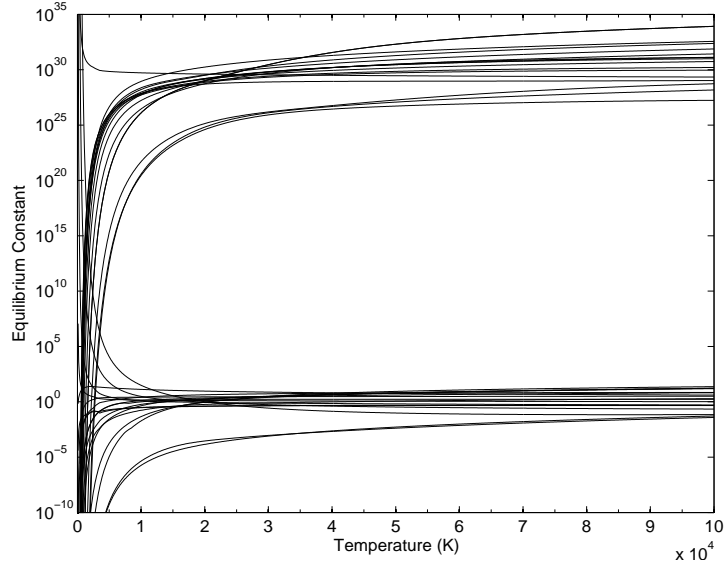}
\caption{\small{Equilibrium Constants for the G\"{o}k\c{c}en rate dataset; T=100--100,000\kelvin}}%
\label{fig:Keq_f}
\end{figure}

In the scope of this work, we have selected two complementary numerical suites. The \emph{Cantera} chemical kinetics suite \cite{Cantera}, coded in MATLAB language, has been firstly utilized for the production of chemical equilibrium concentrations in the range T=100--10,000\kelvin. Then, we have considered a recent calculation of N$_2$--CH$_4$ equilibrium concentrations provided by P. Andr\'{e} from LAEPT \cite{Andre:2010}, and valid in the range T=1,000--50,000\kelvin, with a minimum threshold of about 10$^{-23}$ for the species molar fractions. The two chemical equilibrium datasets have been merged and extrapolated to yield equilibrium constants in the required extended range (T=100--10,000\kelvin). These are reported in Fig. \ref{fig:Keq_f}.

\subsection{Detailed Balancing for Nonequilibrium Flows}
\label{sec:detbal}

Here we will write a short summary describing the adequate application of the detailed blance principle in the case of nonequilibium flows. The question of the applicability of the equilibrium constants $K_{eq}$ is generally invoked in the scope of applying the detailed balance principle in the scope of dissociation/recombination processes. This question arises whenever one is applying one of the multiple multiquantum theories for dissociation (mostly vibrationally-specific) such that:

\begin{equation}
K_{diss}(T,T_{int})=K_{diss}(T)Z(T,T_{int})
\end{equation}

and there is the need to find out the exact relationship for the equilibrium constant:

\begin{equation}
K_{eq}(T,T_{int})=\frac{K_{diss}(T)Z(T,T_{int})}{K_{rec}(T)}
\end{equation}

Several approaches for the calculation of $K_{eq}(T,T_{int})$ have been proposed in the past, and have been sumarized by Willian \cite{William:1999}. This question has been considered in a past work by the authors \cite{LinodaSilva:2009-1}, and is summarized in this subsection.\\

We will start by defining a vibrationally-specific microscopic transitions such that:

\begin{equation}
\label{eq:VD}
\textrm{AB(v)\ +\ M}\rightarrow \textrm{A\ +\ B\ +\ M}.
\end{equation}

Eq. \ref{eq:Keq} may be written in this case:

\begin{equation}
\label{eq:VDKeq}
K_{eq}(T,v)=\frac{K_{diss}(T,v)}{K_{rec}(T,v)}=\frac{[N_A(T)]_{eq}[N_B(T)]_{eq}}{[N_{AB}(T,v)]_{eq}}.
\end{equation}

As it is well known, under thermodynamic equilibrium the distribution of the internal levels of a molecular species is proportional to the individual partition functions of such levels

\begin{equation}
\frac{N_i}{N}=\frac{Q_i(T)}{Q_{tot}}=\frac{g_i\exp\left(-\frac{E_i}{k_B T}\right)}{Q_{tot}}
\end{equation}

with $g_i$ and $E_i$ denoting the multiplicity and the energy of a given $i$ state. For the particular vibrationally-specific reaction (Eq. \ref{eq:VD}), we obtain for Eq. \ref{eq:VDKeq}:

\begin{equation}
\label{eq:VDKeq1}
K_{eq}(T,v)=\frac{\left (Q_{tr}(T)\right)_A\left(Q_{tr}(T)\right)_B}{\left(Q_{tr}(T)Q_{v}(T)\right)_{AB}},
\end{equation}
%

Departing from the state-resolved dissociation-recombination process, as defined by Eq. \ref{eq:VD}, one may easily determine the corresponding total dissociation and recombination rates such that:

\begin{itemize}
\item The total dissociation rate coefficient is the weighted sum of the state-resolved rate coefficients

\begin{equation}
\label{eq:KdSTS}
K_{diss}^*(T) \; = \; \sum_v \delta_v \; K_{diss}(T,v) \; .
\end{equation}

where $\delta_v$ = $[N_v] / N$, with $\sum_v [N_v]$ = $N$, denotes the actual fractional vibrational population in every $v-$th level.

\item The total recombination rate coefficient is the simple sum of the state-resolved recombination rate coefficients

\begin{equation}
\label{eq:KrSTS}
K_{rec}^*(T) \; = \; \sum_v K_{rec}(T,v) \; .
\end{equation}

\end{itemize}

Obviously, the total dissociation rate coefficient obtained
at equilibrium conditions cannot be extended to nonequilibrium
conditions, whereas the total recombination rate can. In most
nonequilibrium situations, the distribution of populations among
the various vibrational levels is not proportional to the partition
function. In certain cases it may occur that, due to high
vibration-vibration (V$-$V) energy exchange collisions, the manifold
of vibrational levels also presents a form of a Boltzmann-like
distribution, but with a characteristic vibrational temperature
$T_v$ $\neq$ $T$. For this case we have $\delta_v$ =
$\exp(-E_v/(k_B T_v))$ in Eq. \ref{eq:KdSTS}.

In the absence of any equilibration of the vibrational manifold to a
Boltzmann distribution, there is no particular advantage in defining
total dissociation and recombination rates, and a state-to-state
approach for the problem is mandatory.
However, in the situation where a characteristic vibrational temperature
$T_v$ $\neq$ $T$ may be defined, we obtain from Eqs. (\ref{eq:VDKeq},\ref{eq:VDKeq1},\ref{eq:KdSTS},\ref{eq:KrSTS}),
after some simple algebraic manipulation, the total equilibrium rate for
the dissociation-recombination reaction (Eq. \ref{eq:VD}):

\begin{equation}
\label{eq:18}
K_{eq}^*(T,T_v) \; = \; K_{eq}^*(T) \; \frac{ \sum_v Q_v(T) }{\sum_v Q_v(T_v)}
\; \frac{\sum_v Q_v(T_v) \; K_{diss}(T,v)}{\sum_v Q_v(T) \; K_{diss}(T,v)} \; ,
\end{equation}

where $K_{eq}^*(T)$ denotes the total equilibrium constant under fully
equilibrium conditions (Eq. \ref{eq:VDKeq1}).\\

By inspection of Eq. \ref{eq:18} we verify that in the case of a two-temperature
description, the two terms involving
the state-specific dissociation rate $K_{diss}(T,v)$ do not cancel each other,
meaning that the definition of a total two-temperature equilibrium constant
does not present any particular advantage, as it relies on a given set
of state-specific dissociation rates.
Such a conclusion is important in what regards two-temperature ($T,T_v$) models
\cite{LinodaSilva:2007-3}, which propose simplified alternatives to state-to-state
models, with the aim of reproducing nonequilibrium phenomena in gases and
plasmas without the often prohibitive computational overheads associated to
state-resolved models. Indeed, taking into account Eqs. (\ref{eq:KdSTS},\ref{eq:18}), we may conclude that one should not strive to obtain a total equilibrium constant of the kind of $K_{eq}^*(T,T_v)$, but instead should define directly a total recombination rate coefficient $K_{rec}^*(T)$, using Eq. \ref{eq:KrSTS}, as such expression is independent from the total dissociation constant being written as $K_{diss}^*(T)$ or $K_{diss}^*(T,T_v)$. In truth, the total recombination rate in nonequilibrium conditions remains expressed in terms of a sole temperature $T$.

\section{Update of the G\"{o}k\c{c}en Neutral Reactions Dataset}
\label{sec:neutral}

Obtaining reliable equilibrium constants of the G\"{o}k\c{c}en dataset enables the verification of the consistency of the proposed rates, for the temperature range that has been self-imposed in the scope of this work (T=100--100,000\kelvin). As discussed in the introduction, it is not realistically possible to strive for a validation of the dataset for such an extended temperature range. It is however possible to check if the dataset is at least physically consistent. As we will see more ahead, this is far from true if one simply considers the dataset of Table \ref{tab:Titan_set}, regardless of the temperature range.

When on speaks about the physical consistency of the dataset under examination, one is specifically considering whether the chemical reaction rates do not exceed the physical limit which is represented by the gas-kinetic collision rate (in other terms, one is checking if the rates do not correspond to transition probabilities larger than unity during the collision between two species, at a given energy.

\subsection{Limiting Gas-Kinetic Rates}

The definition of the gas-kinetic rates for each species collisional pair is outlined in this section. We will firstly write the relationships that have been utilized in this work for defining the two-body gas-kinetic rate $Z$, and the three-body gas-kinetic rate $Z_3$.\bigskip

For a two-body encounter, the collisonal rate is written as

\begin{equation}
Z=\sigma\sqrt{\frac{8k_{B}T}{\pi\widetilde{m}}},
\end{equation}

where $\sigma$ represents the two-species collisional cross-section, and $\widetilde{m}$ the collisional reduced mass.\bigskip

For a three-body encounter (the case of recombination reactions) the calculation of an exact expression, involving the three species interaction radius, is slightly more complicated \cite{Bernshtein:2004}. However, for the purpose of this verification work, it is sufficient to roughly approximate the exact three-body collision rate. We consider that during a binary collision, a third body may interact if it crosses near the vicinity of the two-bodies during the time where these are interacting. We then introduce the cross-section diameter to mean free path ratio such that:

\begin{equation}
\frac{d}{l}=\sqrt{\frac{8\sigma^3}{\pi}}n, 
\end{equation}

and the three-body gas-kinetic rate can then be approximated by the expression

\begin{equation}
Z_3=Z\times\frac{d}{l}.
\end{equation}
\bigskip

All that is left is the calculation of the collisional cross-sections. These are obtained taking into account the interaction diameter of the colliding species such that

\begin{equation}
\sigma=\pi\left(\frac{d_{AB}+d_{CD}}{2}\right)^2
\end{equation}

The interaction diameters for the neutral species of the N$_2$--CH$_4$ kinetic scheme are taken from the compilation report of Svehla when available, or otherwise estimated according to the relationships proposed by the same author \cite{Svehla:1962}. Table \ref{tab:radius} lists the interaction diameter for each of these species.\bigskip

\begin{table*}[!htbp]
\caption{Compiled interaction diameters for the neutral species of a N$_2$--CH$_4$ plasma}%
\label{tab:radius}%
\centering%
\begin{tabular}{r l}
\toprule%
\midrule%
Species & Int. Diameter, $\angstrom$\\%
\midrule%
H     & 2.708\\
C     & 3.385\\     
N     & 3.298\\     
H$_2$ & 2.827\\     
CH    & 3.37 \\     
C$_2$ & 3.913\\     
CN    & 3.856\\     
NH    & 3.312$^a$\\ 
N$_2$ & 3.798\\     
HCN   & 3.84$^b$ \\ 
CH$_2$& 3.5$^c$  \\ 
CH$_3$& 3.63$^c$ \\ 
CH$_4$& 3.758\\     
\midrule%
\bottomrule%
\multicolumn{2}{l}{\footnotesize{a: $1/2(r_{\textrm{N}_2}+r_{\textrm{H}_2})$}}\\
\multicolumn{2}{l}{\footnotesize{b: $5/12(r_{\textrm{H}_2}+r_{\textrm{C}_2}+r_{\textrm{N}_2})-0.55$}}\\
\multicolumn{2}{l}{\footnotesize{c: linear average of CH and CH$_4$}}\\
\end{tabular}
\end{table*}


The calculated gas kinetic rates can then be considered for placing a physical limit to the forward $K_f$ and reverse $K_b$ rates. We can for example consider a Zeldovich-type reaction AB+CD$\rightleftarrows$AC+BD with forward $K_f$ and reverse $K_b$ rates. We can then place two limiting caps to these rates in the form of:

\begin{align}
K_f\leq Z_{AB-CD},\\
K_b\leq Z_{AC-BD}.
\end{align}

However, as we further have $K_f/K_b=K_{eq}$, we may simply write:

\begin{equation}
K_f\leq\left\{Z_{AB-CD} , K_{eq}\times Z_{AC-BD}\right\},
\label{eq:ZcheckKf}
\end{equation}

with a similar treatment applied for the other type of neutral reactions (dissociation, recombination, etc...)\footnote{For reactions involving collisions of the type $A+M$, where $M$ is an arbitrary gas molecule, we consider the interaction diameter of N$_2$.}. 

\subsection{Verification of the G\"{o}k\c{c}en Neutral Rates}

We have carried the verification outlined in Eq. \ref{eq:ZcheckKf}, for the overall 28 neutral rates of the G\"{o}k\c{c}en dataset in Table \ref{tab:Titan_set}. Most of the obtained rates have been found to be physically consistent in the overall T=100--100,000\kelvin\ range. Nevertheless, in the case of 8 specific reactions, inconsistencies have been encountered, mandating the update of the corresponding kinetic rates. In the selection of the appropriate rates, care has been exerted in selecting new physically consistent rates which match the G\"{o}k\c{c}en rates in its validity range (2,000--3,500\kelvin). Replacement rates have been mostly looked up in the NIST chemistry web-book \cite{NIST}.\bigskip

\begin{figure}[!htbp]
\centering
\includegraphics[width=.6\textwidth]{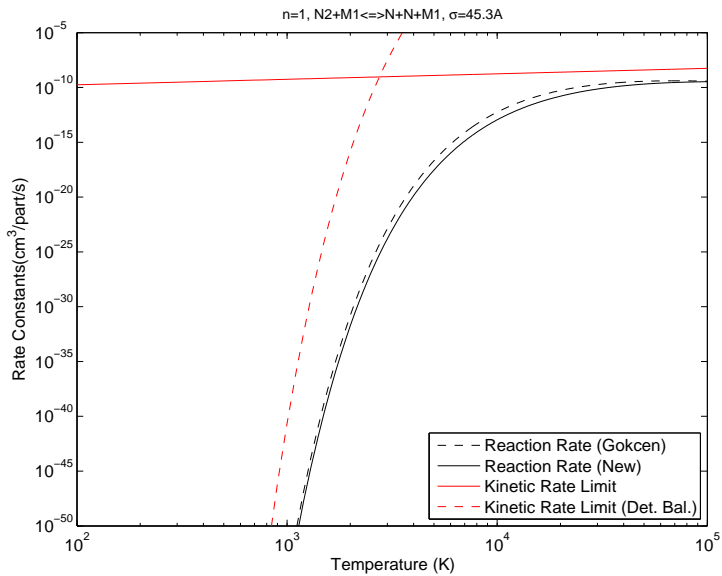}
\caption{\small{Update of rate 1}}%
\label{fig:rate01}
\end{figure}

\begin{figure}[!htbp]
\centering
\includegraphics[width=.6\textwidth]{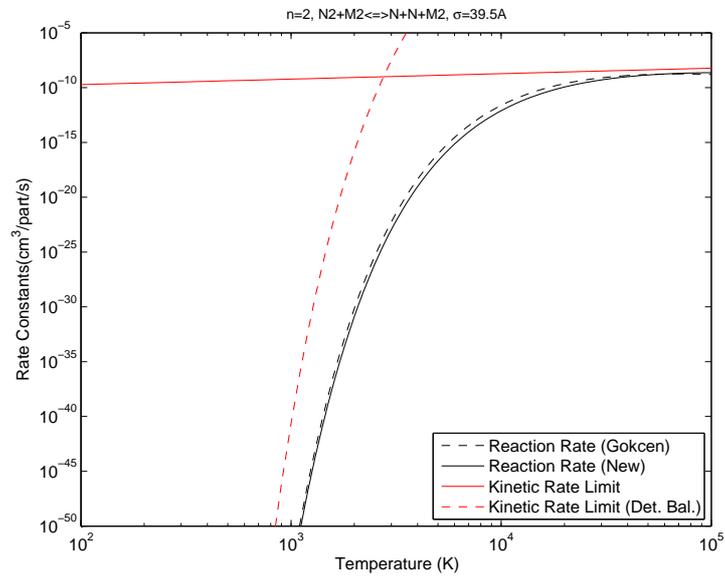}
\caption{\small{Update of rate 2}}%
\label{fig:rate02}
\end{figure}

In the case of Nitrogen dissociation rates (rates 1,2 in Table \ref{tab:Titan_set}), we have replaced the macroscopic rates of the G\"{o}k\c{c}en dataset by the sets of state-specific rates for N$_2$--N$_2$ and N$_2$--N collisions, respectively obtained considering our FHO model \cite{LinodaSilva:2007-1,LinodaSilva:2007-2,LinodaSilva:2009}, and the QCT models from the University of Bari \cite{Capitelli:2004,Esposito:2006}. However, upon verification of the rates proposed by G\"{o}k\c{c}en, we verify that these tend to peak around 70,000\kelvin\ slightly decreasing afterwards. This unphysical behavior results of the extrapolation of such rates largely outside of their initial range for the temperature fit, and it is recommended that such rates are superseded by the corresponding mono-temperature rates derived from the state specific FHO and QCT rates already utilized in previous works. Figs \ref{fig:rate01} and \ref{fig:rate02} show the comparison between these rates and the former rates from the G\"{o}k\c{c}en dataset, along with the gas-kinetic limits of Eq. \ref{eq:ZcheckKf}.\bigskip

\begin{figure}[!htbp]
\centering
\includegraphics[width=.6\textwidth]{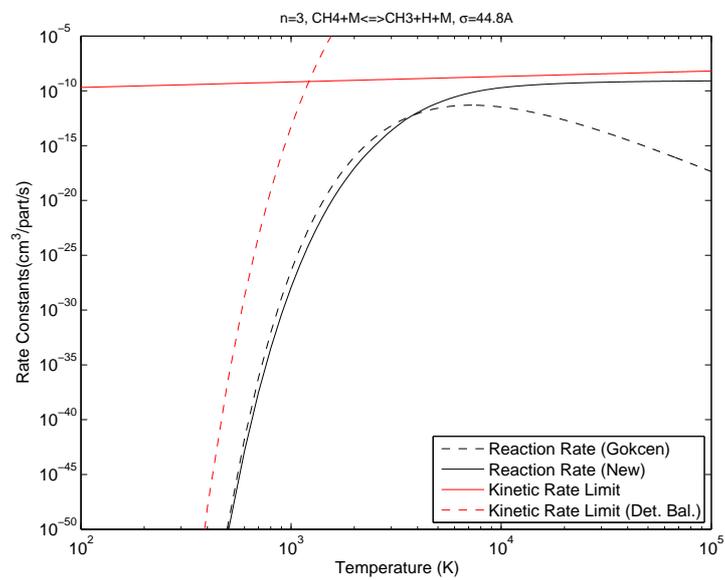}
\caption{\small{Update of rate 3}}%
\label{fig:rate03}
\end{figure}

The examination of rate 3 (CH$_4$+M$\rightarrow$CH$_3$+H+M) shows an unrealistic peak at 7,000\kelvin, with a marked decrease at higher temperatures. This leads to a severe underestimation of the CH$_4$ dissociation rates (2 orders of magnitude at 10,000\kelvin, 8 orders of magnitude at 100,000\kelvin). Most of the other rates in the literature showed similar high-temperature behavior, so we chose the detailed balancing of the measured rate for CH$_3$+H+M recombination, as proposed by Warnatz \cite{Warnatz:1984}. The comparison for the two rates and the limiting kinetic rates is presented in Fig. \ref{fig:rate03}.\bigskip

\begin{figure}[!htbp]
\centering
\includegraphics[width=.6\textwidth]{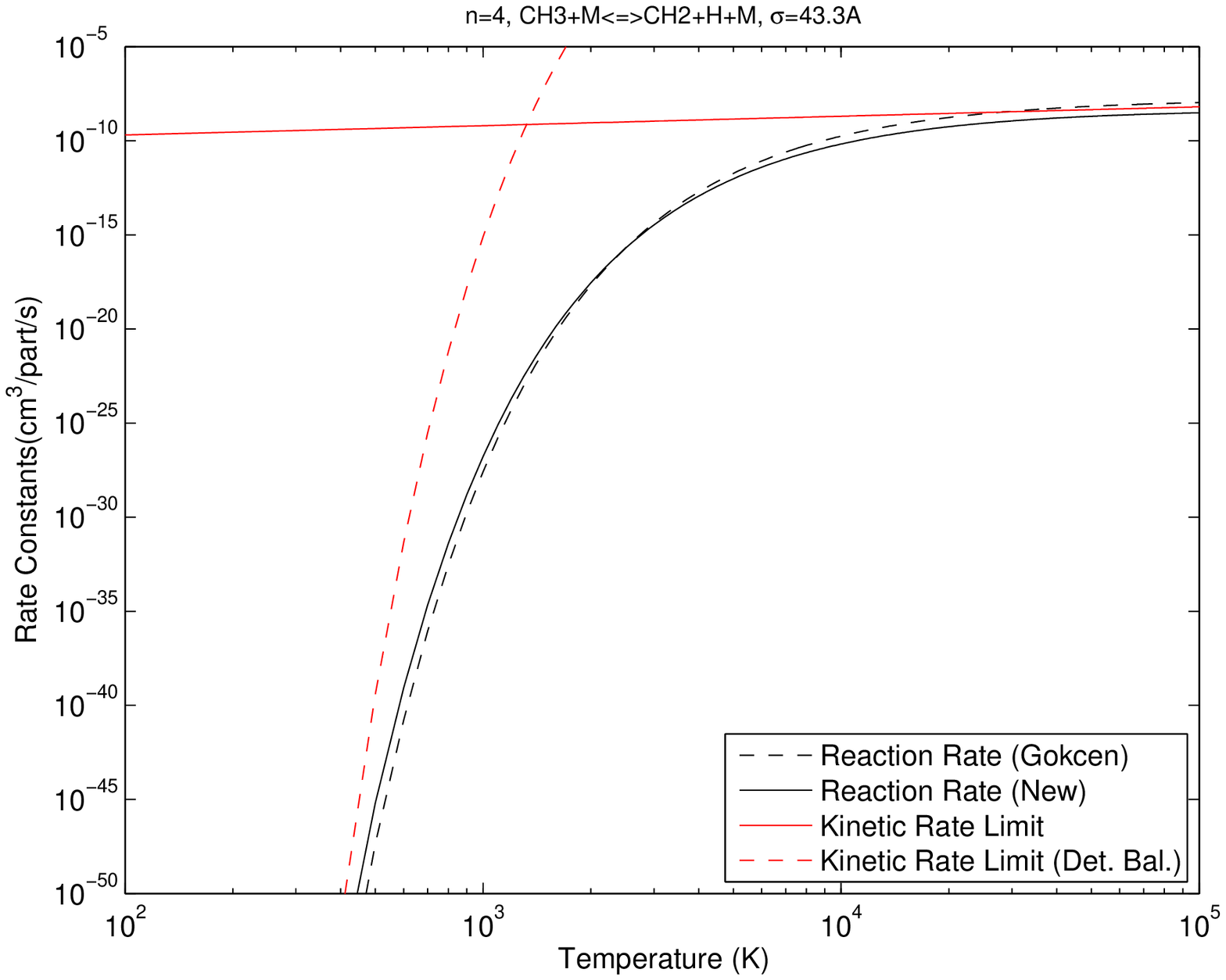}
\caption{\small{Update of rate 4}}%
\label{fig:rate04}
\end{figure}

For the case of rate 4 (dissociation of CH$_3$; CH$_3$+M$\rightarrow$CH$_3$+H+M), the rate slightly exceeds the gas kinetic rate limit at high temperatures. Accordingly, the rate has been replaced by the rate proposed by Lim \cite{Lim:1994}, which has a more satisfactory high temperature behavior (see Fig. \ref{fig:rate04}).\bigskip

\begin{figure}[!htbp]
\centering
\includegraphics[width=.6\textwidth]{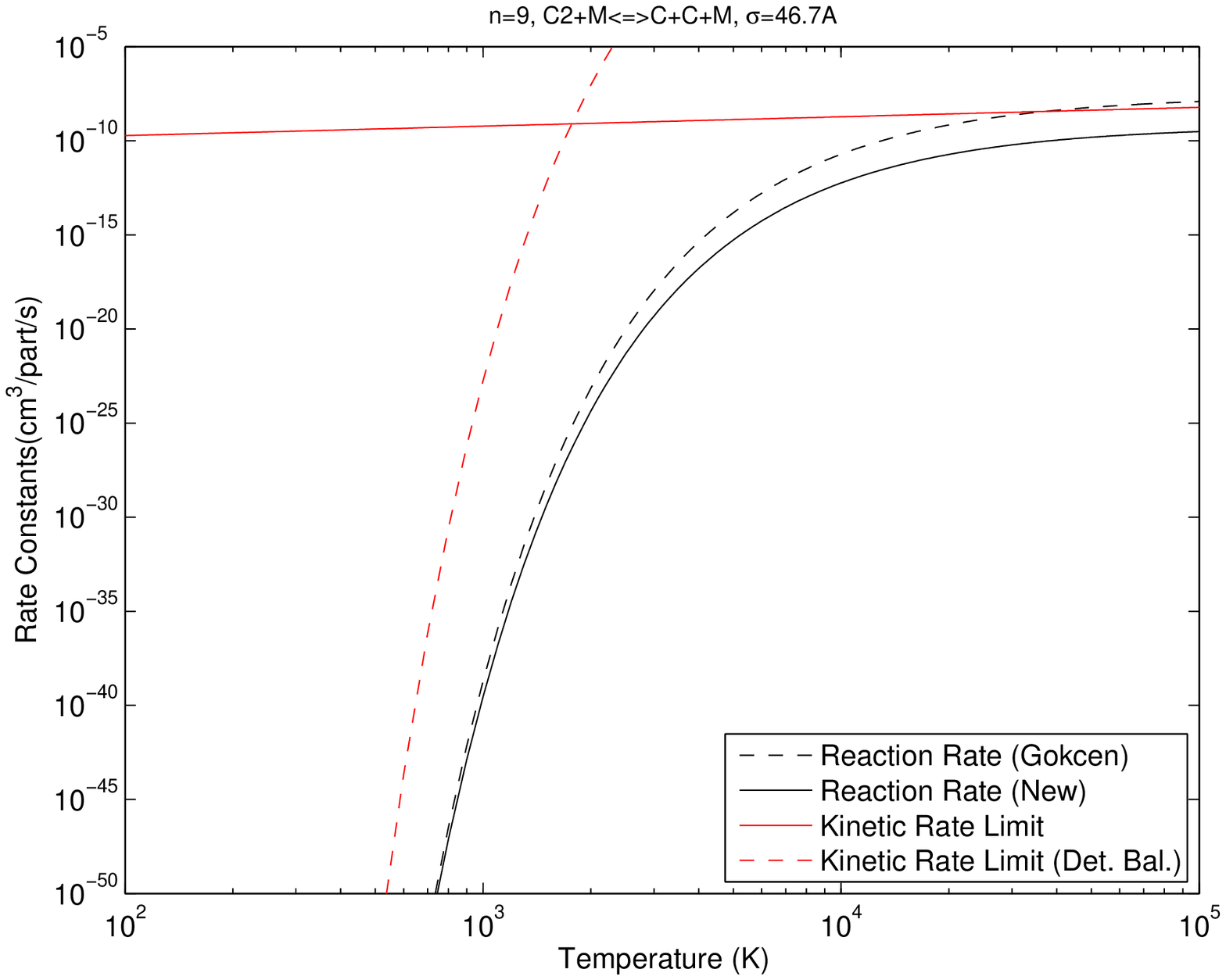}
\caption{\small{Update of rate 9}}%
\label{fig:rate09}
\end{figure}

The similar situation occurs for rate 9 (dissociation of C$_2$; C$_2$+M$\rightarrow$C+C+M), and the G\"{o}k\c{c}en dissociation rate has been replaced by the rate proposed by Beck \cite{Beck:1975} (see Fig \ref{fig:rate09}).\bigskip

\begin{figure}[!htbp]
\centering
\includegraphics[width=.6\textwidth]{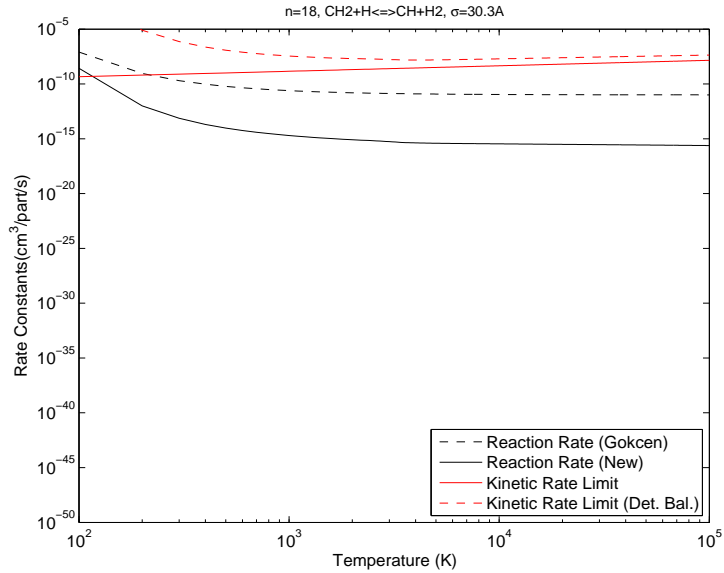}
\caption{\small{Update of rate 18}}%
\label{fig:rate18}
\end{figure}

Rate No. 18 (CH$_2$+H$\rightarrow$CH+H$_2$) is exothermic, and the characteristic temperature of the G\"{o}k\c{c}en rate ($\theta$=$-$900\kelvin) is consistent with the balance of the formation enthalpies for the species involved in the reaction (386.39 \kilo\joule\per\mole, 218 \kilo\joule\per\mole, 0 \kilo\joule\per\mole, and 594.13 \kilo\joule\per\mole\ respectively for CH$_2$, H, H$_2$, CH, yielding an activation energy of $-$1233\kelvin). However, the comparison with the gas kinetic rates (see Fig. \ref{fig:rate18}) shows that the G\"{o}k\c{c}en rate exceeds the limiting gas-kinetic rate at lower temperatures (below 200\kelvin). It has been replaced by the detailed balancing for the rate from Rohrig \cite{Rohrig:1997}, obtained in the temperature range T=2,200--2,600\kelvin). The obtained rate remains physically consistent at lower temperatures, down to almost 100\kelvin\ (see Fig. \ref{fig:rate18} for a comparison with the former G\"{o}k\c{c}en rate). The Arrhenius fit to the new rate yields an activation energy of $-$1,560\kelvin, showing that the pre-exponential factor $A$ of the former G\"{o}k\c{c}en rate was over-estimated.\bigskip

\begin{figure}[!htbp]
\centering
\includegraphics[width=.6\textwidth]{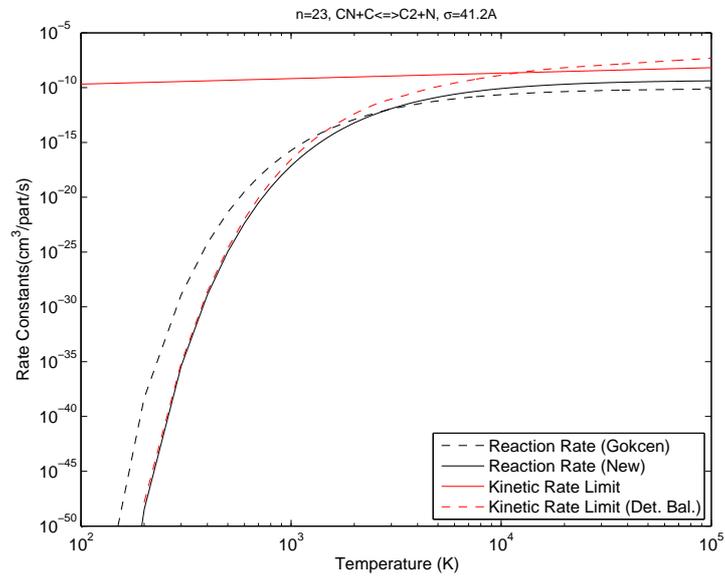}
\caption{\small{Update of rate 23}}%
\label{fig:rate23}
\end{figure}

In the case of reaction 23 (CN+C$\rightarrow$C$_2$+N), the rate exceeds the detailed balance of the reverse gas-kinetic rate, which means that the reverse reaction will be overestimated in the lower temperature ranges (in a way similar to rate 18). The rate has been replaced by the one proposed by Slack \cite{Slack:1976}. Fig. \ref{fig:rate23} shows a comparison between these different rates.\bigskip

\begin{figure}[!htbp]
\centering
\includegraphics[width=.6\textwidth]{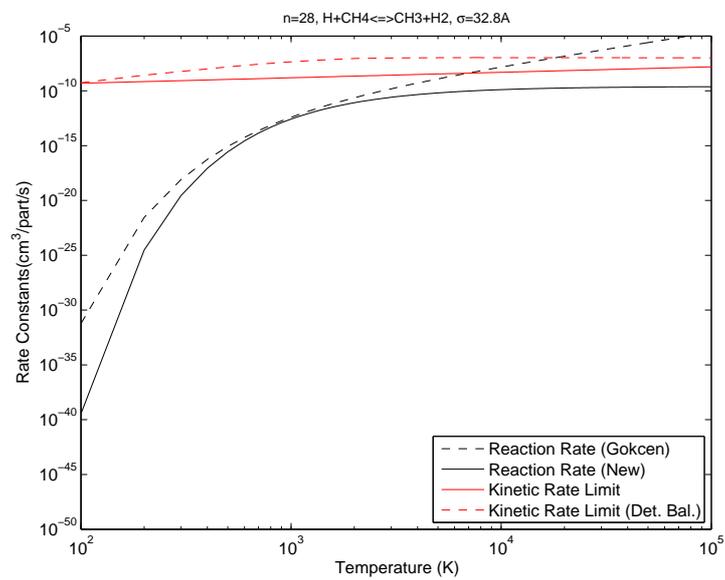}
\caption{\small{Update of rate 28}}%
\label{fig:rate28}
\end{figure}

The rate for reaction 28 (H+CH$_4\rightarrow$CH$_3$+H$_2$) has abnormally high values in the high-temperature limit, and has been replaced by the rate proposed by Bryukov \cite{Bryukov:2001}. Fig. \ref{fig:rate28} shows a comparison between the different rates.\bigskip

\section{Update of the G\"{o}k\c{c}en Ionized Reactions Dataset}
\label{sec:ions}

As discussed in the introduction, an entirely new dataset for ionized reactions has been produced in this work, given that G\"{o}k\c{c}en \cite{Gokcen:2005} didn't update the previous ionized rates of Nelson \cite{Nelson:1991}. The need for update of this rates has been further emphasized by preliminary calculations of high-speed N$_2$--CH$_4$ flows carried out by the authors, whose results evidenced peculiar behaviors directly linked to the evolution of the ionized species concentrations. Notably, simulations evidenced several local maxima for the CN Violet system radiation, a behavior which has not been found in any of the experimental data available from the literature.\bigskip

The verification of ionized rates is less straightforward than for the case of neutral rates. Indeed, the complex interactions between electrostatic forces precludes a straightforward definition of an interaction radius for ionized species. Fortunately, a large wealth of experimental works exists, who propose energy-dependent cross-sections for a large number of ionized reactions. A first task is to determine the requirement for the maximum range of the published cross-sections, so that the integration of cross-sections with the Maxwellian distribution function includes a majority of the velocity probabilities:

\begin{equation}
\label{eq:X2R}
K(T)=\int_{0}^{\infty}E\ \sigma(E)f(E,T)dE\backsimeq \int_{0}^{E_{max}}E\ \sigma(E)f(E,T)dE
\end{equation}

\begin{figure}[!htbp]
\centering
\includegraphics[width=.6\textwidth]{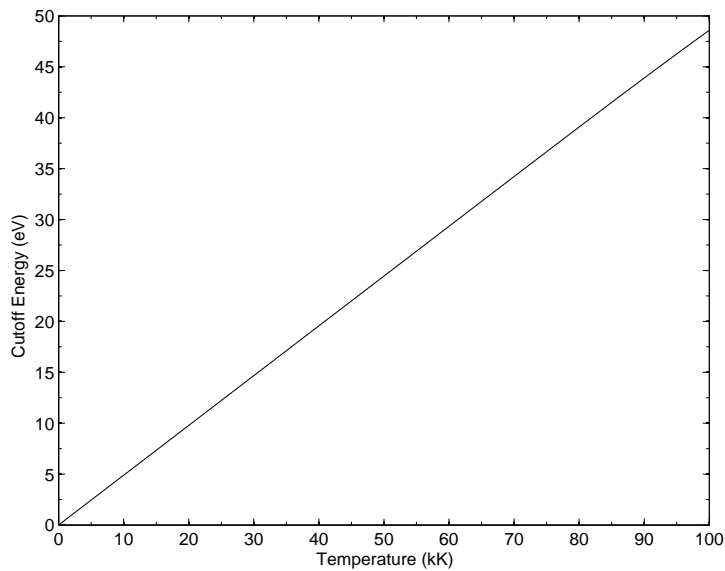}
\caption{\small{Maximum energy needed for containing 99\% of the temperature-dependent Maxwellian probability distribution function}}%
\label{fig:maxwell}
\end{figure}

In this work, we consider that the cutoff for the cross-section $\sigma(E)$ may be carried out at 99\% of the energy for the Maxwellian distribution function. This will naturally depend on the temperature $T$ under consideration. Fig. \ref{fig:maxwell} plots the temperature-dependence of this cutoff value.\bigskip

For $T=100,000$\kelvin, the cutoff value is $E=49\electronvolt$, which sets the minimum enforced range ($E=0-49\electronvolt$ for the published cross-sections selected for this work.\bigskip

\begin{figure}[!htbp]
\centering
\includegraphics[width=.6\textwidth]{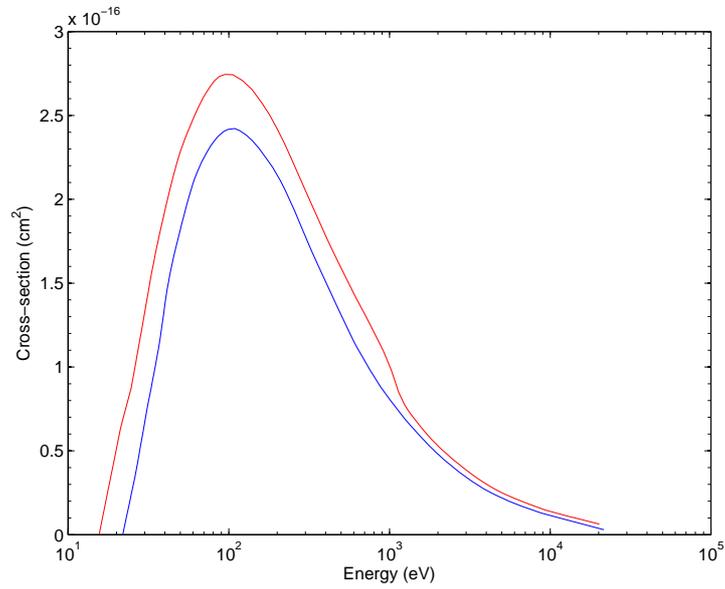}
\caption{\small{Minimum and maximum bounds for the published cross-sections for electron-impact N$_2$ ionization.}}%
\label{fig:N2X_minmax}
\end{figure}

A second important topic is what is the accuracy that might be expected to be achieved from the different published cross-sections. One first comment can be brought in that an extensive effort has been produced in providing such cross-sections, either by experimental or numerical methods. The second comment is that there is a significant amount of published cross-sections which can then be compared for obtaining a general idea of the scattering of these several cross-sections. As an example, we may examine the review of the different published electron-impact ionization cross-sections, as carried out in Ref. \cite{Deutsch:2000}, Fig. 14. Ten different cross-sections are reported, providing lower and upper-bounds for all the calculations that have been carried for this reaction. These are reported in Fig. \ref{fig:N2X_minmax}.\bigskip

\begin{figure}[!htbp]
\centering
\includegraphics[width=.6\textwidth]{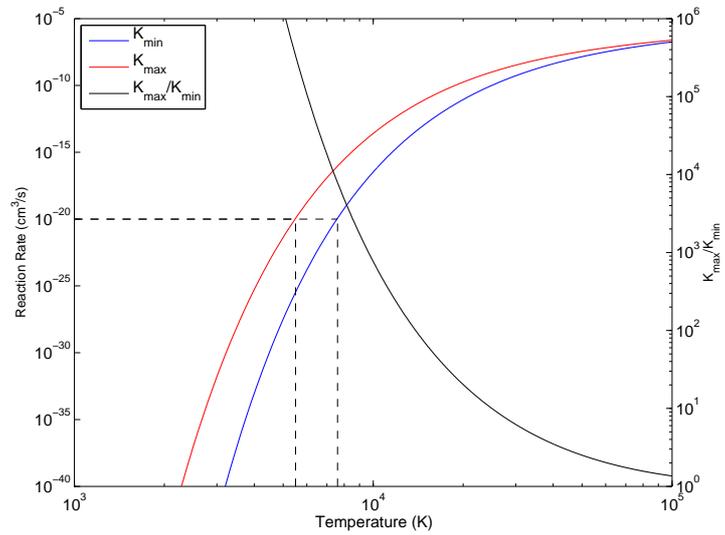}
\caption{\small{Corresponding rates for the cross-sections of Fig. \ref{fig:N2X_minmax}.}}%
\label{fig:N2R_minmax}
\end{figure}

Upon integration of Eq. \ref{eq:X2R}, the minimum and maximum bounds for the rates are then obtained (see Fig. \ref{fig:N2R_minmax}).\bigskip

\begin{figure}[!htbp]
\centering
\includegraphics[width=.6\textwidth]{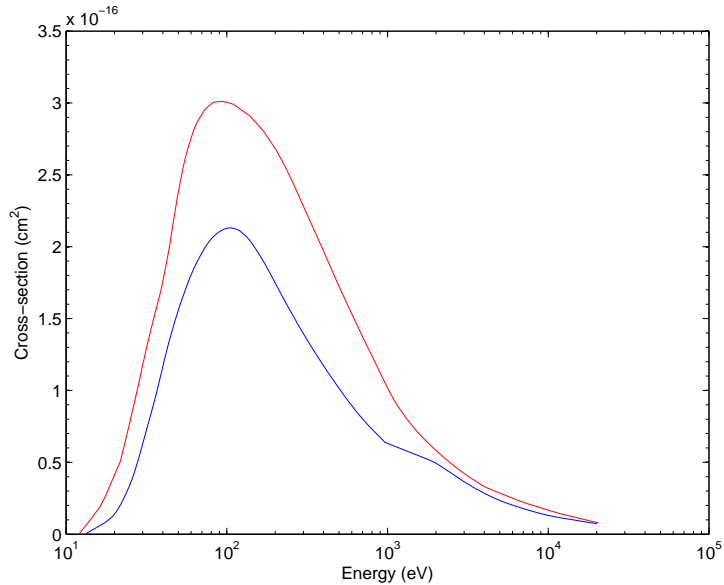}
\caption{\small{Minimum and maximum bounds for the published cross-sections for electron-impact O$_2$ ionization.}}%
\label{fig:O2X_minmax}
\end{figure}

\begin{figure}[!htbp]
\centering
\includegraphics[width=.6\textwidth]{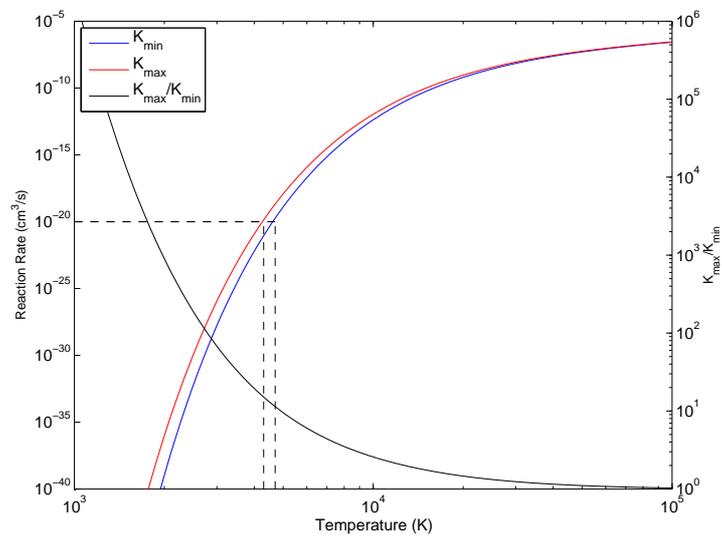}
\caption{\small{Corresponding rates for the cross-sections of Fig. \ref{fig:O2X_minmax}.}}%
\label{fig:O2R_minmax}
\end{figure}

Fairly close cross-sections are produced. However, the location of the cross-section threshold has a big impact on the lower-temperature rate (as it would be expectable). If we arbitrarily define 10$^{-20}$ as the minimum threshold where the rate is ``relevant'', we verify that we have a 5 orders of magnitude difference (vertically) or 2000\kelvin\ difference (horizontally) in the two rates. This comparison is however slightly skewed, as only one of the ten rates has a threshold which significantly differs from the others, meaning that the calculated cross-sections may probably be discarded. We have therefore carried out the same comparison for electron-impact ionization reactions of O$_2$ (Fig. 15 of Ref. \cite{Deutsch:2000}). The minimum and maximum bounds for the ionization cross-sections are presented in Fig. \ref{fig:O2X_minmax}, and the corresponding rates are presented in Fig. \ref{fig:O2R_minmax}.\bigskip

The rates are fairly similar, although the cross-section minimum and maximum bound differ more in the case of O$_2$ (Fig. \ref{fig:O2X_minmax}) than in the case of N$_2$ (Fig. \ref{fig:N2X_minmax}). However, at the position where the rates are equivalent to $K=10^{-20}$, we have less than an order of magnitude difference (vertically) or 200\kelvin\ difference (horizontally). This means that differences in the obtained rates will be essentially derived from the corresponding cross-sections thresholds, but also than most of the data available in the literature is accurate enough than most of the produced ionized rates can be considered to be less than an order of magnitude accurate in their range of relevance ($K>10^{-20}$\centi\metre\cubed\per\second).

\subsection{Literature Study of Available Cross-Sections, and Production of the Corresponding Rates}

Here we will summarize the rates obtained from a literature study of the published cross-sections. For the sake of brevity we will only cite the reference of the selected cross-section. This is generally selected from the latest published works, generally taking into account agreement with previous studies, as discussed in the cited references.

\clearpage

\subsubsection{Associative Ionization/Dissociative Recombination Reactions}

Here we will discuss selected rates of the type

\begin{equation}
\textrm{A\ +\ B}\ \rightleftarrows\ \textrm{AB}^{+}\ +\ \textrm{e}^{-}
\end{equation}

For such rates, the literature exclusively proposes cross-sections for the dissociative recombination reaction $\textrm{AB}^{+}+\textrm{e}^{-}\rightarrow\textrm{A\ +\ B}$. The reverse rate has then to be determined through the detailed balance principle.\bigskip

Several of the electron-ion dissociation rates are available from the compilation of Ref. \cite{Florescu:2006}. The rates are summarized in Table \ref{tab:EIonDiss}. The reverse rate has been obtained through detailed balance.

\begin{table*}[!htbp]
\caption{Electron-Ion Dissociation Rates}%
\label{tab:EIonDiss}%
\centering%
\begin{tabular}{r l l}
\toprule%
\midrule%
\multicolumn{1}{l}{Reaction} & Rate (\centi\cubic\metre\per\second) & Reference\\%
\midrule%
H$_{2}^{+}$\ +\ e$^{-}\rightarrow$ H\ +\ H & $2.3\times 10^{-7}\times\left(300/T\right)^{0.41}$ & \cite{Florescu:2006,Auerbach:1977}\\
CH$^{+}$\ +\ e$^{-}\rightarrow$ C\ +\ H    & $1.5\times 10^{-7}\times\left(300/T\right)^{0.42}$ & \cite{Florescu:2006,Mitchell:1978}\\
C$_{2}^{+}$\ +\ e$^{-}\rightarrow$ C\ +\ C & $3.0\times 10^{-7}\times\left(300/T\right)^{0.5}$  & \cite{Florescu:2006,Mul:1980}\\
NH$^{+}$\ +\ e$^{-}\rightarrow$ N\ +\ H    & $4.3\times 10^{-8}\times\left(300/T\right)^{0.5}$  & \cite{Florescu:2006,Mul:unp}\\
\midrule%
\bottomrule%
\end{tabular}
\end{table*}

\begin{figure}[!htbp]
\centering
\includegraphics[width=.6\textwidth]{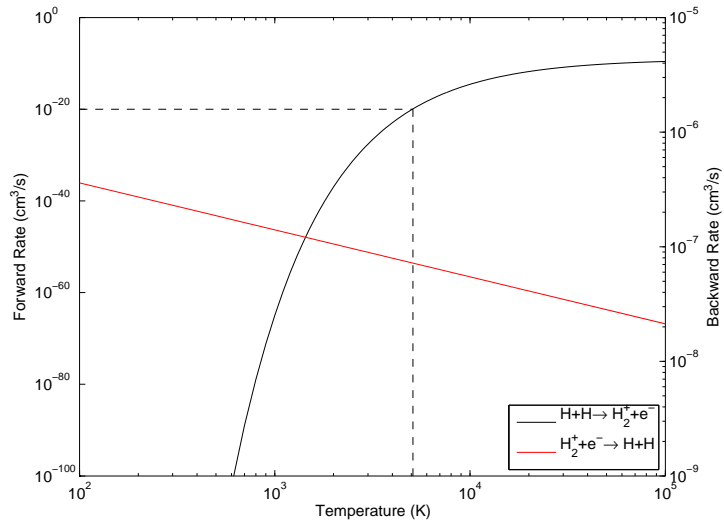}
\caption{\small{Forward and backward rates for the reaction $\textrm{H+H}\rightleftarrows\textrm{H}_{2}^{+}+\textrm{e}^{-}$. The limit where the rate reaches the value $10^{-20}$ is reported}}%
\label{fig:rateA10}
\end{figure}

Sample rates for the $\textrm{H+H}\rightleftarrows\textrm{H}_{2}^{+}+\textrm{e}^{-}$ reaction are plotted in Fig. \ref{fig:rateA10}.\bigskip

\begin{figure}[!htbp]
\centering
\includegraphics[width=.6\textwidth]{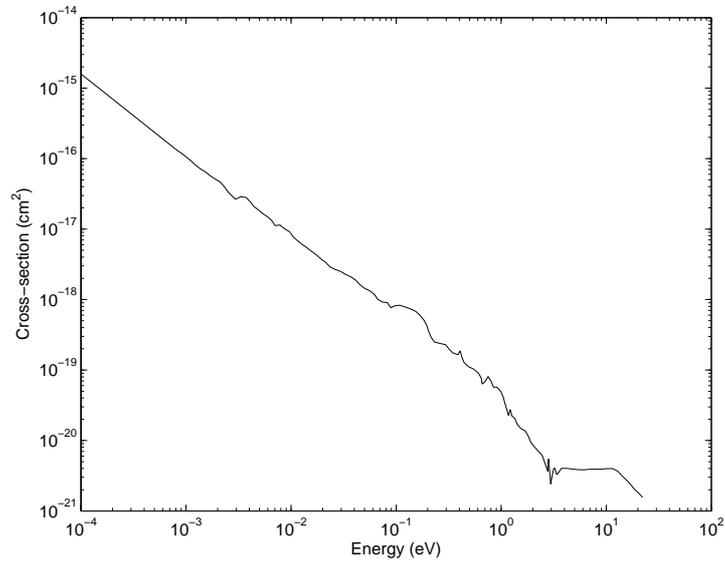}
\caption{\small{Cross-section for the reaction $\textrm{CN}^{+}+\textrm{e}^{-}\rightarrow\textrm{C+N}$}}%
\label{fig:xsec30}
\end{figure}

\begin{figure}[!htbp]
\centering
\includegraphics[width=.6\textwidth]{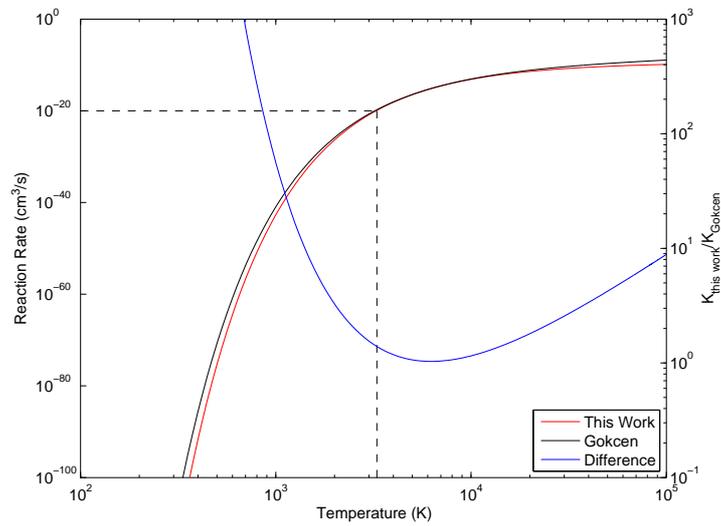}
\caption{\small{Forward rate for the reaction $\textrm{C+N}\rightleftarrows\textrm{CN}^{+}+\textrm{e}^{-}$, and comparison with the G\"{o}k\c{c}en rate. The limit where the rate reaches the value $10^{-20}$ is reported.}}%
\label{fig:rate30}
\end{figure}

Published cross-sections for the reaction $\textrm{CN}^{+}+\textrm{e}^{-}\rightarrow\textrm{C+N}$ are taken from Ref. \cite{LePadellec:1999}. The cross-section is plotted in Fig. \ref{fig:xsec30}. The corresponding reaction and a comparison with the reaction of the G\"{o}k\c{c}en dataset is presented in Fig. \ref{fig:rate30}.\bigskip

\begin{figure}[!htbp]
\centering
\includegraphics[width=.6\textwidth]{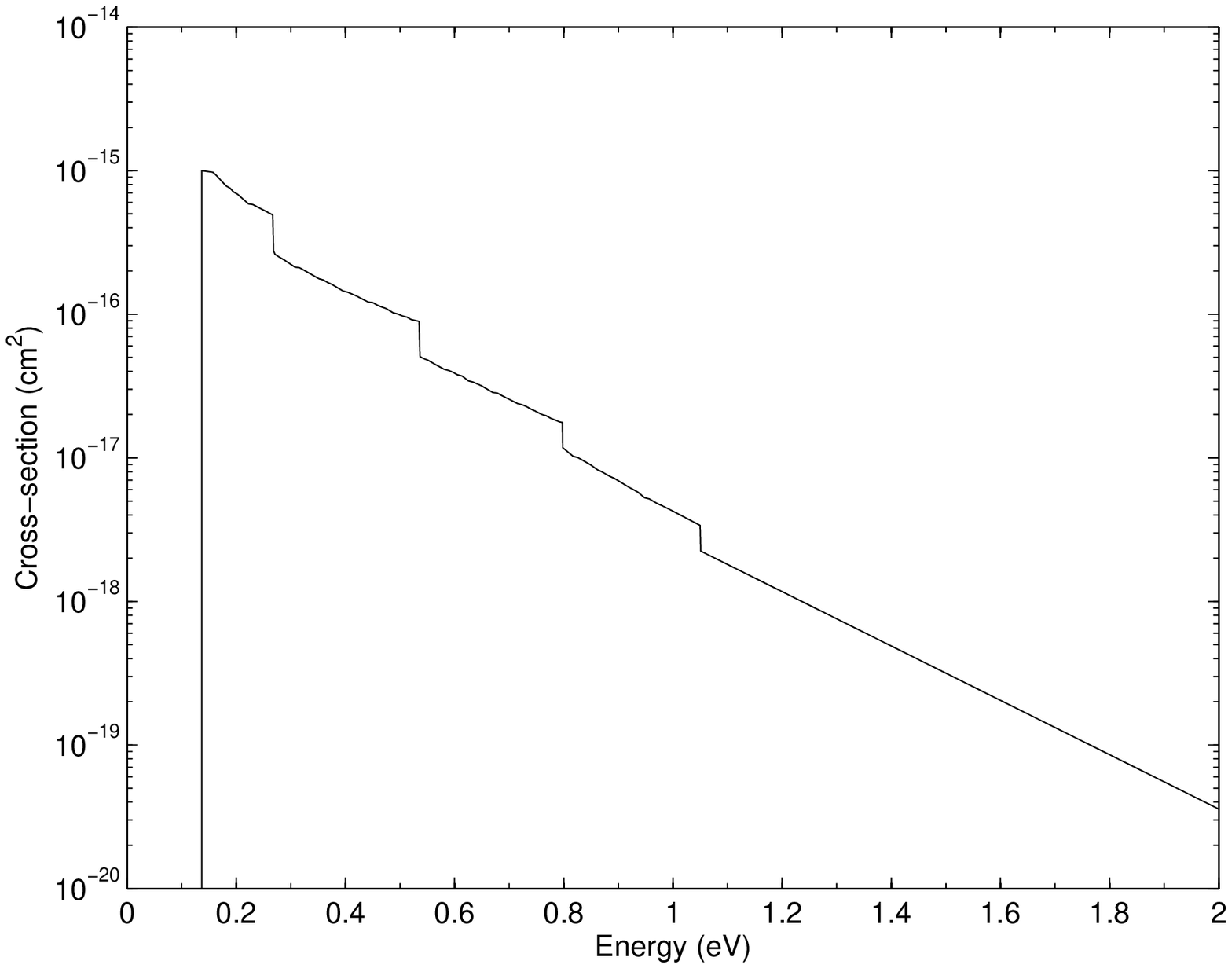}
\caption{\small{Cross-section for the reaction $\textrm{N}_2^{+}+\textrm{e}^{-}\rightarrow\textrm{N+N}$}}%
\label{fig:xsec29}
\end{figure}

\begin{figure}[!htbp]
\centering
\includegraphics[width=.6\textwidth]{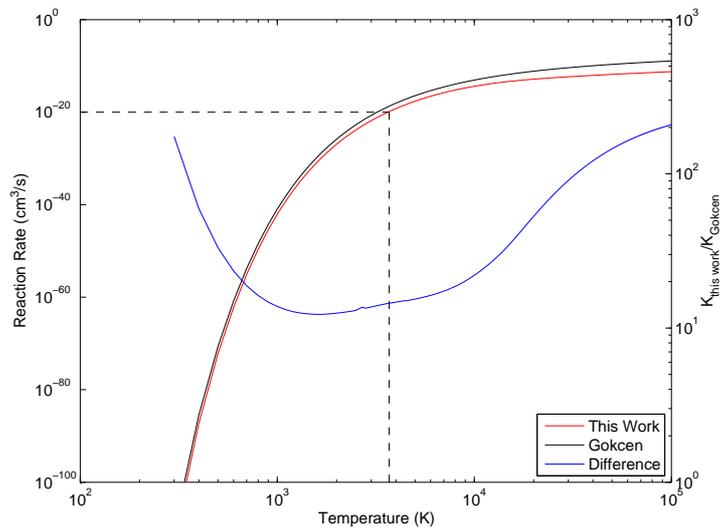}
\caption{\small{Forward rate for the reaction $\textrm{N}+\textrm{N}\rightleftarrows\textrm{N}_{2}^{+}+\textrm{e}^{-}$, and comparison with the G\"{o}k\c{c}en rate. The limit where the rate reaches the value $10^{-20}$ is reported}}%
\label{fig:rate29}
\end{figure}

Published cross-sections for the reaction $\textrm{N}_2^{+}+\textrm{e}^{-}\rightarrow\textrm{N+N}$ are taken from Ref. \cite{Guberman:2009}. The cross-section is plotted in Fig. \ref{fig:xsec29}. The corresponding reaction and a comparison with the reaction of the G\"{o}k\c{c}en dataset is presented in Fig. \ref{fig:rate29}.\bigskip

\clearpage

\subsubsection{Electron-Atom Ionization Reactions}

Here we will discuss selected rates of the type

\begin{equation}
\textrm{A}\ +\ \textrm{e}^{-}\ \rightleftarrows\ \textrm{A}^{+}\ +\ 2\textrm{e}^{-}
\end{equation}

\begin{figure}[!htbp]
\centering
\includegraphics[width=.6\textwidth]{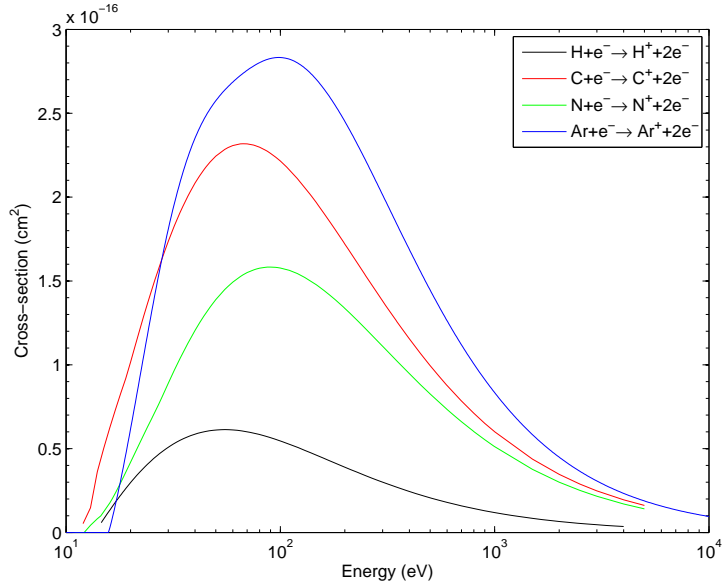}
\caption{\small{Cross-sections for the electron-impact ionization of H, C, N, and Ar}}%
\label{fig:xsec31234}
\end{figure}

\begin{figure}[!htbp]
\centering
\includegraphics[width=.6\textwidth]{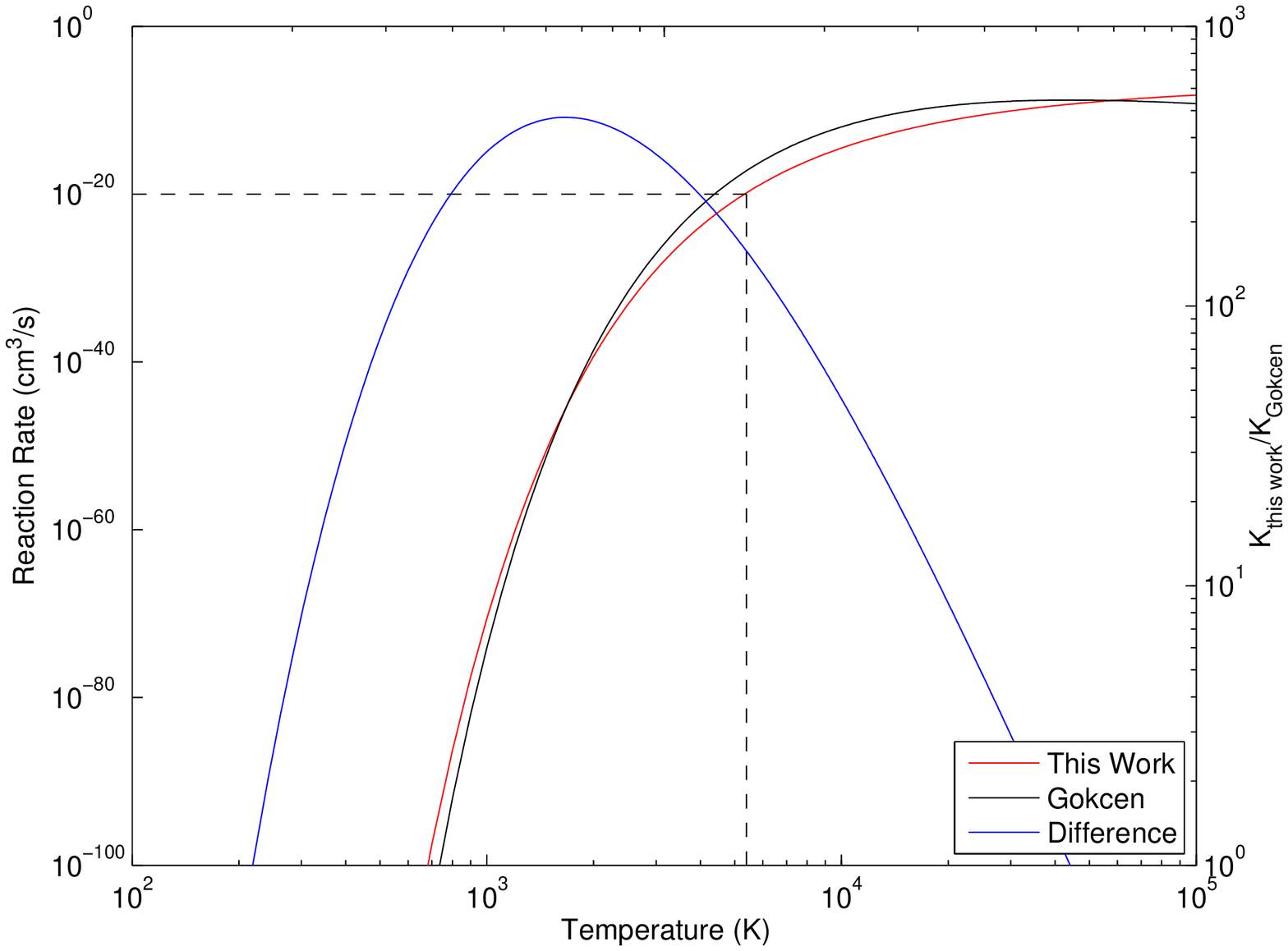}
\caption{\small{Rate for Electron-Impact Ionization of atomic H, and comparison with the G\"{o}k\c{c}en rate. The limit where the rate reaches the value $10^{-20}$ is reported.}}%
\label{fig:rate31}
\end{figure}

\begin{figure}[!htbp]
\centering
\includegraphics[width=.6\textwidth]{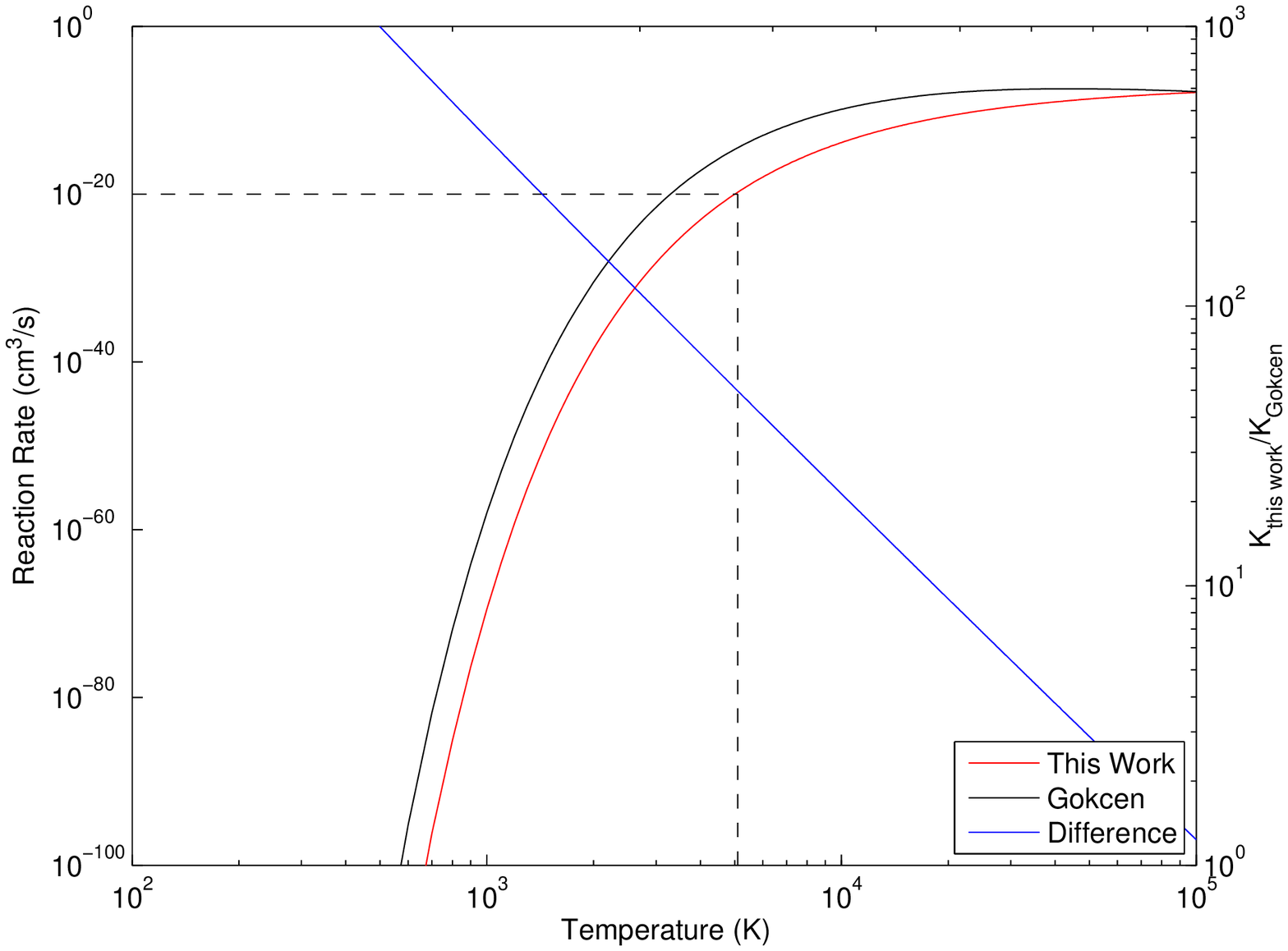}
\caption{\small{Rate for Electron-Impact Ionization of atomic C, and comparison with the G\"{o}k\c{c}en rate. The limit where the rate reaches the value $10^{-20}$ is reported.}}%
\label{fig:rate32}
\end{figure}

\begin{figure}[!htbp]
\centering
\includegraphics[width=.6\textwidth]{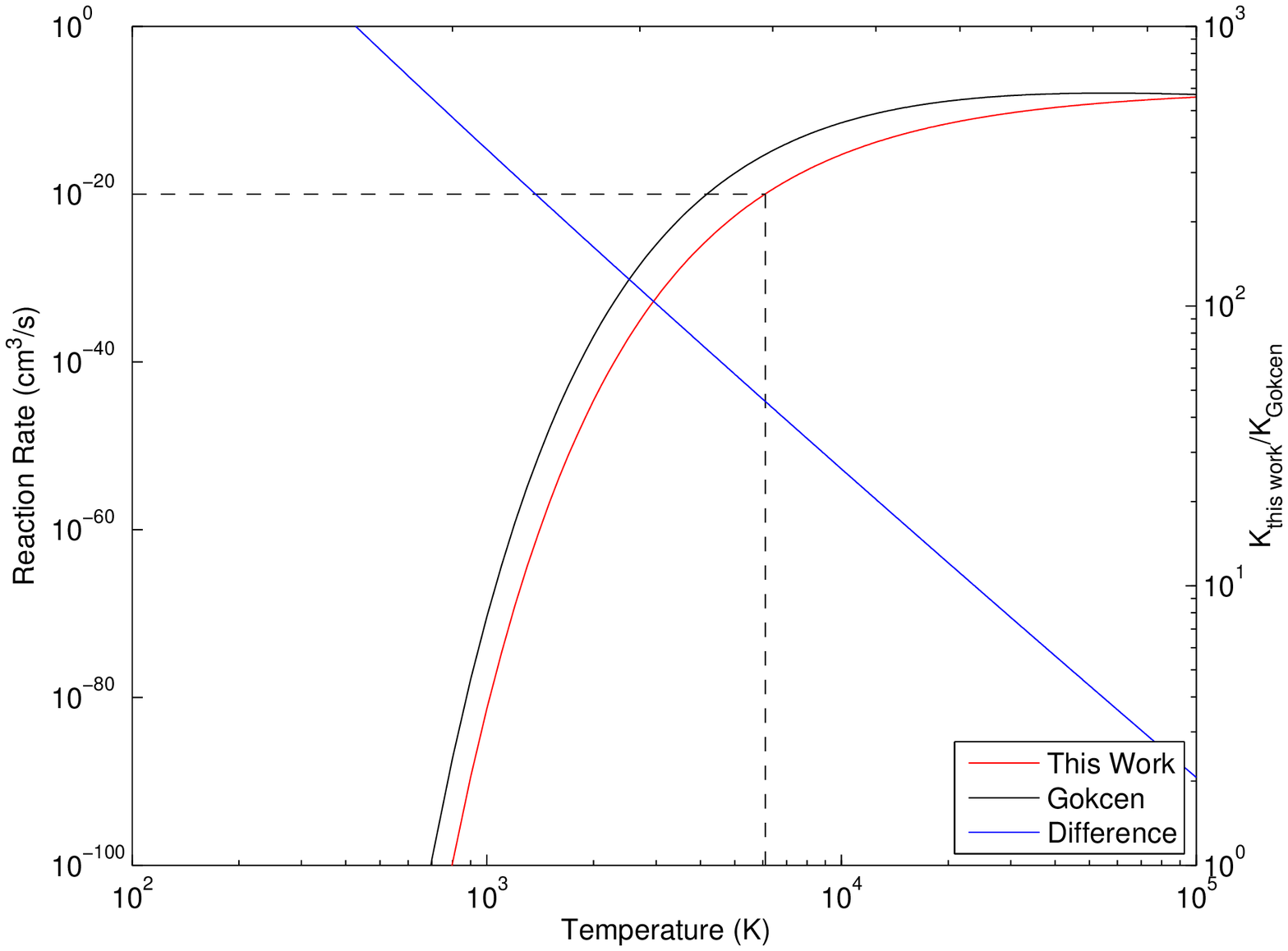}
\caption{\small{Rate for Electron-Impact Ionization of atomic N, and comparison with the G\"{o}k\c{c}en rate. The limit where the rate reaches the value $10^{-20}$ is reported.}}%
\label{fig:rate33}
\end{figure}

\begin{figure}[!htbp]
\centering
\includegraphics[width=.6\textwidth]{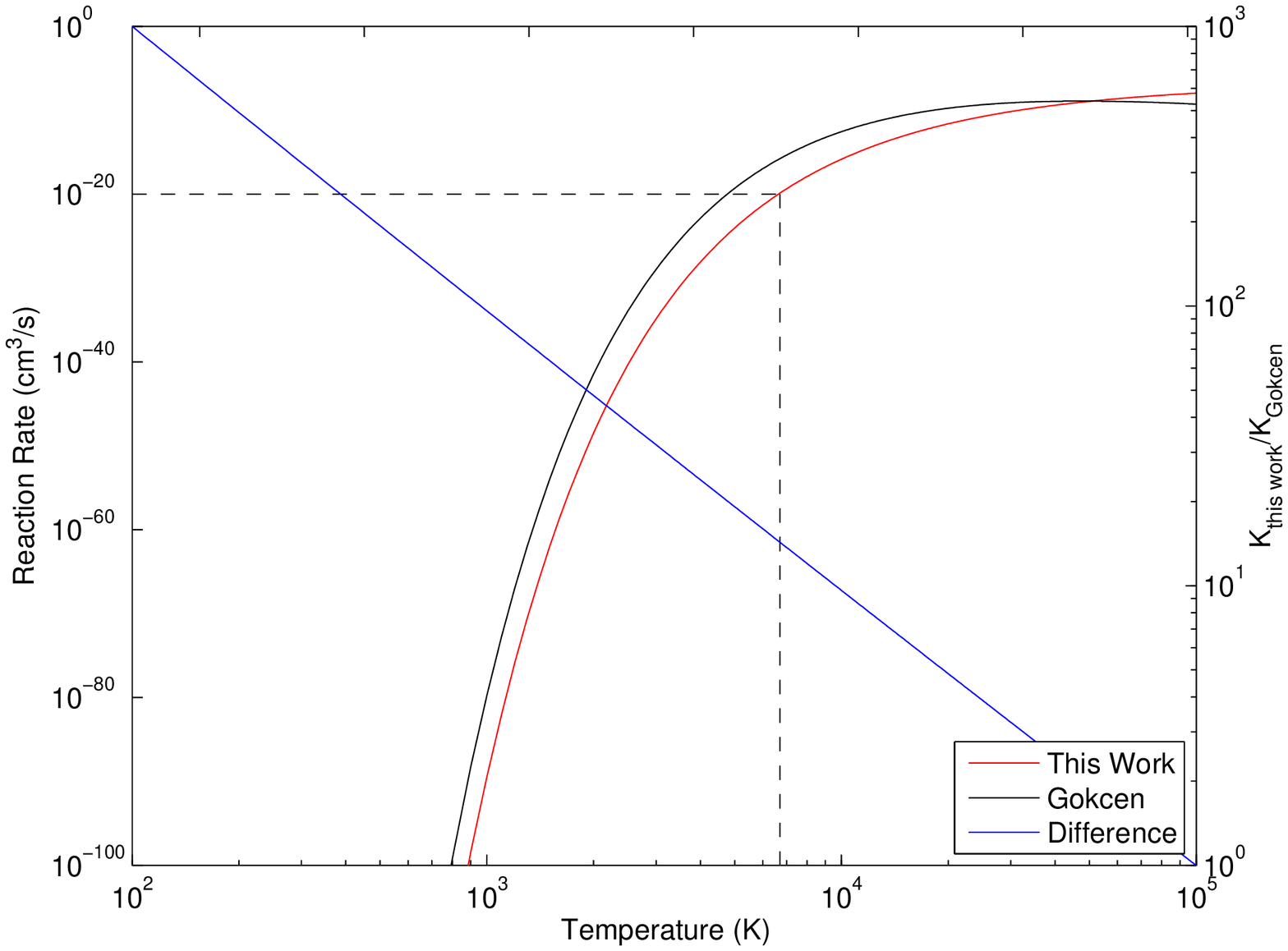}
\caption{\small{Rate for Electron-Impact Ionization of atomic Ar, and comparison with the G\"{o}k\c{c}en rate. The limit where the rate reaches the value $10^{-20}$ is reported.}}%
\label{fig:rate34}
\end{figure}

Cross-sections for electron-impact ionization reactions of H, C, N, and Ar, are respectively taken from Ref. \cite{Kim:1994}, Ref. \cite{Kim:2002}, Ref. \cite{Kim:2002}, and  Refs. \cite{Rapp:1965,Schram:1966}. They are reported in Fig. \ref{fig:xsec31234}. The comparison of the corresponding rates with the previous G\"{o}k\c{c}en dataset is presented in Figs. \ref{fig:rate31} to \ref{fig:rate34}. Contrarily to the new Associative Ionization rates, which remain close to the previous G\"{o}k\c{c}en rates, the new Electron-Atom Ionization reactions differ significantly from the previous rates by G\"{o}k\c{c}en.

\subsubsection{Electron-Molecule Ionization Reactions}

Here we will discuss selected rates of the type

\begin{equation}
\textrm{AB}\ +\ \textrm{e}^{-}\ \rightleftarrows\ \textrm{AB}^{+}\ +\ 2\textrm{e}^{-}.
\end{equation}

\begin{figure}[!htbp]
\centering
\includegraphics[width=.6\textwidth]{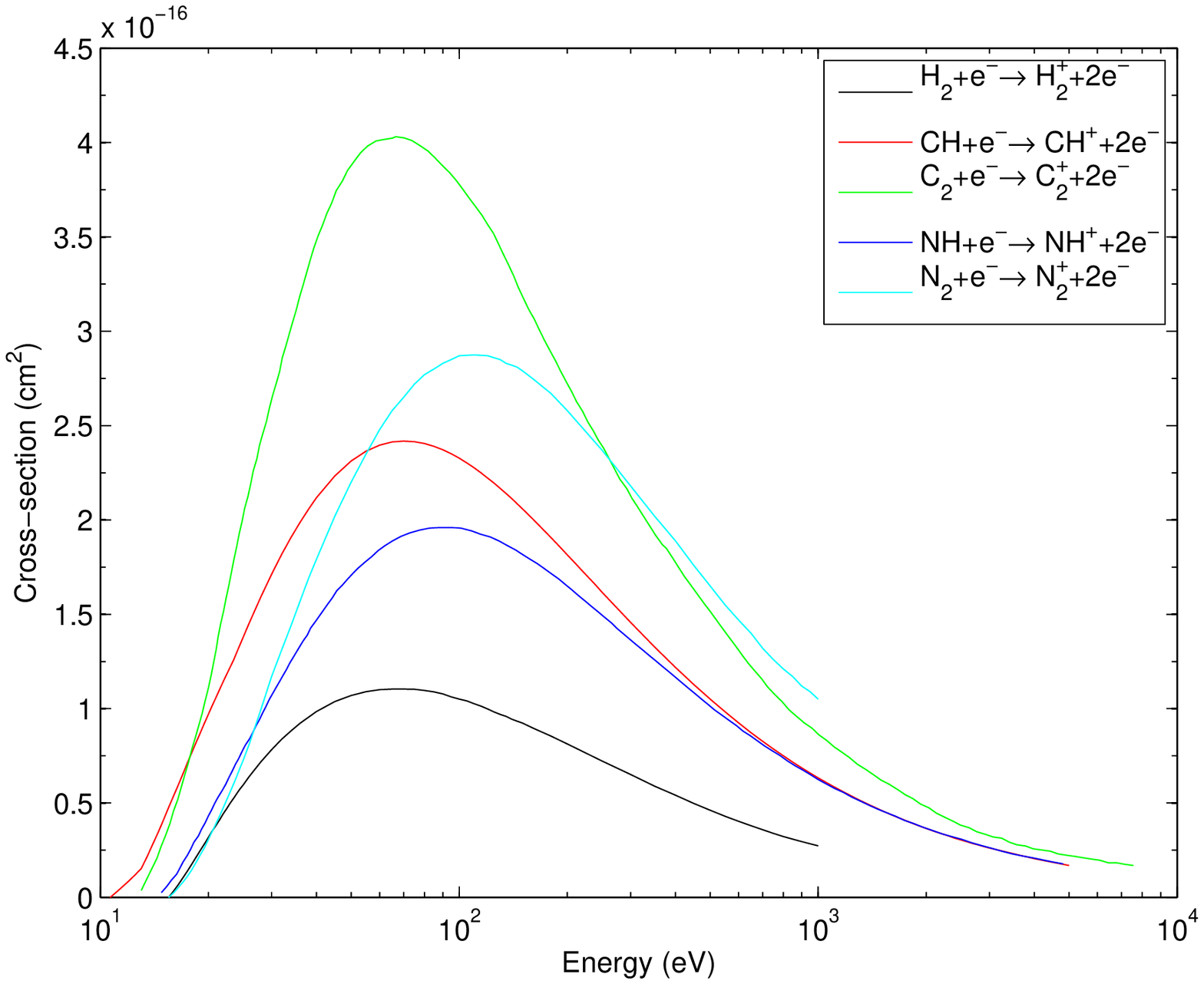}
\caption{\small{Cross-sections for the electron-impact ionization of H$_2$, CH, C$_2$, NH, and N$_2$}}%
\label{fig:xsecA26}
\end{figure}

Cross-sections for the ionization reactions of H$_2$, CH, C$_2$, NH, and N$_2$ have been respectively obtained from Refs. \cite{Rapp:1965}, \cite{Kim:2000}, \cite{Deutsch:2000}, \cite{Rajvanshi:2010}, and \cite{Rapp:1965}. No data for the electron-impact ionization of CN has been found. These rates were not initially included in the G\"{o}k\c{c}en dataset for ionized reactions. The cross-sections for these different molecules are reported in Fig. \ref{fig:xsecA26}.

\subsubsection{Other Ionized Reactions}

\begin{figure}[!htbp]
\centering
\includegraphics[width=.6\textwidth]{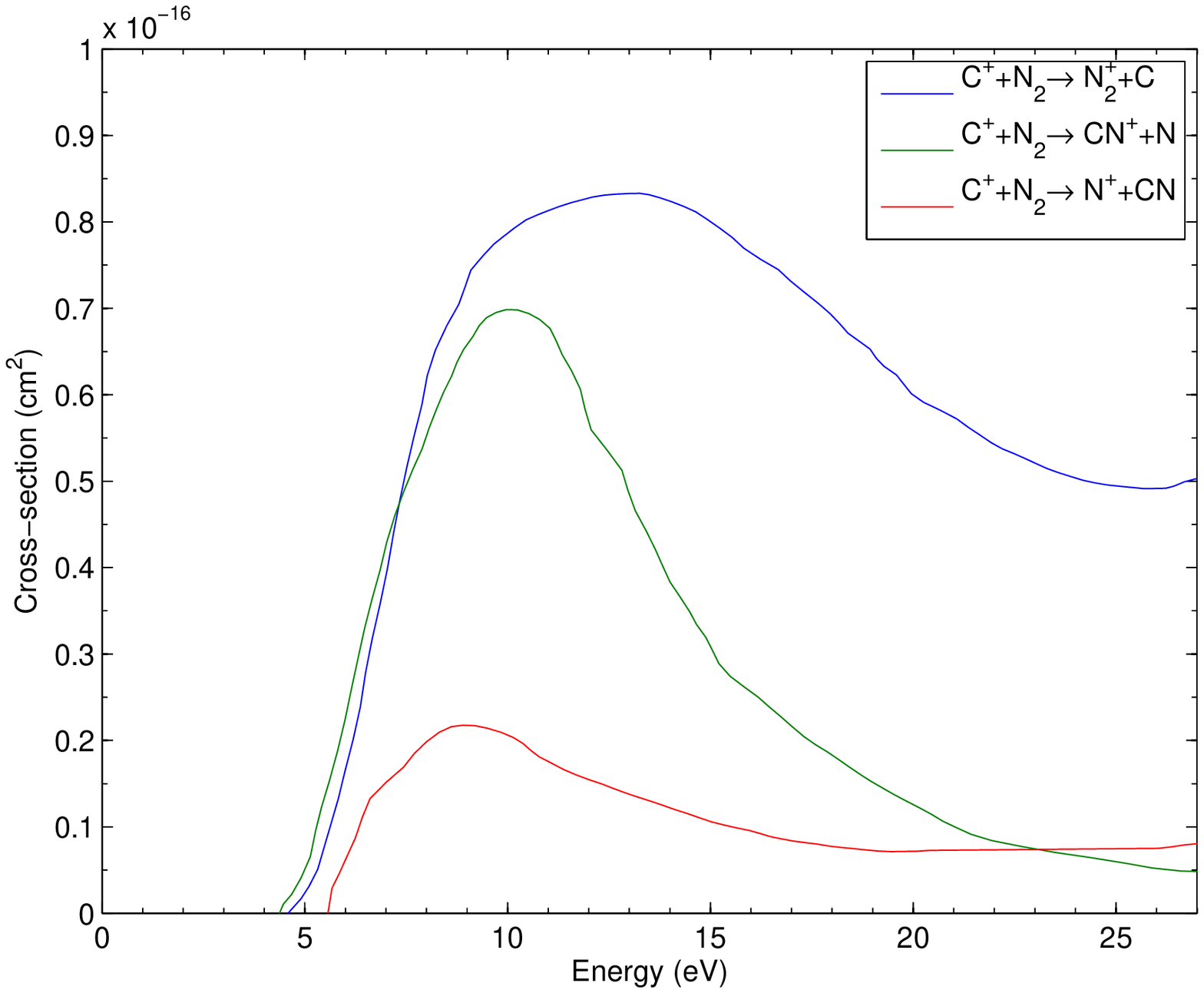}
\caption{\small{Cross-section for the reactions $\textrm{C}^++\textrm{N}_2\rightarrow \textrm{N}_2^++\textrm{C}$, $\textrm{C}^++\textrm{N}_2\rightarrow \textrm{CN}^++\textrm{N}$, and $\textrm{C}^++\textrm{N}_2\rightarrow \textrm{N}^++\textrm{CN}$}}%
\label{fig:xsec36}
\end{figure}

\begin{figure}[!htbp]
\centering
\includegraphics[width=.6\textwidth]{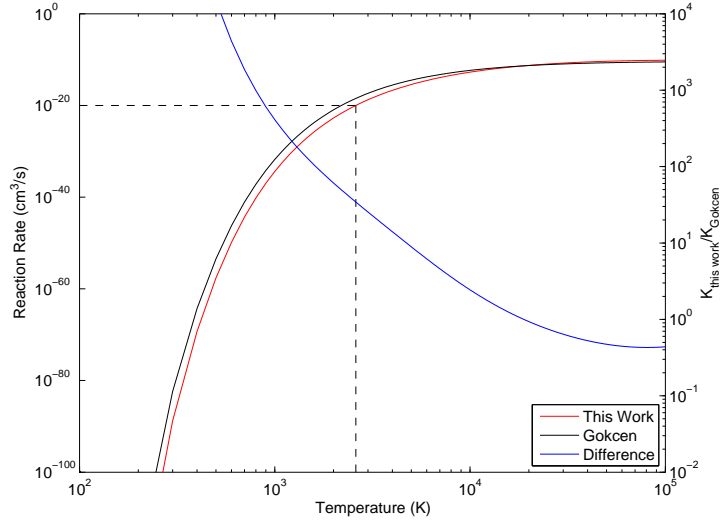}
\caption{\small{Forward rate for the reaction $\textrm{C}^++\textrm{N}_2\rightleftarrows \textrm{N}_2^++\textrm{C}$, and comparison with the G\"{o}k\c{c}en rate. The limit where the rate reaches the value $10^{-20}$ is reported.}}%
\label{fig:rate36}
\end{figure}

Cross-sections for the reactions $\textrm{C}^++\textrm{N}_2\rightarrow \textrm{N}_2^++\textrm{C}$, $\textrm{C}^++\textrm{N}_2\rightarrow \textrm{CN}^++\textrm{N}$, and $\textrm{C}^++\textrm{N}_2\rightarrow \textrm{N}^++\textrm{CN}$ are taken from Ref. \cite{Burley:1991}, and are plotted in Fig. \ref{fig:xsec36}. The comparison with the G\"{o}k\c{c}en rate for the reaction $\textrm{C}^++\textrm{N}_2\rightarrow \textrm{N}_2^++\textrm{C}$ is presented in Fig. \ref{fig:rate36}.\bigskip

\begin{figure}[!htbp]
\centering
\includegraphics[width=.6\textwidth]{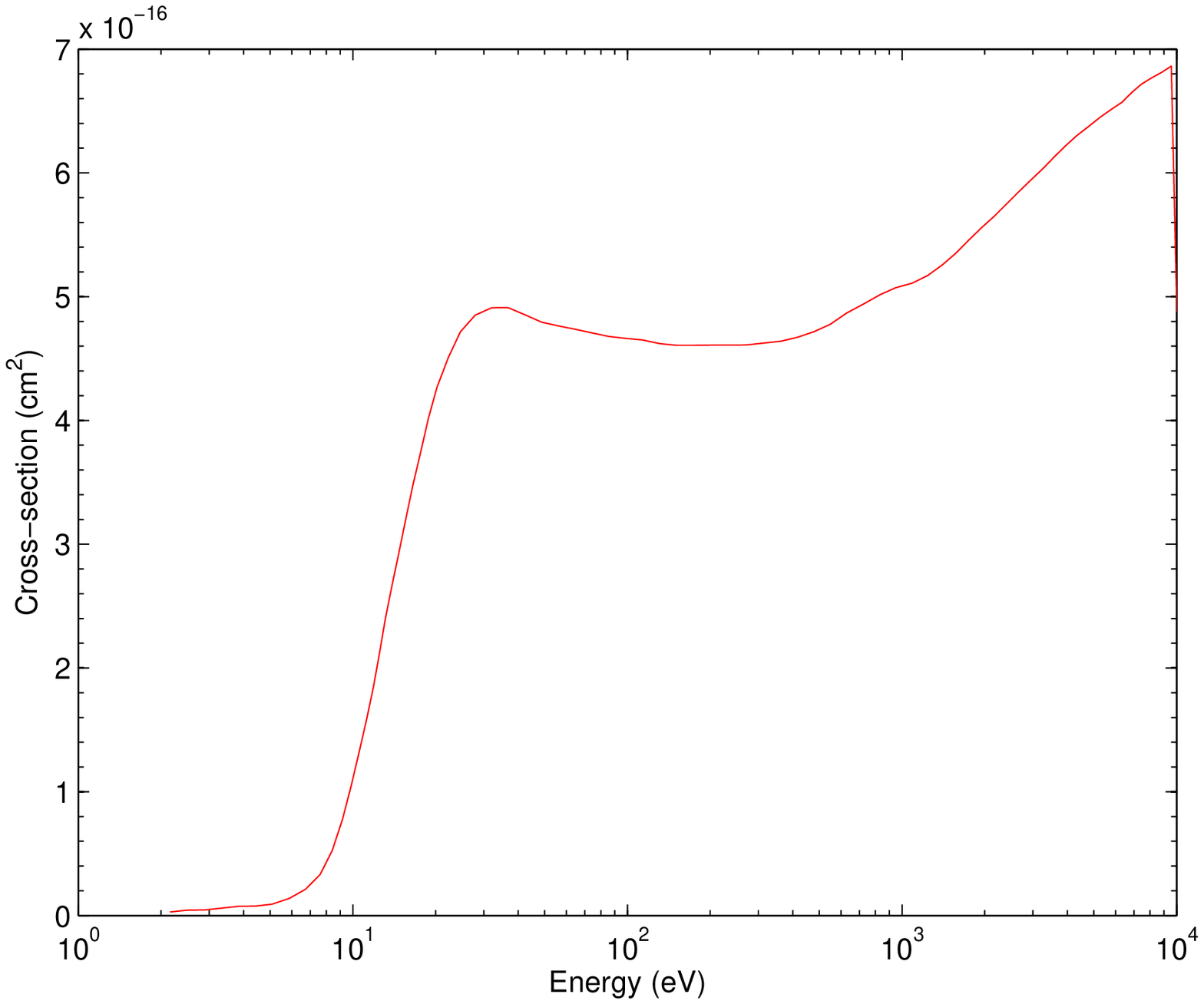}
\caption{\small{Cross-section for the reaction $\textrm{N}^++\textrm{N}_2\rightarrow \textrm{N}_2^++\textrm{N}$}}%
\label{fig:xsecA7}
\end{figure}

Cross-sections for the reaction $\textrm{N}^++\textrm{N}_2\rightarrow \textrm{N}_2^++\textrm{N}$ are taken from Ref. \cite{Luna:2003}, and are plotted in Fig. \ref{fig:xsecA7}.\bigskip

Several other rates for electron-impact dissociation and double ionization of ions have been compiled, and the corresponding rates have been produced. These are shortly reported here in tabular form \ref{tab:EIonIon}, as the sensitivity study presented in Section \ref{sec:sensitivity} whill show that these reactions can mostly be neglected.\bigskip

\begin{table*}[!htbp]
\caption{Electron-Ion Ionization Rates}%
\label{tab:EIonIon}%
\centering%
\begin{tabular}{l l l}
\toprule%
\midrule%
Reaction & Fitted Arrhenius Rate (\centi\cubic\metre\per\mole\per\second) & Reference\\%
\midrule%
H$_{2}^{+}$\phantom{H} + e$^{-}$          $\rightleftarrows$ H$^{+\phantom{+}}$ + H$^{+}$ + 2e$^{-}$                     & $2.28\times 10^{15}T^{0.03}\exp\left(-355,960/T\right)$ & \cite{Peart:1973,Kim:2000}\\%
CH$^{+}$ + e$^{-}$                        $\rightleftarrows$ C$^{+\phantom{+}}$ + H$^{\phantom{+}}$ + \phantom{2}e$^{-}$ & $1.23\times 10^{11}T^{1.07}\exp\left(-\phantom{0}62,080/T\right)$  & \cite{Lecointre:2007}\\%
CH$^{+}$ + e$^{-}$                        $\rightleftarrows$ C\phantom{$^{++}$} + H$^{+}$ + \phantom{2}e$^{-}$           & $8.53\times 10^{09}T^{1.23}\exp\left(-\phantom{0}51,400/T\right)$  & \cite{Lecointre:2007}\\%
CH$^{+}$ + e$^{-}$                        $\rightleftarrows$ C$^{+\phantom{+}}$ + H$^{+}$ + 2e$^{-}$                     & $3.92\times 10^{09}T^{1.30}\exp\left(-247,590/T\right)$ & \cite{Lecointre:2007}\\%
CH$^{+}$ + e$^{-}$                        $\rightleftarrows$ C$^{++}$           + H$^{\phantom{+}}$ + 2e$^{-}$           & $6.47\times 10^{05}T^{1.83}\exp\left(-339,470/T\right)$ & \cite{Lecointre:2007}\\%
CN$^{+}$ + e$^{-}$                        $\rightleftarrows$ C$^{+\phantom{+}}$ + N$^{\phantom{+}}$ + \phantom{2}e$^{-}$ & $3.83\times 10^{10}T^{0.13}\exp\left(-\phantom{0}44,060/T\right)$ & \cite{LePadellec:1999}\\%
CN$^{+}$ + e$^{-}$                        $\rightleftarrows$ C$\phantom{^{++}}$ + N$^{+}$ + \phantom{2}e$^{-}$           & $1.13\times 10^{08}T^{0.60}\exp\left(-\phantom{0}66,750/T\right)$ & \cite{LePadellec:1999}\\%
NH$^{+}$ + e$^{-}$                        $\rightleftarrows$ N$^{+\phantom{+}}$ + H$^{\phantom{+}}$ + \phantom{2}e$^{-}$ & $1.20\times 10^{14}T^{0.42}\exp\left(-229,740/T\right)$ & \cite{Lecointre:2010}\\%
NH$^{+}$ + e$^{-}$                        $\rightleftarrows$ N$^{+\phantom{+}}$ + H$^{+}$ + 2e$^{-}$                     & $1.25\times 10^{13}T^{0.56}\exp\left(-\phantom{0}27,350/T\right)$ & \cite{Lecointre:2010}\\%
N$_{2}^{+}$\phantom{H} + e$^{-}$          $\rightleftarrows$ N$_{2}^{++}$ \phantom{+\ H$^{+}$} + 2e$^{-}$                & $3.84\times 10^{12}T^{0.64}\exp\left(-289,260/T\right)$ & \cite{Bahati:2001}\\%
N$_{2}^{+}$\phantom{H} + e$^{-}$          $\rightleftarrows$ N$^{+\phantom{+}}$ + N$^{\phantom{+}}$ + \phantom{2}e$^{-}$ & $7.47\times 10^{11}T^{0.84}\exp\left(-\phantom{0}80,600/T\right)$ & \cite{Bahati:2001}\\%
N$_{2}^{+}$\phantom{H} + e$^{-}$          $\rightleftarrows$ N$^{+\phantom{+}}$ + N$^{+}$ + 2e$^{-}$                     & $1.47\times 10^{13}T^{0.44}\exp\left(-335,900/T\right)$ & \cite{Bahati:2001}\\%
\midrule%
\bottomrule%
\end{tabular}
\end{table*}

\subsection{Rates Reduction Through a Sensitivity Study}
\label{sec:sensitivity}

A sensitivity study of the updated ionized rates has been carried out for achieving a more compact set of ionized reactions. Changes in neutral rates have not been considered in this sensitivity study, as the conclusions from the previous study of G\"{o}k\c{c}en \cite{Gokcen:2005} remain valid. As discussed in the Introduction, such improved ionized rates are mainly relevant for high-speed shocked flows. As such, we have considered two different cases for this sensitivity study: The 9\kilo\metre\per\second\ experimental point from the X2 Shock-tube (see Tab. \ref{tab:postshock}), and an extra theoretical condition at v=11\kilo\metre\per\second\ and the same pre-shock pressure (13 \pascal). The choice for this specific velocity is interesting because it corresponds to the maximum initial entry velocity for Earth--Titan orbital trajectories. A linear rate sensitivity analysis has been therefore carried out for these two high-speed N$_2$--CH$_4$ shocked flows.\\

We define the normalized speed of reaction $R_r$ as:

\begin{equation}
\label{eq:speed}
R_{r}=K_r\prod_{i=1}^{n}N_{i},
\end{equation}

and the normalized sensitivity coefficient of parameter $X$, with respect to reaction $r$:

\begin{equation}
\label{eq:SXr}
S_{X,r}=\frac{K_r}{X_0}\frac{\partial X}{\partial K_r}.
\end{equation}

The normalization is carried against the maximum time-value of $S_{X,r}$.\\

For this specific category of shocked flows, the choice of parameters $X$ is reasonably straightforward. We restrict ourselves to the analysis on the rates impact for the production and destruction of CN, given that radiation from the CN Violet system remains ubiquitous in the near-VUV to near-IR spectral region. We further select the electrons e$^+$ as a parameter, given that these high-speed flows will be significantly ionized (above 10\%), and we finally select the temperature $T$ as the last parameter.\bigskip

Two batches of calculations at shock speeds of v=9\kilo\metre\per\second\ and v=11\kilo\metre\per\second\ have been carried out, either considering the corrected macroscopic model of G\"{o}k\c{c}en, or a combination of the corrected macroscopic model of G\"{o}k\c{c}en and the FHO state-to-state model for N$_2$ dissociation processes. Both cases consider the updated 32 ionization rates described in Section \ref{sec:ions}, and the collisional-radiative (CR) model described in the previous report \cite{report1}. For lack of time, a consistency check of the CR model has not been carried out in this work. Also, results from the sensitivity study using the FHO model could not be processed in due time, and are not presented here.\bigskip

\begin{figure}[!htbp]
\centering
\includegraphics[width=.49\textwidth]{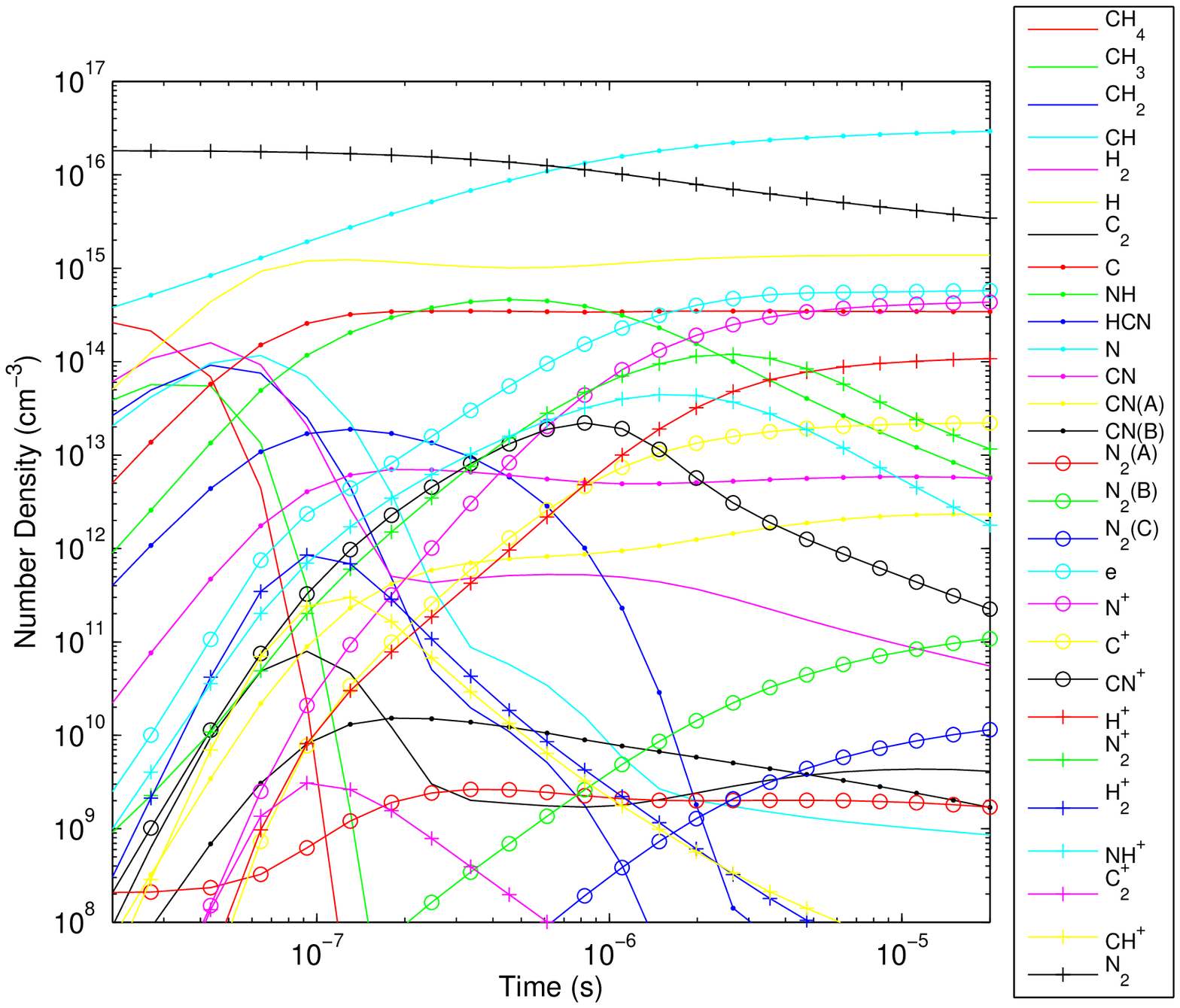}
\includegraphics[width=.49\textwidth]{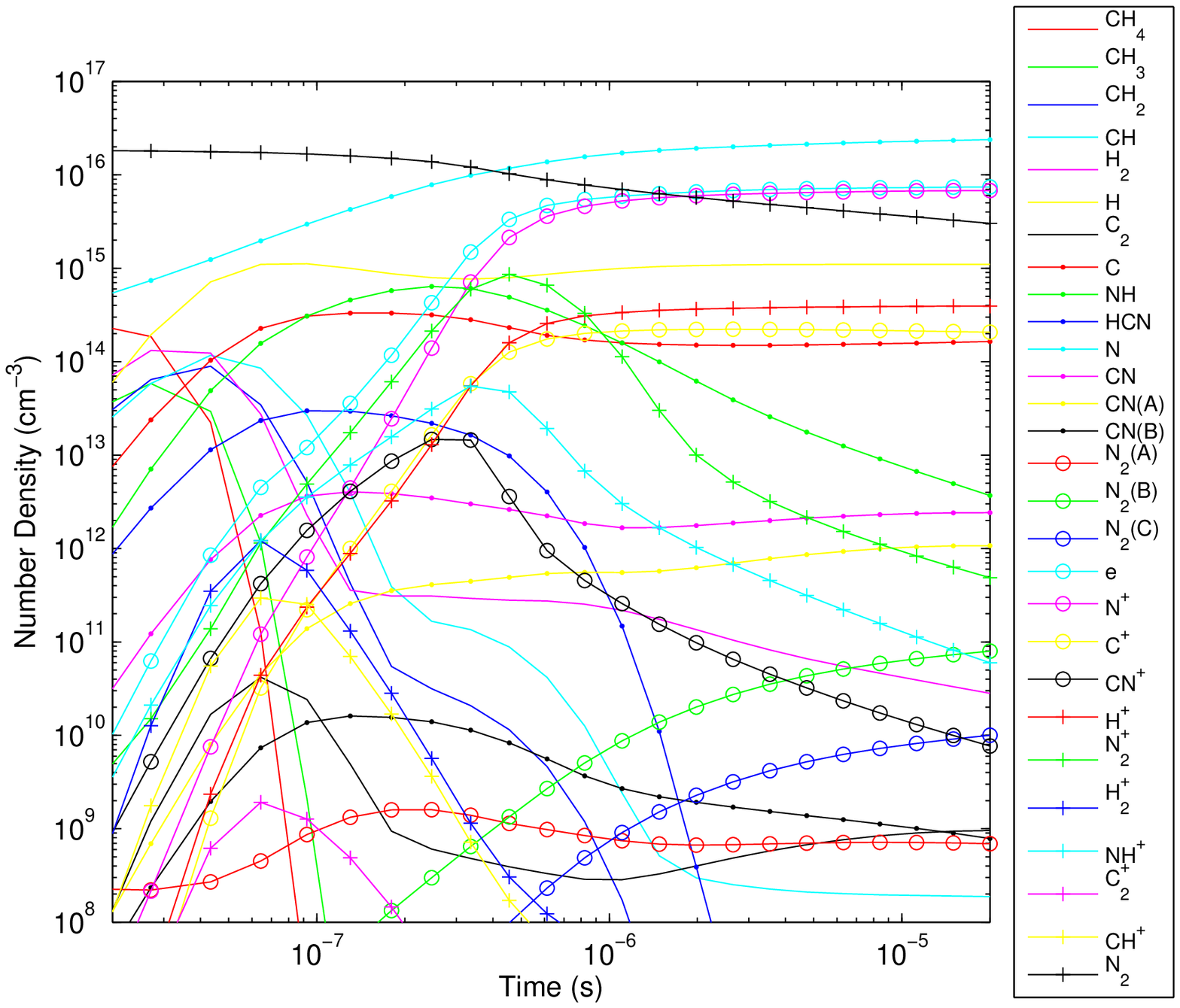}
\includegraphics[width=.49\textwidth]{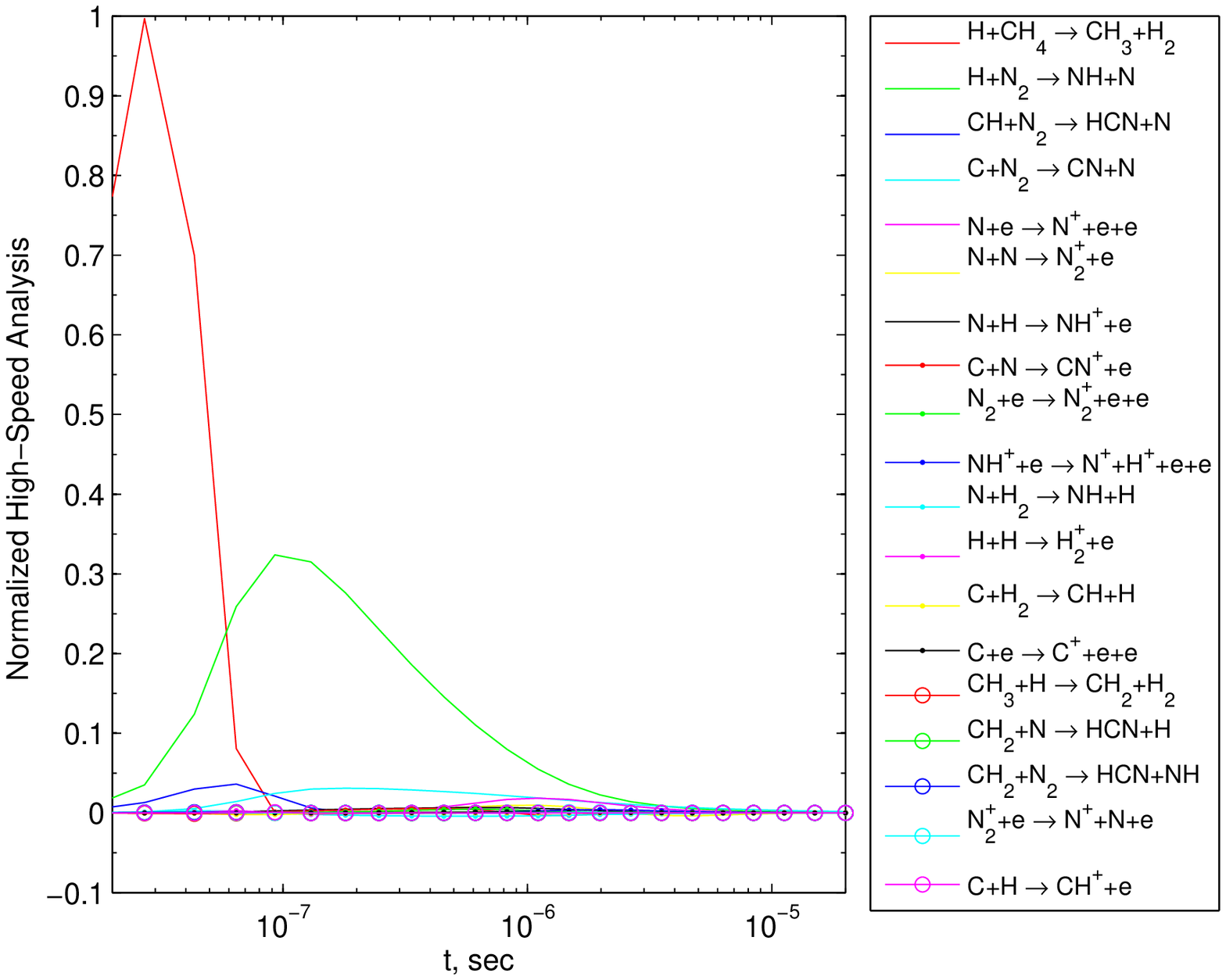}
\includegraphics[width=.49\textwidth]{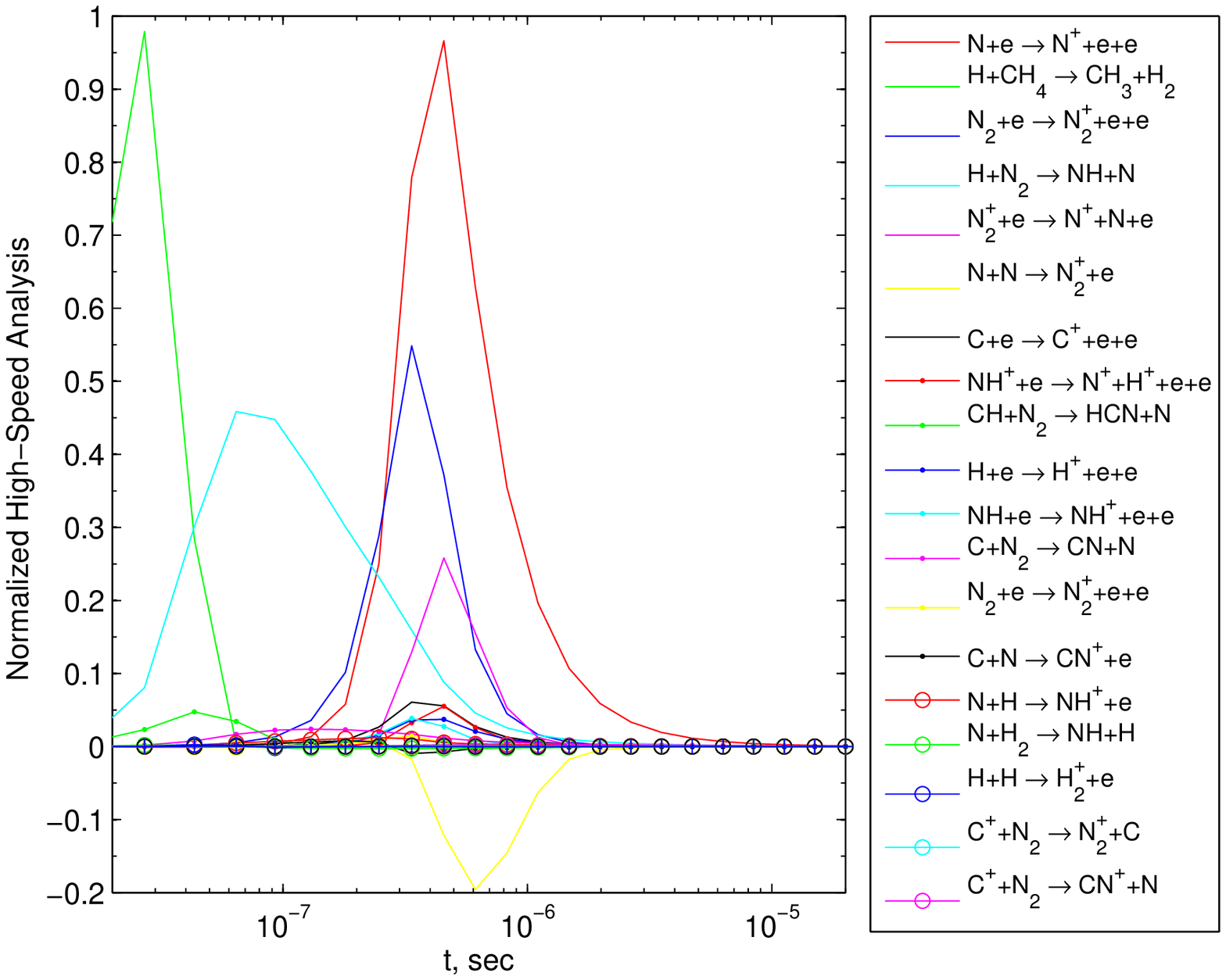}
\caption{\small{Time-evolution of the post-shock species concentrations (top) for v=9\kilo\metre\per\second\ (left) and v=11\kilo\metre\per\second\ (right), and corresponding speed analysis (bottom).}}%
\label{fig:conc}
\end{figure}

\begin{figure}[!htbp]
\centering
\includegraphics[width=.49\textwidth]{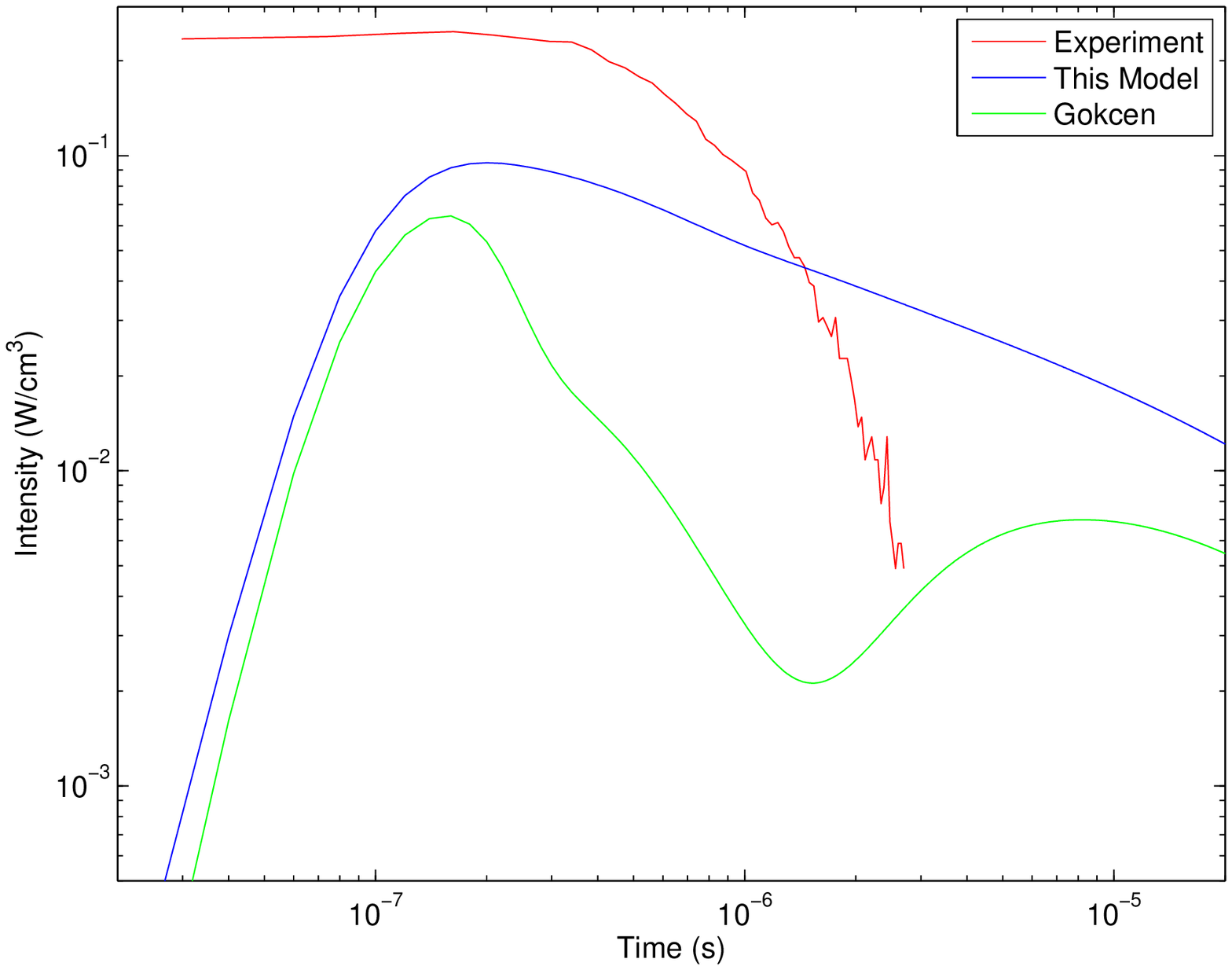}
\includegraphics[width=.49\textwidth]{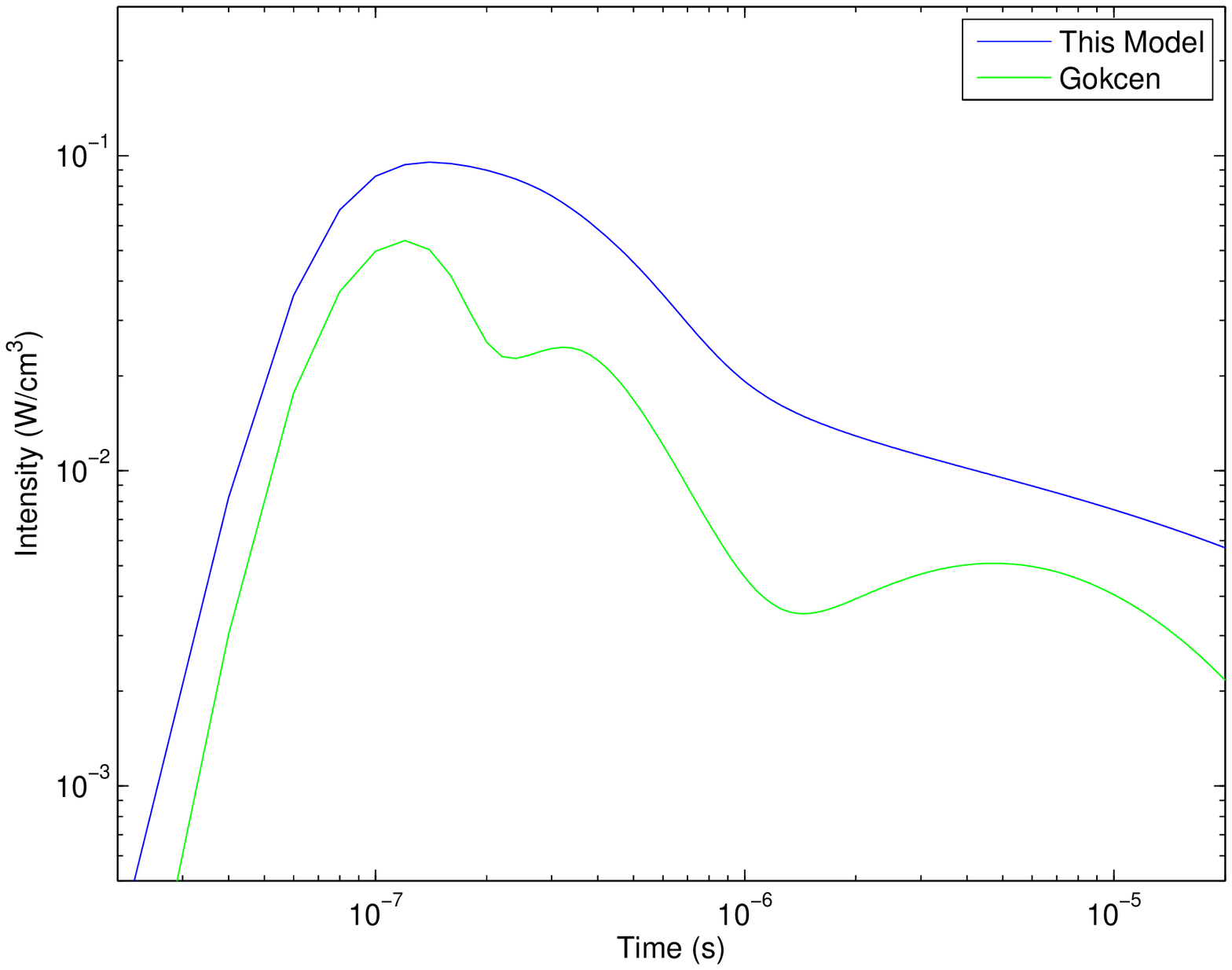}
\caption{\small{Simulated time-evolution of the CN Violet radiative intensity for v=9\kilo\metre\per\second\ (left) and v=11\kilo\metre\per\second\ (right). A comparison with the X2 experiment at v=9\kilo\metre\per\second\ is also presented in the left figure}}%
\label{fig:ICNB}
\end{figure}

Fig. \ref{fig:conc} shows the time evolution of the post-shock species concentrations for the two shock velocities, and Fig. \ref{fig:ICNB} shows the time evolution from the CN Violet system radiation. In this latter figure, the same simulation has been carried out simply considering the G\"{o}k\c{c}en dataset, without any of the updates of this work (plotted green). For the shock at v=9\kilo\metre\per\second, the experimental data from the X2 shock-tube facility is also reported (in red). We may verify that although the macroscopic model underpredicts the experimental solution (as discussed in detail in the previous report; Ref. \cite{report1}), the updated dataset provides slighter higher radiative intensities, closer to the experimental results at v=9\kilo\metre\per\second. More importantly the marked double intensity maxima seen at v=9\kilo\metre\per\second, as well as the less marked triple intensity maxima at v=11\kilo\metre\per\second\, are no longer occurring. These results point at a significant quality increase of simulated results (using macroscopic models or otherwise), as a result of the improved kinetic model proposed in this work.\bigskip

We have then sought to reduce the number of ionized reactions (from a grand total of 32) to a more compact set of reactions, without any significant loss of accuracy. While it is immediately apparent that some reactions (such as the one involving doubly ionized species, or reactions with a large characteristic temperature $\theta_r$) can be straightforwardly neglected, a sensitivity analysis remains mandatory for a more consistent decrease in the number of rates.\bigskip

\begin{figure}[!htbp]
\centering
\includegraphics[width=.49\textwidth]{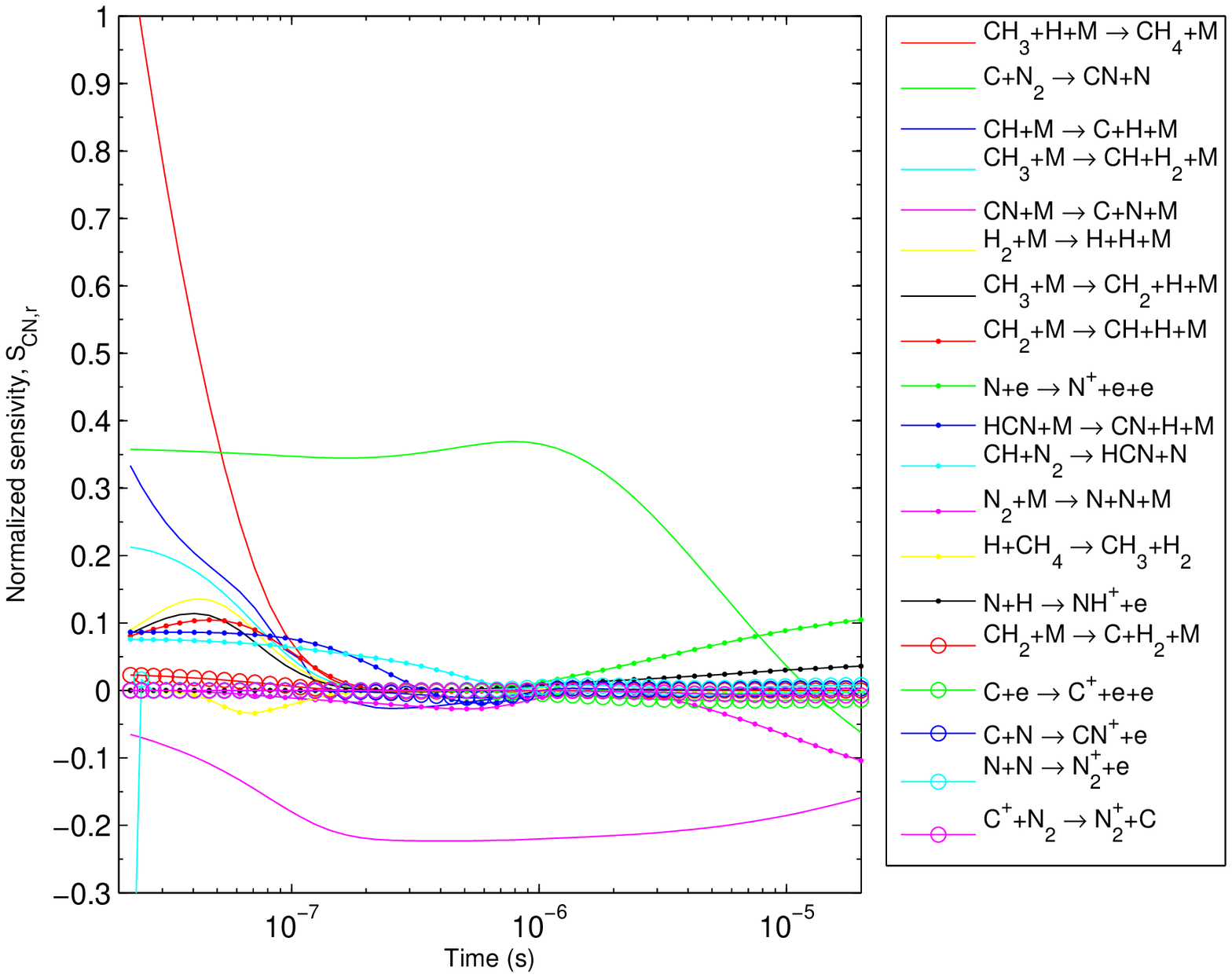}
\includegraphics[width=.49\textwidth]{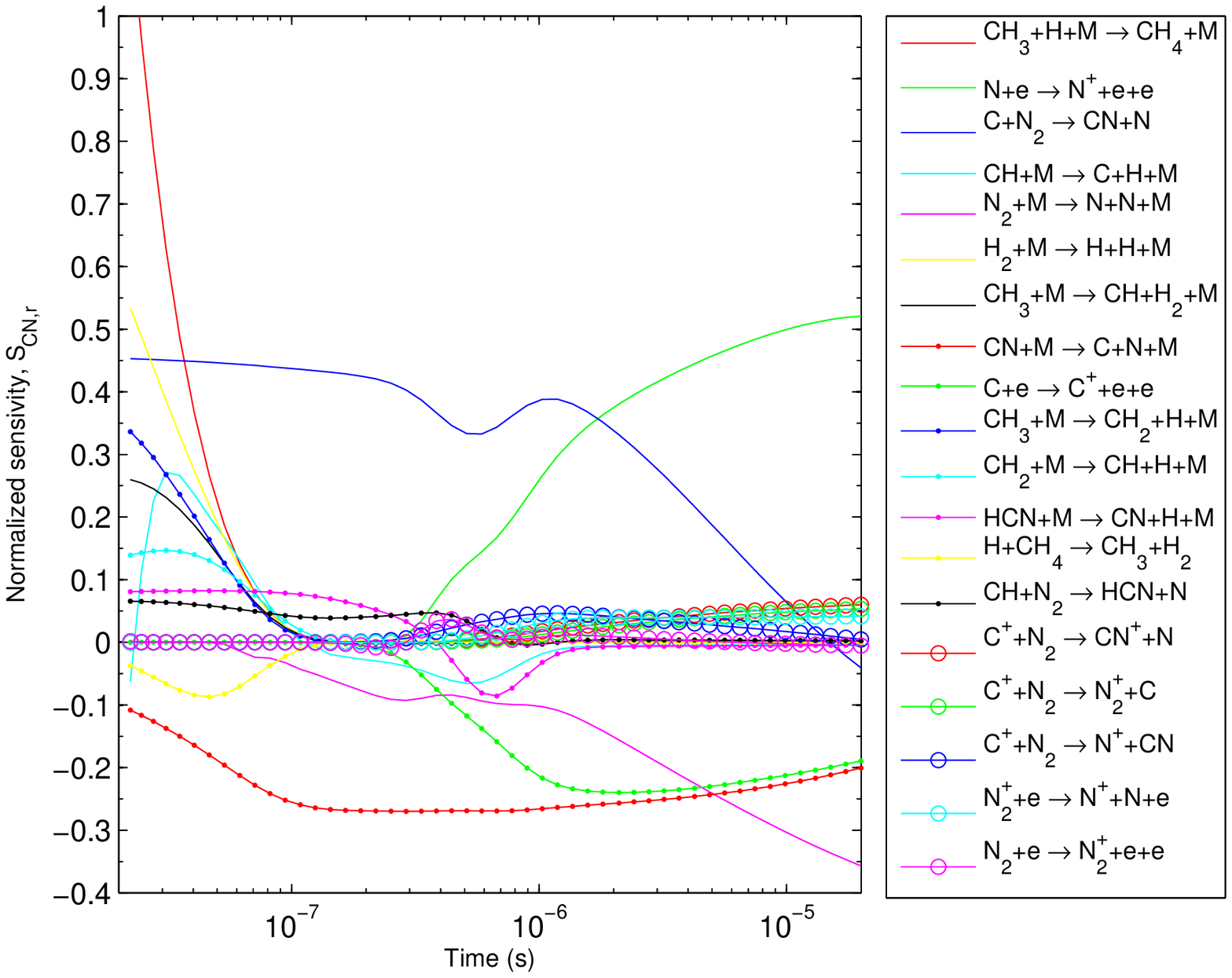}
\includegraphics[width=.49\textwidth]{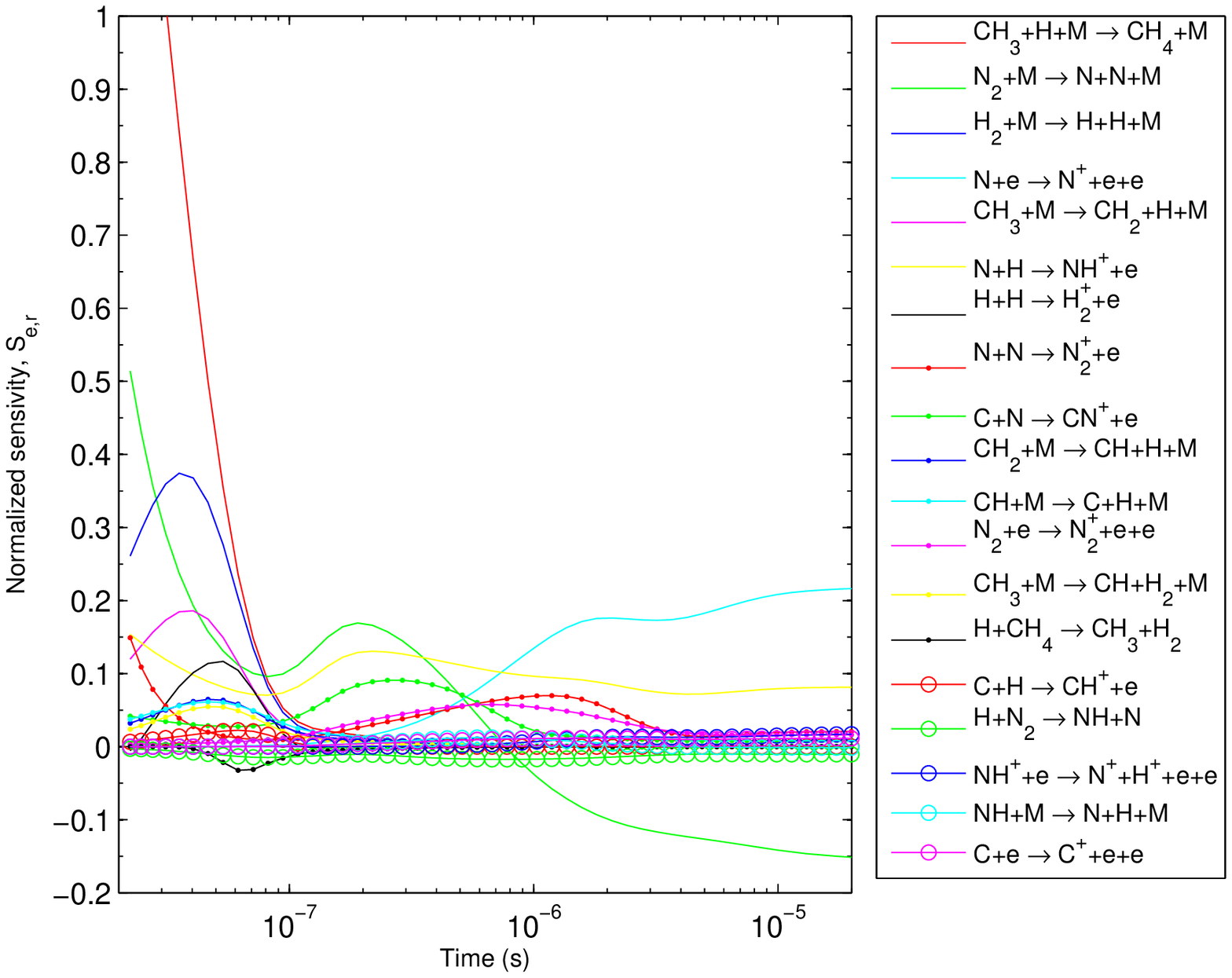}
\includegraphics[width=.49\textwidth]{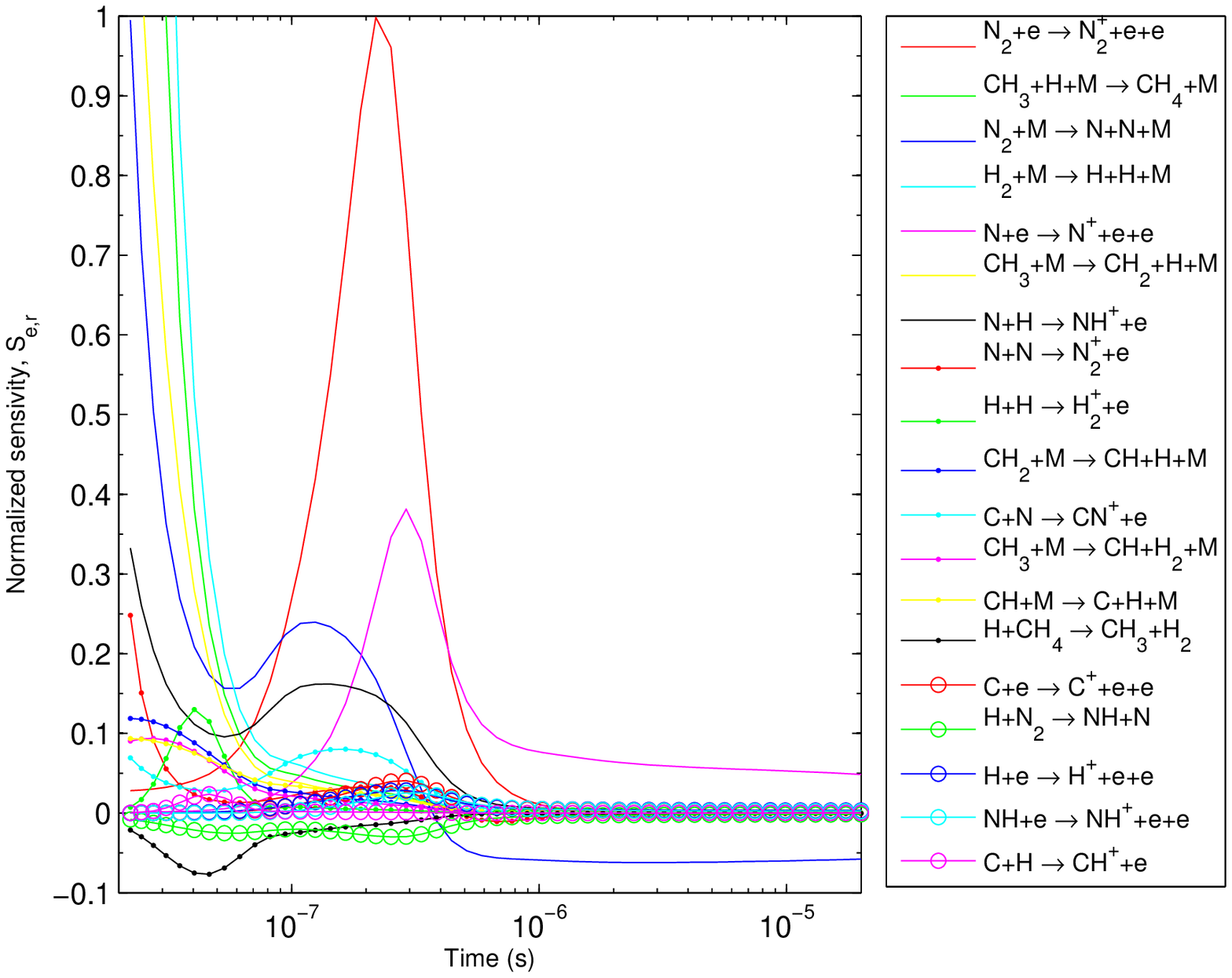}
\includegraphics[width=.49\textwidth]{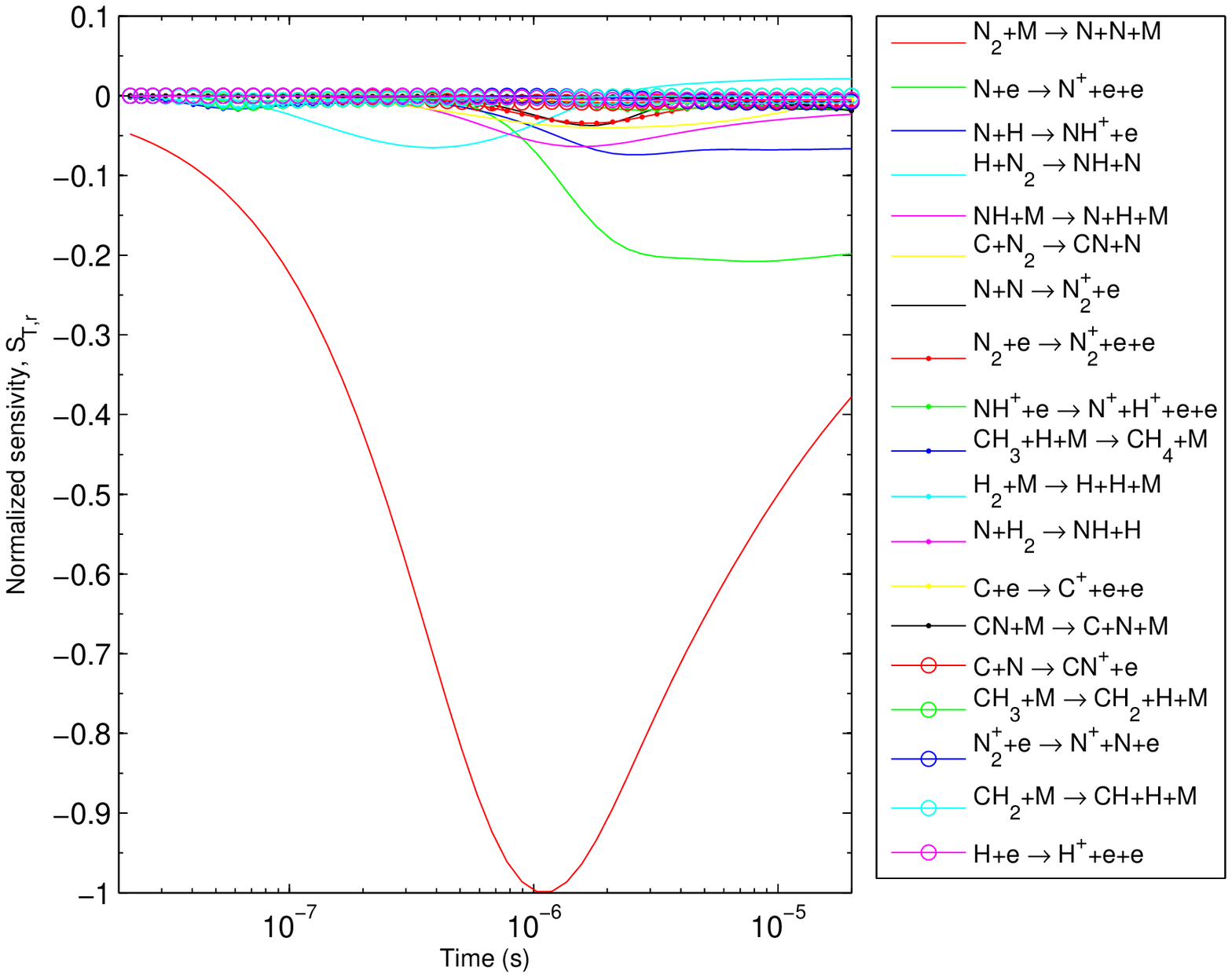}
\includegraphics[width=.49\textwidth]{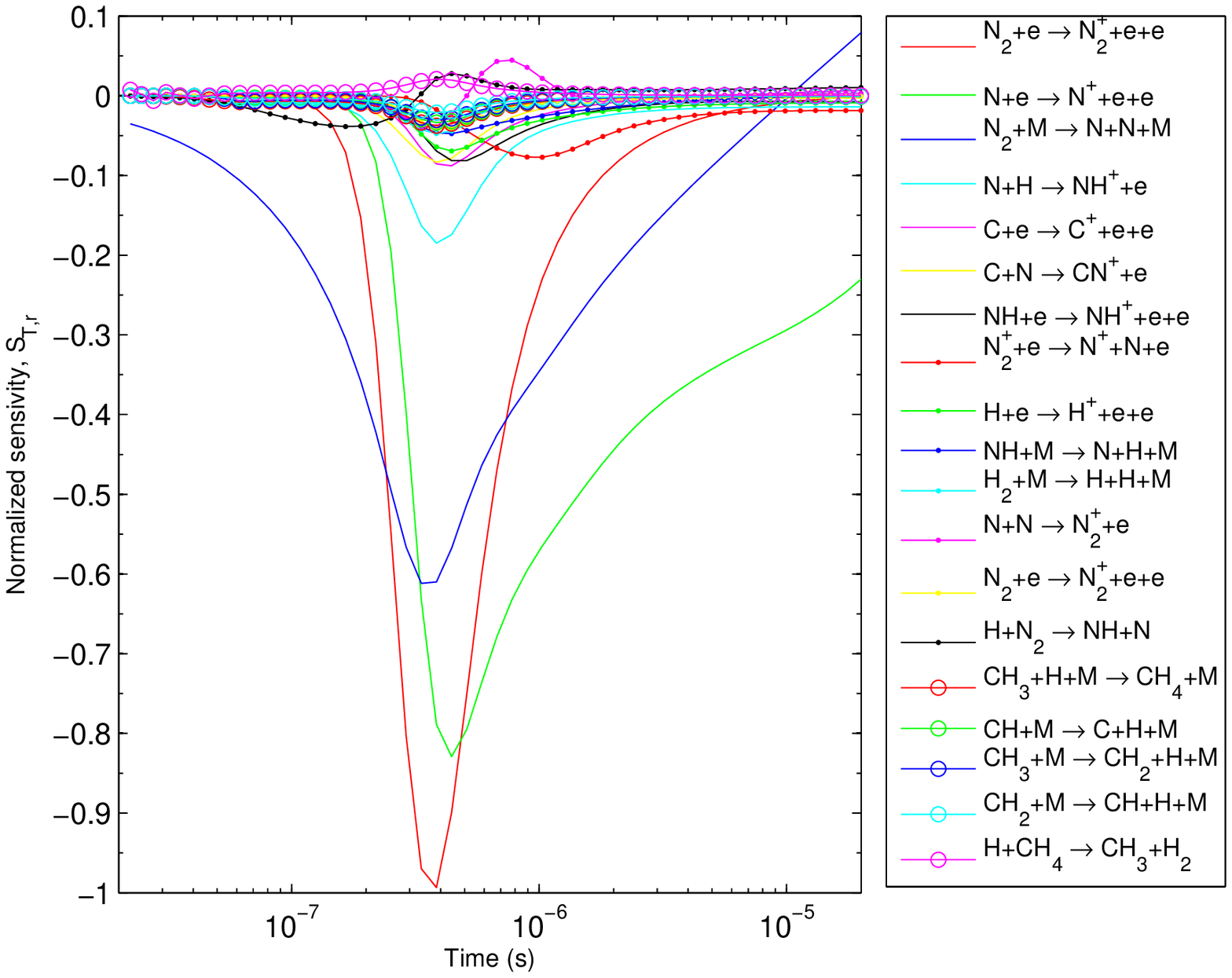}
\caption{\small{Sensitivity analysis for CN and e$^-$ concentrations and temperature $T$ (top to bottom), for v=9\kilo\metre\per\second\ (left) and v=11\kilo\metre\per\second\ (right).}}%
\label{fig:sensitivity}
\end{figure}

The sensitivity analysis of these simulated results has then been carried for the speed of reaction $R_{r}$ (Eq. \ref{eq:speed}), and for the normalized sensitivity coefficient of parameter $S_{X,r}$, with respect to reaction CN and e$^-$ concentrations, and temperature $T$ (Eq. \ref{eq:SXr}). The obtained results are presented in Fig. \ref{fig:sensitivity}, for the two v=9\kilo\metre\per\second\ and v=11\kilo\metre\per\second\ cases. For each sensitivity analysis, the more important 19 reactions are reported.\bigskip

An analysis of the sensitivity results shows that most of the recombination ionization reactions A+B$\rightleftarrows$ AB$^+$+e$^-$ are significantly important,  with the reactions producing NH$^+$, H$_2^+$, N$_2^+$, and CN$^+$ ranking 6th to 9th for the production of electrons for v=9\kilo\metre\per\second, and 7th--9th and 11th at v=11\kilo\metre\per\second. Electron-impact reactions of the type (AB; A)+e$^-\rightleftarrows$ (AB; A)$^+$+e$^-$+e$^-$ are also important, with electron-impact ionization of N, N$_2$ and C ranking as 4th, 12th, and 19th for the production of electrons for v=9\kilo\metre\per\second, and 5th, 1st and 15th at v=11\kilo\metre\per\second, with  electron-impact ionization of C appearing as 17th in this latter case. Further, ion exchange reactions of the type $C^++N_2$ also specifically affect the production of the CN species.\bigskip

This sensitivity analysis highlights the importance of recombination ionization and electron-impact ionization reactions. Therefore, it has been chosen to retain all the reactions of this kind, with the notable exception of reactions involving C$_2$, which is a species who is not present in sufficient concentrations to warrant inclusion in the reduced dataset. Additionally to these reactions, all the other ionized reactions appearing in the top 19 of the analysis summarized in Fig. \ref{fig:sensitivity} have been included in the reduced dataset.\bigskip

The summary of the full ionized rates dataset, complete with the Arrhenius fit from this work, is presented in Table \ref{tab:TotalIon}. The ionized rates that appear among the top-ranking 19 rates in any of the 6 results of Fig. \ref{fig:sensitivity} are highlighted in red. In addition to the highlighted rates, we select rates 10, 11 and 13 for inclusion in the reduced ionization dataset.\bigskip

\begin{table*}[!htbp]
\caption{Overall Ionized Species Rates}%
\label{tab:TotalIon}%
\centering%
\begin{tabular}{l r@{}l c l}
\toprule%
\midrule%
No. & \multicolumn{2}{c}{Reaction} & Rate (\centi\cubic\metre\per\mole\per\second) & Ref.\\%
\midrule%
\color{red}\phantom{0}1 & \color{red}H\ +\ H\phantom{$_{2}$} &\color{red}\ $\rightleftarrows$ \phantom{C}H$_{2}^{+}$\ +\ e$^{-}$                              & $1.13\times 10^{15}T^{-0.06}\exp(-129,060/T)$ & \cite{Florescu:2006,Auerbach:1977}\notag\\%
\color{red}\phantom{0}2 & \color{red}C\ +\ H\phantom{$_{2}$} &\color{red}\ $\rightleftarrows$ CH$^{+}$\ +\ e$^{-}$                                            & $9.95\times 10^{11}T^{\phantom{+}0.52}\exp(-\phantom{0}84,830/T)$ & \cite{Florescu:2006,Mitchell:1978}\notag\\%
\phantom{0}3 & C\ +\ C\phantom{$_{2}$} &\ $\rightleftarrows$ \phantom{C}C$_{2}^{+}$\ +\ e$^{-}$                              & $3.19\times 10^{14}T^{-0.24}\exp(-\phantom{0}70,690/T)$ & \cite{Florescu:2006,Mul:1980}\notag\\%
\color{red}\phantom{0}4 & \color{red}C\ +\ N\phantom{$_{2}$} &\color{red}\ $\rightleftarrows$ CN$^{+}$\ +\ e$^{-}$                                            & $3.80\times 10^{12}T^{\phantom{+}0.33}\exp(-\phantom{0}74,810/T)$ & \cite{LePadellec:1999}\notag\\%
\color{red}\phantom{0}5 & \color{red}N\ +\ H\phantom{$_{2}$} &\color{red}\ $\rightleftarrows$ NH$^{+}$\ +\ e$^{-}$                                            & $2.99\times 10^{14}T^{-0.06}\exp(-118,760/T)$ & \cite{Florescu:2006,Mul:unp}\notag\\%
\color{red}\phantom{0}6 & \color{red}N\ +\ N\phantom{$_{2}$} &\color{red}\ $\rightleftarrows$ \phantom{C}N$_{2}^{+}$\ +\ e$^{-}$                              & $2.13\times 10^{10}T^{\phantom{+}0.48}\exp(-\phantom{0}69,190/T)$ & \cite{Guberman:2009}\notag\\%
\midrule%
\color{red}\phantom{0}7 & \color{red}H\ +\ e$^{-}$           &\color{red}\ $\rightleftarrows$ \phantom{C}H$^{+}$\ +\ e$^{-}$\hspace{1.2mm} +\ e$^{-}$         & $1.36\times 10^{15}T^{\phantom{+}0.18}\exp(-169,000/T)$ & \cite{Kim:1994}\notag\\%
\color{red}\phantom{0}8 & \color{red}C\ +\ e$^{-}$           &\color{red}\ $\rightleftarrows$ \phantom{C}C$^{+}$\ +\ e$^{-}$\hspace{1.2mm} +\ e$^{-}$         & $1.24\times 10^{15}T^{\phantom{+}0.28}\exp(-142,700/T)$ & \cite{Kim:2002}\notag\\%
\color{red}\phantom{0}9 & \color{red}N\ +\ e$^{-}$           &\color{red}\ $\rightleftarrows$ \phantom{C}N$^{+}$\ +\ e$^{-}$\hspace{1.2mm} +\ e$^{-}$         & $1.67\times 10^{13}T^{\phantom{+}0.59}\exp(-143,220/T)$ & \cite{Kim:2002}\notag\\%
10 & Ar\ +\ e$^{-}$          &\ $\rightleftarrows$ \hspace{1.2mm}Ar$^{+}$\ +\ e$^{-}$ \hspace{1.2mm}+\ e$^{-}$     & $5.52\times 10^{13}T^{\phantom{+}0.58}\exp(-186,210/T)$ & \cite{Rapp:1965,Schram:1966}\notag\\%
\midrule%
11 & H$_{2}$\ +\ e$^{-}$     &\ $\rightleftarrows$ \phantom{C}H$_{2}^{+}$\ +\ e$^{-}$\hspace{1.2mm}\ +\ e$^{-}$    & $4.05\times 10^{13}T^{\phantom{+}0.52}\exp(-180,767/T)$ & \cite{Rapp:1965}\notag\\%
12 & H$_{2}^{+}$\ +\ e$^{-}$ &\ $\rightleftarrows$ \phantom{C}H$^{+}$\ +\ H$^{+}$\ +\ e$^{-}$ +\ e$^{-}$           & $2.28\times 10^{15}T^{\phantom{+}0.03}\exp\left(-355,960/T\right)$ & \cite{Peart:1973,Kim:2000}\\%
13 & CH\ +\ e$^{-}$          &\ $\rightleftarrows$ CH$^{+}$\ +\ e$^{-}$\hspace{1.2mm} +\ e$^{-}$                   & $1.15\times 10^{12}T^{\phantom{+}0.87}\exp(-123,430/T)$ & \cite{Kim:2000}\notag\\%
14 & CH$^{+}$\ +\ e$^{-}$    &\ $\rightleftarrows$ \phantom{C}C$^{+}$\ +\ H\phantom{$^{+}$} +\ e$^{-}$             & $1.23\times 10^{11}T^{\phantom{+}1.07}\exp\left(-\phantom{0}62,080/T\right)$  & \cite{Lecointre:2007}\\%
15 & CH$^{+}$\ +\ e$^{-}$    &\ $\rightleftarrows$ \phantom{C}C\phantom{$^{+}$}\ +\ H$^{+}$ +\ e$^{-}$             & $8.53\times 10^{09}T^{\phantom{+}1.23}\exp\left(-\phantom{0}51,400/T\right)$  & \cite{Lecointre:2007}\\%
16 & CH$^{+}$\ +\ e$^{-}$    &\ $\rightleftarrows$ \phantom{C}C$^{+}$\ +\ H$^{+}$\ +\ e$^{-}$\ +\ e$^{-}$          & $3.92\times 10^{09}T^{\phantom{+}1.30}\exp\left(-247,590/T\right)$ & \cite{Lecointre:2007}\\%
17 & CH$^{+}$\ +\ e$^{-}$    &\ $\rightleftarrows$ \phantom{C}C$^{++}$\ +\ H\ +\ e$^{-}$\hspace{0.2mm}\ +\ e$^{-}$ & $6.47\times 10^{05}T^{\phantom{+}1.83}\exp\left(-339,470/T\right)$ & \cite{Lecointre:2007}\\%
18 & C$_{2}$\ +\ e$^{-}$     &\ $\rightleftarrows$ \phantom{C}C$_{2}^{+}$\ +\ e$^{-}$\hspace{1.2mm}\ +\ e$^{-}$    & $1.83\times 10^{13}T^{\phantom{+}0.68}\exp(-151,110/T)$ & \cite{Deutsch:2000}\notag\\%
19 & CN$^{+}$\ +\ e$^{-}$    &\ $\rightleftarrows$ \phantom{C}C$^{+}$\ +\ N\phantom{$^{+}$}\ +\ e$^{-}$            & $3.83\times 10^{10}T^{\phantom{+}0.13}\exp\left(-\phantom{0}44,060/T\right)$ & \cite{LePadellec:1999}\\%
20 & CN$^{+}$\ +\ e$^{-}$    &\ $\rightleftarrows$ \phantom{C}C\phantom{$^{+}$}\ +\ N$^{+}$\ +\ e$^{-}$            & $1.13\times 10^{08}T^{\phantom{+}0.60}\exp\left(-\phantom{0}66,750/T\right)$ & \cite{LePadellec:1999}\\%
\color{red}21 & \color{red}NH\ +\ e$^{-}$          &\color{red}\ $\rightleftarrows$ NH$^{+}$\ +\ e$^{-}$\hspace{1.2mm}\ +\ e$^{-}$                  & $2.39\times 10^{13}T^{\phantom{+}0.59}\exp(-172,430/T)$ & \cite{Rajvanshi:2010}\notag\\%
22 & NH$^{+}$\ +\ e$^{-}$    &\ $\rightleftarrows$ \phantom{C}N$^{+}$\ +\ H\phantom{$^{+}$}\ +\ e$^{-}$            & $1.20\times 10^{14}T^{\phantom{+}0.42}\exp\left(-229,740/T\right)$ & \cite{Lecointre:2010}\\%
\color{red}23 & \color{red}NH$^{+}$\ +\ e$^{-}$    &\color{red}\ $\rightleftarrows$ \phantom{C}N$^{+}$\ +\ H$^{+}$\ +\ e$^{-}$\ +\ e$^{-}$          & $1.25\times 10^{13}T^{\phantom{+}0.56}\exp\left(-\phantom{0}27,350/T\right)$ & \cite{Lecointre:2010}\\%
\color{red}24 & \color{red}N$_{2}$\ +\ e$^{-}$     &\color{red}\ $\rightleftarrows$ \phantom{C}N$_{2}^{+}$\ +\ e$^{-}$\hspace{1.2mm}\ +\ e$^{-}$    & $5.92\times 10^{11}T^{\phantom{+}0.92}\exp(-178,630/T)$ & \cite{Rapp:1965}\notag\\%
25 & N$_{2}^{+}$\ +\ e$^{-}$ &\ $\rightleftarrows$ \phantom{C}N$_{2}^{++}$+\ e$^{-}$\ +\ e$^{-}$                   & $3.84\times 10^{12}T^{\phantom{+}0.64}\exp\left(-289,260/T\right)$ & \cite{Bahati:2001}\\%
\color{red}26 & \color{red}N$_{2}^{+}$\ +\ e$^{-}$ &\color{red}\ $\rightleftarrows$ \phantom{C}N$^{+}$\ +\ N\phantom{$^{+}$}\ +\ e$^{-}$            & $7.47\times 10^{11}T^{\phantom{+}0.84}\exp\left(-\phantom{0}80,600/T\right)$ & \cite{Bahati:2001}\\%
27 & N$_{2}^{+}$\ +\ e$^{-}$ &\ $\rightleftarrows$ \phantom{C}N$^{+}$\ +\ N$^{+}$\ +\ e$^{-}$\ +\ e$^{-}$          & $1.47\times 10^{13}T^{\phantom{+}0.44}\exp\left(-335,900/T\right)$ & \cite{Bahati:2001}\\%
\midrule%
\color{red}28 & \color{red}C$^{+}$\ +\ N$_{2}$     &\color{red}\ $\rightleftarrows$ \phantom{C}N$_{2}^{+}$\ +\ C                                    & $1.01\times 10^{11}T^{\phantom{+}0.60}\exp(-\phantom{0}53,830/T)$ & \cite{Burley:1991}\notag\\%
\color{red}29 & \color{red}C$^{+}$\ +\ N$_{2}$     &\color{red}\ $\rightleftarrows$ CN$^{+}$\ +\ N                                                  & $1.32\times 10^{12}T^{\phantom{+}0.33}\exp(-\phantom{0}51,430/T)$ & \cite{Burley:1991}\notag\\%
\color{red}30 & \color{red}C$^{+}$\ +\ N$_{2}$     &\color{red}\ $\rightleftarrows$ \phantom{C}N$^{+}$\ +\ CN                                       & $8.93\times 10^{13}T^{-0.14}\exp(-\phantom{0}65,260/T)$ & \cite{Burley:1991}\notag\\%
31 & N$^{+}$\ +\ N$_{2}$     &\ $\rightleftarrows$ \phantom{C}N$_{2}^{+}$\ +\ N                                    & $1.63\times 10^{06}T^{\phantom{+}1.65}\exp(-\phantom{0}21,240/T)$ & \cite{Luna:2003}\notag\\%
32 & CN$^{+}$\ +\ N\phantom{$_{2}$} &\ $\rightleftarrows$ CN\phantom{$^{+}$}\ +\ N$^{+}$                           & $9.80\times 10^{12}\phantom{T^{-0.00}}\exp(-\phantom{0}40,700/T)$ & \cite{Gokcen:2005}\notag\\%
\midrule%
\bottomrule%
\end{tabular}
\end{table*}

\section{Conclusions}

A new dataset has been produced, for the simulation of N$_2$--CH$_4$ flows up to 100,000\kelvin. The rates have been verified to be physically consistent, and to have similar values to the G\"{o}k\c{c}en dataset, in the lower temperature range where it has been validated. The ionized rates of this dataset have been completely updated, considering the most recent and accurate sets of published cross-sections from which rates have been obtained. Additional ionized reaction rates have been added after a sensitivity study. It has been found that many ionized rates have significant differences, compared to the ones proposed in the G\"{o}k\c{c}en dataset. As these latter ones have just been reported from earlier works without any specific validation work, it is expected that this new set of rates will improve the accuracy of simulation of nonequilibrium, high-temperature N$_2$--CH$_4$ flows. The new proposed dataset for the simulation of high-temperature N$_2$--CH$_4$ plasmas is reported in Table \ref{tab:Titan_set_new} Updates to the G\"{o}k\c{c}en dataset are reported in red. The fit for the corresponding equilibrium constants is presented in Appendix \ref{sec:Keq}.

\begin{table*}[!htbp]
\vspace{-1cm}
\caption{New Chemical Dataset for N$_2$--CH$_4$ flows}%
\label{tab:Titan_set_new}%
\centering%
\begin{tabular}{l r@{}l c l}
\toprule%
\midrule%
No. & \multicolumn{2}{c}{Reaction} & Rate (\centi\cubic\metre\per\mole\per\second) & Ref.\\%
\midrule%
\phantom{0}1  & N$_{2}$\ +\ Mol.&\ $\rightleftarrows$ N\ +\ N\ +\ Mol                    & \color{red}$1.72\times 10^{18}T^{-0.89}\exp(-111,700/T)$ & \cite{LinodaSilva:2007-1}$^a$\notag\\%
\phantom{0}2  & N$_{2}$\ +\ Atom&\ $\rightleftarrows$ N\ +\ N\ +\ Atom                   & \color{red}$1.21\times 10^{19}T^{-0.89}\exp(-111,700/T)$ & \cite{Capitelli:2004}$^a$\notag\\%
\phantom{0}3  & CH$_{4}$\ +\ M\phantom{$_{2}$}&\ $\rightleftarrows$ CH$_{3}$\ +\ H\ +\ M & \color{red}$1.06\times 10^{22}T^{-1.46}\exp(-\phantom{0}49,990/T)$ &  \cite{Warnatz:1984}$^{b,c}$\notag\\%
\phantom{0}4  & CH$_{3}$\ +\ M\phantom{$_{2}$}&\ $\rightleftarrows$ CH$_{2}$\ +\ H\ +\ M & \color{red}$2.82\times 10^{14}\phantom{T^{-0.00}}\exp(-\phantom{0}42,460/T)$ & \cite{Lim:1994}\notag\\%
\phantom{0}5  & CH$_{3}$\ +\ M\phantom{$_{2}$}&\ $\rightleftarrows$ CH\ +\ H$_{2}$\ +\ M & $5.00\times 10^{15}\phantom{T^{-0.00}}\exp(-\phantom{0}42,800/T)$ & \cite{Gokcen:2005}\notag\\%
\phantom{0}6  & CH$_{2}$\ +\ M\phantom{$_{2}$}&\ $\rightleftarrows$ CH\ +\ H\ +\ M       & $4.00\times 10^{15}\phantom{T^{-0.00}}\exp(-\phantom{0}41,800/T)$ & \cite{Gokcen:2005}\notag\\%
\phantom{0}7  & CH$_{2}$\ +\ M\phantom{$_{2}$}&\ $\rightleftarrows$ C\ +\ H$_{2}$\ +\ M  & $1.30\times 10^{14}\phantom{T^{-0.00}}\exp(-\phantom{0}29,700/T)$ & \cite{Gokcen:2005}\notag\\%
\phantom{0}8  & CH\ +\ M\phantom{$_{2}$}&\ $\rightleftarrows$ C\ +\ H\ +\ M              & $1.90\times 10^{14}\phantom{T^{-0.00}}\exp(-\phantom{0}33,700/T)$ & \cite{Gokcen:2005}\notag\\%
\phantom{0}9  & C$_{2}$\ +\ M\phantom{$_{2}$}&\ $\rightleftarrows$ C\ +\ C\ +\ M         & \color{red}$3.72\times 10^{14}\phantom{T^{-0.00}}\exp(-\phantom{0}69,800/T)$ & \cite{Beck:1975}\notag\\%
10 & H$_{2}$\ +\ M\phantom{$_{2}$}&\ $\rightleftarrows$ H\ +\ H\ +\ M         & $2.23\times 10^{14}\phantom{T^{-0.00}}\exp(-\phantom{0}48,350/T)$ & \cite{Gokcen:2005}\notag\\%
11 & CN\ +\ M\phantom{$_{2}$}&\ $\rightleftarrows$ C\ +\ N\ +\ M              & $2.53\times 10^{14}\phantom{T^{-0.00}}\exp(-\phantom{0}71,000/T)$ & \cite{Gokcen:2005}\notag\\%
12 & NH\ +\ M\phantom{$_{2}$}&\ $\rightleftarrows$ N\ +\ H\ +\ M              & $1.80\times 10^{14}\phantom{T^{-0.00}}\exp(-\phantom{0}37,600/T)$ & \cite{Gokcen:2005}\notag\\%
13 & HCN\ +\ M\phantom{$_{2}$}&\ $\rightleftarrows$ CN\ +\ H\ +\ M            & $3.57\times 10^{26}T^{-2.60}\exp(-\phantom{0}62,845/T)$ & \cite{Gokcen:2005}\notag\\%
\midrule%
14 & CH$_{3}$\ +\ N\phantom{$_{2}$}&\ $\rightleftarrows$ HCN\ +\ H\ +\ H      & $7.00\times 10^{13}\phantom{T^{-0.00}\exp(-000,000/T)}$ & \cite{Gokcen:2005}\notag\\%
15 & CH$_{3}$\ +\ H\phantom{$_{2}$}&\ $\rightleftarrows$ CH$_{2}$\ +\ H$_{2}$ & $6.03\times 10^{13}\phantom{T^{-0.00}}\exp(-\phantom{00}7,600/T)$ & \cite{Gokcen:2005}\notag\\%
16 & CH$_{2}$\ +\ N$_{2}$&\ $\rightleftarrows$ HCN\ +\ NH                     & $4.82\times 10^{12}\phantom{T^{-0.00}}\exp(-\phantom{0}18,000/T)$ & \cite{Gokcen:2005}\notag\\%
17 & CH$_{2}$\ +\ N\phantom{$_{2}$}&\ $\rightleftarrows$ HCN\ +\ H            & $5.00\times 10^{13}\phantom{T^{-0.00}\exp(-000,000/T)}$ & \cite{Gokcen:2005}\notag\\%
18 & CH$_{2}$\ +\ H\phantom{$_{2}$}&\ $\rightleftarrows$ CH\ +\ H$_{2}$       & \color{red}$4.21\times 10^{\phantom{0}8}T^{-0.09}\exp(+\phantom{00}1,560/T)$ & \cite{Rohrig:1997}$^b$\notag\\%
19 & CH\ +\ N$_{2}$&\ $\rightleftarrows$ HCN\ +\ N                            & $4.40\times 10^{12}\phantom{T^{-0.00}}\exp(-\phantom{0}11,060/T)$ & \cite{Gokcen:2005}\notag\\%
20 & CH\ +\ C\phantom{$_{2}$}&\ $\rightleftarrows$ C$_{2}$\ +\ H              & $2.00\times 10^{14}\phantom{T^{-0.00}\exp(-000,000/T)}$ & \cite{Gokcen:2005}\notag\\%
21 & C$_{2}$\ +\ N$_{2}$&\ $\rightleftarrows$ CN\ +\ CN                       & $1.50\times 10^{13}\phantom{T^{-0.00}}\exp(-\phantom{0}21,000/T)$ & \cite{Gokcen:2005}\notag\\%
22 & CN\ +\ H$_{2}$&\ $\rightleftarrows$ HCN\ +\ H                            & $2.95\times 10^{05}\phantom{T^{-0.00}}\exp(-\phantom{00}1,130/T)$ & \cite{Gokcen:2005}\notag\\%
23 & CN\ +\ C\phantom{$_{2}$}&\ $\rightleftarrows$ C$_{2}$\ +\ N              & \color{red}$3.00\times 10^{14}\phantom{T^{-0.00}}\exp(-\phantom{0}18,040/T)$ & \cite{Slack:1976}\notag\\%
24 & N\ +\ H$_{2}$&\ $\rightleftarrows$ NH\ +\ H                              & $1.60\times 10^{14}\phantom{T^{-0.00}}\exp(-\phantom{0}12,650/T)$ & \cite{Gokcen:2005}\notag\\%
25 & C\ +\ N$_{2}$&\ $\rightleftarrows$ CN\ +\ N                              & $5.24\times 10^{13}\phantom{T^{-0.00}}\exp(-\phantom{0}22,600/T)$ & \cite{Gokcen:2005}\notag\\%
26 & C\ +\ H$_{2}$&\ $\rightleftarrows$ CH\ +\ H                              & $4.00\times 10^{14}\phantom{T^{-0.00}}\exp(-\phantom{0}11,700/T)$ & \cite{Gokcen:2005}\notag\\%
27 & H\ +\ N$_{2}$&\ $\rightleftarrows$ NH\ +\ N                              & $3.00\times 10^{12}T^{-0.50}\exp(-\phantom{0}71,400/T)$ & \cite{Gokcen:2005}\notag\\%
28 & H\ +\ CH$_{4}$&\ $\rightleftarrows$ CH$_{3}$\ +\ H$_{2}$                 & \color{red}$1.54\times 10^{14}\phantom{T^{-0.00}}\exp(-\phantom{00}6,874/T)$ & \cite{Bryukov:2001}\notag\\%
\midrule%
\color{red}29 & \color{red}H\ +\ H\phantom{$_{2}$} &\color{red}\ $\rightleftarrows$ \phantom{C}H$_{2}^{+}$\ +\ e$^{-}$                              & \color{red}$1.13\times 10^{15}T^{-0.06}\exp(-129,060/T)$ & \cite{Florescu:2006,Auerbach:1977}$^a$\notag\\%
\color{red}30 & \color{red}C\ +\ H\phantom{$_{2}$} &\color{red}\ $\rightleftarrows$ CH$^{+}$\ +\ e$^{-}$                                            & \color{red}$9.95\times 10^{11}T^{\phantom{+}0.52}\exp(-\phantom{0}84,830/T)$ & \cite{Florescu:2006,Mitchell:1978}$^a$\notag\\%
\color{red}31 & \color{red}C\ +\ N\phantom{$_{2}$} &\color{red}\ $\rightleftarrows$ CN$^{+}$\ +\ e$^{-}$                                            & \color{red}$3.80\times 10^{12}T^{\phantom{+}0.33}\exp(-\phantom{0}74,810/T)$ & \cite{LePadellec:1999}$^a$\notag\\%
\color{red}32 & \color{red}N\ +\ H\phantom{$_{2}$} &\color{red}\ $\rightleftarrows$ NH$^{+}$\ +\ e$^{-}$                                            & \color{red}$2.99\times 10^{14}T^{-0.06}\exp(-118,760/T)$ & \cite{Florescu:2006,Mul:unp}$^a$\notag\\%
\color{red}33 & \color{red}N\ +\ N\phantom{$_{2}$} &\color{red}\ $\rightleftarrows$ \phantom{C}N$_{2}^{+}$\ +\ e$^{-}$                              & \color{red}$2.13\times 10^{10}T^{\phantom{+}0.48}\exp(-\phantom{0}69,190/T)$ & \cite{Guberman:2009}$^a$\notag\\%
\midrule%
\color{red}34 & \color{red}H\ +\ e$^{-}$           &\color{red}\ $\rightleftarrows$ \phantom{C}H$^{+}$\ +\ e$^{-}$\hspace{1.2mm} +\ e$^{-}$         & \color{red}$1.36\times 10^{15}T^{\phantom{+}0.18}\exp(-169,000/T)$ & \cite{Kim:1994}$^a$\notag\\%
\color{red}35 & \color{red}C\ +\ e$^{-}$           &\color{red}\ $\rightleftarrows$ \phantom{C}C$^{+}$\ +\ e$^{-}$\hspace{1.2mm} +\ e$^{-}$         & \color{red}$1.24\times 10^{15}T^{\phantom{+}0.28}\exp(-142,700/T)$ & \cite{Kim:2002}$^a$\notag\\%
\color{red}36 & \color{red}N\ +\ e$^{-}$           &\color{red}\ $\rightleftarrows$ \phantom{C}N$^{+}$\ +\ e$^{-}$\hspace{1.2mm} +\ e$^{-}$         & \color{red}$1.67\times 10^{13}T^{\phantom{+}0.59}\exp(-143,220/T)$ & \cite{Kim:2002}$^a$\notag\\%
\color{red}37 & \color{red}Ar\ +\ e$^{-}$          &\color{red}\ $\rightleftarrows$ \hspace{1.2mm}Ar$^{+}$\ +\ e$^{-}$ \hspace{1.2mm}+\ e$^{-}$     & \color{red}$5.52\times 10^{13}T^{\phantom{+}0.58}\exp(-186,210/T)$ & \cite{Rapp:1965,Schram:1966}$^a$\notag\\%
\midrule%
\color{red}38 & \color{red}H$_{2}$\ +\ e$^{-}$     &\color{red}\ $\rightleftarrows$ \phantom{C}H$_{2}^{+}$\ +\ e$^{-}$\hspace{1.2mm}\ +\ e$^{-}$    & \color{red}$4.05\times 10^{13}T^{\phantom{+}0.52}\exp(-180,767/T)$ & \cite{Rapp:1965}$^a$\notag\\%
\color{red}39 & \color{red}CH\ +\ e$^{-}$          &\color{red}\ $\rightleftarrows$ CH$^{+}$\ +\ e$^{-}$\hspace{1.2mm} +\ e$^{-}$                   & \color{red}$1.15\times 10^{12}T^{\phantom{+}0.87}\exp(-123,430/T)$ & \cite{Kim:2000}$^a$\notag\\%
\color{red}40 & \color{red}NH\ +\ e$^{-}$          &\color{red}\ $\rightleftarrows$ NH$^{+}$\ +\ e$^{-}$\hspace{1.2mm}\ +\ e$^{-}$                  & \color{red}$2.39\times 10^{13}T^{\phantom{+}0.59}\exp(-172,430/T)$ & \cite{Rajvanshi:2010}$^a$\notag\\%
\color{red}41 & \color{red}N$_{2}$\ +\ e$^{-}$     &\color{red}\ $\rightleftarrows$ \phantom{C}N$_{2}^{+}$\ +\ e$^{-}$\hspace{1.2mm}\ +\ e$^{-}$    & \color{red}$5.92\times 10^{11}T^{\phantom{+}0.92}\exp(-178,630/T)$ & \cite{Rapp:1965}$^a$\notag\\%
\color{red}42 & \color{red}NH$^{+}$\ +\ e$^{-}$    &\color{red}\ $\rightleftarrows$ \phantom{C}N$^{+}$\ +\ H$^{+}$\ +\ e$^{-}$\ +\ e$^{-}$          & \color{red}$1.25\times 10^{13}T^{\phantom{+}0.56}\exp\left(-\phantom{0}27,350/T\right)$ & \cite{Lecointre:2010}$^a$\\%
\color{red}43 & \color{red}N$_{2}^{+}$\ +\ e$^{-}$ &\color{red}\ $\rightleftarrows$ \phantom{C}N$^{+}$\ +\ N\phantom{$^{+}$}\ +\ e$^{-}$            & \color{red}$7.47\times 10^{11}T^{\phantom{+}0.84}\exp\left(-\phantom{0}80,600/T\right)$ & \cite{Bahati:2001}$^a$\\%
\midrule%
\color{red}44 & \color{red}C$^{+}$\ +\ N$_{2}$     &\color{red}\ $\rightleftarrows$ \phantom{C}N$_{2}^{+}$\ +\ C                                    & \color{red}$1.01\times 10^{11}T^{\phantom{+}0.60}\exp(-\phantom{0}53,830/T)$ & \cite{Burley:1991}$^a$\notag\\%
\color{red}45 & \color{red}C$^{+}$\ +\ N$_{2}$     &\color{red}\ $\rightleftarrows$ CN$^{+}$\ +\ N                                                  & \color{red}$1.32\times 10^{12}T^{\phantom{+}0.33}\exp(-\phantom{0}51,430/T)$ & \cite{Burley:1991}$^a$\notag\\%
\color{red}46 & \color{red}C$^{+}$\ +\ N$_{2}$     &\color{red}\ $\rightleftarrows$ \phantom{C}N$^{+}$\ +\ CN                                       & \color{red}$8.93\times 10^{13}T^{-0.14}\exp(-\phantom{0}65,260/T)$ & \cite{Burley:1991}$^a$\notag\\%
\midrule%
\bottomrule%
\multicolumn{5}{l}{\footnotesize{a: fit from this work.}}\\
\multicolumn{5}{l}{\footnotesize{b: detailed balancing of published reverse rate, followed by a fit to an Arrhenius form.}}\\
\multicolumn{5}{l}{\footnotesize{c: rate slightly departs from an Arrhenius form, fit is approximate.}}\\
\multicolumn{5}{l}{\footnotesize{\phantom{c:} More accurate fit for T=500--20,000\kelvin: $K_f=1.22\times 10^{28}T^{-2.91}\exp(-54,000/T)$}}\\
\multicolumn{5}{l}{\footnotesize{\phantom{c:} More accurate fit for T=10,000--100,000\kelvin: $K_f=2.28\times 10^{15}T^{-0.12}\exp(-17,960/T)$}}\\
\end{tabular}
\end{table*}

\subsubsection*{Acknowledgements}
\small

The authors would like to thank D. Tsyhanou for assistance with developping and running the chemistry sensitivity codes.
\vspace{-0.08cm}

\appendix

\section{Equilibrium Constants for the Final N$_{2}$--CH$_{4}$ Chemical Dataset}
\label{sec:Keq}

The  Equilibrium constants $K_{eq}$, calculated for the rates presented in Table \ref{tab:Titan_set_new}, have been individually fitted to the following expression:

\begin{eqnarray*}
K_{eq}(i)=\exp\left(
\begin{array}{l}
\phantom{+}a_1^i\left(\frac{T}{1000}\right)^{-3}+a_2^i\left(\frac{T}{1000}\right)^{-2}+a_3^i\left(\frac{T}{1000}\right)^{-1}\vspace{3mm}\\
+a_4^i+a_5^i\ln\left(\frac{T}{1000}\right)+a_6^i\left(\frac{T}{1000}\right)^{1}+a_7^i\left(\frac{T}{1000}\right)^{2}\vspace{3mm}\\
+a_8^i\left(\frac{T}{1000}\right)^{3}+a_9^i\left(\frac{T}{1000}\right)^{4}
\end{array}
\right)
\end{eqnarray*}

The fitting parameters $a_0^i$ to $a_9^i$ are presented in Table \ref{tab:Keq}.
\begin{landscape}

\begin{table*}[!htbp]
\vspace{-2.5cm}
\small
\caption{Fitting Parameters for the Equilibrium Constants $K_{eq}$ of the New Chemical Dataset for N$_2$--CH$_4$ flows}%
\label{tab:Keq}%
\centering%
\begin{tabular}{r r r r r r r r r r r}
\toprule%
\midrule%
\multicolumn{1}{c}{Rate} & \multicolumn{1}{c}{$a_1$} & \multicolumn{1}{c}{$a_2$} & \multicolumn{1}{c}{$a_3$} & \multicolumn{1}{c}{$a_4$} & \multicolumn{1}{c}{$a_5$} & \multicolumn{1}{c}{$a_6$} & \multicolumn{1}{c}{$a_7$} & \multicolumn{1}{c}{$a_8$} & \multicolumn{1}{c}{$a_9$}\\%
\midrule%
1  & $+$5.850E$-$02 & $-$1.435E$-$01 & $-$1.137E$+$02 & $+$6.954E$+$01 & $+$9.841E$-$01 & $-$1.096E$-$01 & $+$7.632E$-$03 & $-$1.061E$-$04 & $+$4.609E$-$07 & $+$7.166E$+$00 \\%
2  & $+$5.850E$-$02 & $-$1.435E$-$01 & $-$1.137E$+$02 & $+$6.954E$+$01 & $+$9.841E$-$01 & $-$1.096E$-$01 & $+$7.632E$-$03 & $-$1.061E$-$04 & $+$4.609E$-$07 & $+$7.166E$+$00 \\%
3  & $-$4.151E$-$02 & $+$8.111E$-$01 & $-$5.670E$+$01 & $+$7.369E$+$01 & $-$9.367E$-$01 & $+$7.233E$-$02 & $+$6.960E$-$04 & $-$1.540E$-$05 & $+$7.424E$-$08 & $+$1.435E$+$00 \\%
4  & $-$4.654E$-$02 & $+$8.837E$-$01 & $-$5.987E$+$01 & $+$7.279E$+$01 & $-$1.218E$+$00 & $+$6.323E$-$02 & $+$2.470E$-$03 & $-$4.329E$-$05 & $+$2.035E$-$07 & $+$6.218E$+$00 \\%
5  & $-$2.987E$-$02 & $+$6.120E$-$01 & $-$5.702E$+$01 & $+$7.231E$+$01 & $-$6.628E$-$01 & $-$1.000E$-$02 & $+$2.631E$-$03 & $-$3.561E$-$05 & $+$1.508E$-$07 & $+$1.723E$+$00 \\%
6  & $-$8.489E$-$03 & $+$3.219E$-$01 & $-$5.227E$+$01 & $+$6.910E$+$01 & $+$3.023E$-$01 & $-$4.340E$-$02 & $+$2.434E$-$03 & $-$2.940E$-$05 & $+$1.184E$-$07 & $+$7.619E$-$01 \\%
7  & $+$9.485E$-$02 & $-$6.713E$-$01 & $-$3.734E$+$01 & $+$6.591E$+$01 & $+$1.267E$+$00 & $-$1.004E$-$01 & $+$6.481E$-$04 & $+$5.150E$-$07 & $-$1.500E$-$08 & $+$6.309E$-$01 \\%
8  & $-$3.730E$-$02 & $+$5.985E$-$01 & $-$4.311E$+$01 & $+$6.816E$+$01 & $-$5.386E$-$01 & $+$1.418E$-$01 & $-$2.340E$-$03 & $+$1.903E$-$05 & $-$6.081E$-$08 & $+$1.137E$-$01 \\%
9  & $-$1.328E$-$01 & $+$1.506E$+$00 & $-$7.778E$+$01 & $+$7.275E$+$01 & $-$2.055E$+$00 & $+$2.757E$-$01 & $-$4.048E$-$03 & $+$3.147E$-$05 & $-$9.700E$-$08 & $+$4.013E$+$00 \\%
10 & $-$2.985E$-$02 & $+$5.760E$-$01 & $-$5.485E$+$01 & $+$6.925E$+$01 & $-$1.406E$-$01 & $+$6.133E$-$02 & $+$2.414E$-$07 & $-$6.125E$-$06 & $+$3.629E$-$08 & $+$6.426E$-$01 \\%
11 & $-$3.300E$-$02 & $+$7.489E$-$01 & $-$9.381E$+$01 & $+$7.119E$+$01 & $-$1.222E$+$00 & $+$2.174E$-$01 & $-$2.900E$-$03 & $+$1.933E$-$05 & $-$5.067E$-$08 & $+$4.399E$-$01 \\%
12 & $-$2.183E$-$01 & $+$2.375E$+$00 & $-$4.859E$+$01 & $+$7.221E$+$01 & $-$2.578E$+$00 & $+$3.102E$-$01 & $-$4.903E$-$03 & $+$4.172E$-$05 & $-$1.404E$-$07 & $+$2.160E$+$00 \\%
13 & $-$1.300E$-$02 & $+$3.158E$-$01 & $-$6.437E$+$01 & $+$7.008E$+$01 & $+$3.007E$-$01 & $-$6.019E$-$02 & $+$1.662E$-$03 & $-$1.638E$-$05 & $+$5.757E$-$08 & $+$3.946E$-$01 \\%
14 & $-$1.189E$-$02 & $+$2.211E$-$01 & $+$5.439E$+$00 & $+$6.673E$+$01 & $+$7.998E$-$01 & $-$8.102E$-$02 & $+$1.173E$-$03 & $-$9.862E$-$06 & $+$3.320E$-$08 & $+$1.859E$-$01 \\%
15 & $-$7.469E$-$03 & $+$1.603E$-$01 & $-$4.386E$+$00 & $+$2.931E$+$00 & $-$9.402E$-$01 & $+$7.538E$-$02 & $-$1.132E$-$03 & $+$9.690E$-$06 & $-$3.306E$-$08 & $+$3.678E$-$02 \\%
16 & $+$2.180E$-$02 & $-$3.998E$-$01 & $-$8.780E$+$00 & $-$3.855E$+$00 & $+$2.424E$+$00 & $-$1.766E$-$01 & $+$3.436E$-$03 & $-$3.318E$-$05 & $+$1.211E$-$07 & $+$1.065E$+$00 \\%
17 & $+$3.573E$-$02 & $-$5.963E$-$01 & $+$6.488E$+$01 & $-$5.547E$+$00 & $+$1.990E$+$00 & $-$2.482E$-$01 & $+$3.925E$-$03 & $-$3.102E$-$05 & $+$9.681E$-$08 & $+$1.049E$+$00 \\%
18 & $+$4.629E$-$03 & $-$6.330E$-$02 & $+$1.710E$+$00 & $+$4.154E$-$01 & $-$1.937E$-$01 & $+$2.909E$-$02 & $-$7.221E$-$04 & $+$7.946E$-$06 & $-$3.163E$-$08 & $+$3.010E$-$01 \\%
19 & $+$4.136E$-$04 & $-$3.159E$-$02 & $-$3.225E$-$01 & $-$2.535E$+$00 & $+$7.168E$-$01 & $-$2.716E$-$02 & $+$1.698E$-$04 & $-$3.649E$-$07 & $-$1.086E$-$09 & $+$2.427E$-$02 \\%
20 & $+$1.534E$-$02 & $-$1.050E$-$01 & $+$3.176E$+$01 & $-$2.616E$+$00 & $-$2.838E$-$02 & $+$8.946E$-$02 & $-$2.545E$-$03 & $+$2.898E$-$05 & $-$1.159E$-$07 & $+$1.158E$+$00 \\%
21 & $+$1.397E$-$02 & $-$2.712E$-$01 & $-$3.990E$+$00 & $-$3.877E$-$02 & $+$7.334E$-$01 & $-$3.178E$-$02 & $+$4.339E$-$04 & $-$4.015E$-$06 & $+$1.504E$-$08 & $+$5.926E$-$01 \\%
22 & $-$1.938E$-$03 & $+$1.407E$-$03 & $+$1.095E$+$01 & $-$1.947E$+$00 & $+$5.070E$-$01 & $-$8.818E$-$03 & $-$2.514E$-$04 & $+$3.900E$-$06 & $-$1.679E$-$08 & $+$1.503E$-$01 \\%
23 & $+$8.508E$-$02 & $-$5.981E$-$01 & $-$1.668E$+$01 & $-$1.125E$+$00 & $+$4.152E$-$01 & $+$1.893E$-$02 & $-$8.797E$-$04 & $+$1.274E$-$05 & $-$5.840E$-$08 & $+$1.590E$+$00 \\%
24 & $-$3.988E$-$02 & $+$4.209E$-$01 & $-$1.370E$+$01 & $+$2.475E$+$00 & $-$4.692E$-$01 & $+$6.469E$-$03 & $-$1.665E$-$04 & $+$2.641E$-$06 & $-$1.283E$-$08 & $+$9.622E$-$01 \\%
25 & $+$1.074E$-$01 & $-$9.877E$-$01 & $-$2.001E$+$01 & $-$1.618E$+$00 & $+$1.697E$+$00 & $-$1.339E$-$01 & $+$2.505E$-$03 & $-$2.246E$-$05 & $+$7.633E$-$08 & $+$6.593E$-$01 \\%
26 & $-$3.026E$-$02 & $+$2.868E$-$01 & $-$1.254E$+$01 & $+$1.546E$+$00 & $+$1.211E$-$01 & $-$4.671E$-$02 & $+$1.442E$-$03 & $-$1.624E$-$05 & $+$6.424E$-$08 & $+$2.896E$-$01 \\%
27 & $+$1.669E$-$01 & $-$1.327E$+$00 & $-$6.978E$+$01 & $+$8.988E$-$01 & $+$1.349E$+$00 & $-$8.971E$-$02 & $+$4.435E$-$04 & $+$5.121E$-$06 & $-$4.304E$-$08 & $+$2.356E$+$00 \\%
28 & $-$1.556E$-$02 & $+$3.051E$-$01 & $-$2.250E$+$00 & $+$4.823E$+$00 & $-$1.012E$+$00 & $+$3.662E$-$02 & $-$5.565E$-$04 & $+$4.821E$-$06 & $-$1.662E$-$08 & $+$4.696E$-$02 \\%
29 & $+$7.446E$-$01 & $-$5.178E$+$00 & $-$1.134E$+$02 & $-$1.634E$+$01 & $+$6.576E$+$00 & $-$3.690E$-$01 & $-$1.966E$-$05 & $+$4.056E$-$05 & $-$2.413E$-$07 & $+$1.046E$+$01 \\%
30 & $+$8.958E$-$01 & $-$6.458E$+$00 & $-$6.545E$+$01 & $-$2.098E$+$01 & $+$8.774E$+$00 & $-$6.744E$-$01 & $+$1.094E$-$02 & $-$9.106E$-$05 & $+$2.946E$-$07 & $+$3.575E$+$00 \\%
31 & $-$1.923E$-$02 & $-$3.792E$-$01 & $-$6.898E$+$01 & $-$1.398E$+$01 & $+$6.021E$+$00 & $-$6.185E$-$01 & $+$1.048E$-$02 & $-$8.996E$-$05 & $+$2.986E$-$07 & $+$3.746E$+$00 \\%
32 & $+$1.279E$+$00 & $-$8.765E$+$00 & $-$9.470E$+$01 & $-$2.051E$+$01 & $+$9.502E$+$00 & $-$7.342E$-$01 & $+$9.492E$-$03 & $-$6.221E$-$05 & $+$1.540E$-$07 & $+$8.992E$+$00 \\%
33 & $+$5.115E$-$01 & $-$3.765E$+$00 & $-$5.664E$+$01 & $-$1.741E$+$01 & $+$6.166E$+$00 & $-$5.080E$-$01 & $+$8.650E$-$03 & $-$7.443E$-$05 & $+$2.473E$-$07 & $+$1.191E$+$00 \\%
34 & $+$2.465E$-$01 & $-$1.628E$+$00 & $-$1.540E$+$02 & $+$5.480E$+$01 & $+$3.732E$+$00 & $-$7.655E$-$02 & $-$9.365E$-$04 & $+$2.322E$-$05 & $-$1.171E$-$07 & $+$1.507E$+$00 \\%
35 & $+$3.416E$-$01 & $-$2.314E$+$00 & $-$1.249E$+$02 & $+$5.391E$+$01 & $+$4.657E$+$00 & $-$2.291E$-$01 & $+$2.050E$-$03 & $-$8.212E$-$06 & $+$1.231E$-$09 & $+$1.573E$+$00 \\%
36 & $+$8.229E$-$01 & $-$4.310E$+$00 & $-$1.602E$+$02 & $+$5.389E$+$01 & $+$5.129E$+$00 & $-$2.314E$-$01 & $+$1.017E$-$03 & $+$7.697E$-$06 & $-$6.931E$-$08 & $+$1.832E$+$00 \\%
37 & $+$6.983E$+$00 & $-$3.029E$+$01 & $-$2.873E$+$02 & $+$4.576E$+$01 & $+$1.635E$+$01 & $-$1.341E$+$00 & $+$1.765E$-$02 & $-$1.131E$-$04 & $+$2.600E$-$07 & $+$2.185E$+$01 \\%
38 & $+$7.314E$-$02 & $-$6.126E$-$01 & $-$1.768E$+$02 & $+$5.775E$+$01 & $+$4.050E$+$00 & $-$1.942E$-$01 & $+$2.403E$-$03 & $-$5.731E$-$06 & $-$2.915E$-$08 & $+$8.423E$-$01 \\%
39 & $+$3.191E$-$01 & $-$2.240E$+$00 & $-$1.177E$+$02 & $+$5.266E$+$01 & $+$4.628E$+$00 & $-$1.014E$-$01 & $+$1.233E$-$03 & $-$8.389E$-$06 & $+$2.159E$-$08 & $+$1.971E$-$01 \\%
40 & $+$3.945E$-$02 & $-$3.399E$-$01 & $-$1.554E$+$02 & $+$5.785E$+$01 & $+$3.417E$+$00 & $-$1.021E$-$01 & $+$1.814E$-$03 & $-$1.866E$-$05 & $+$6.891E$-$08 & $+$4.991E$-$02 \\%
41 & $+$4.497E$-$01 & $-$3.117E$+$00 & $-$1.728E$+$02 & $+$5.349E$+$01 & $+$5.539E$+$00 & $-$2.268E$-$01 & $+$2.477E$-$03 & $-$2.203E$-$05 & $+$7.317E$-$08 & $+$1.010E$+$00 \\%
42 & $-$3.411E$+$00 & $+$1.359E$+$01 & $-$2.316E$+$02 & $+$7.912E$+$01 & $-$1.999E$+$00 & $+$3.263E$-$01 & $-$3.335E$-$03 & $+$2.117E$-$05 & $-$5.863E$-$08 & $+$5.162E$-$01 \\%
43 & $-$5.565E$-$01 & $+$3.147E$+$00 & $-$1.085E$+$02 & $+$7.344E$+$01 & $-$1.321E$+$00 & $+$8.027E$-$02 & $+$1.811E$-$03 & $-$2.870E$-$05 & $+$1.240E$-$07 & $+$6.387E$-$01 \\%
44 & $+$2.633E$-$01 & $-$1.543E$+$00 & $-$4.657E$+$01 & $-$1.083E$+$00 & $+$1.207E$+$00 & $-$1.581E$-$02 & $+$1.911E$-$04 & $-$2.387E$-$06 & $+$1.216E$-$08 & $+$3.630E$-$01 \\%
45 & $-$8.130E$-$01 & $+$3.981E$+$00 & $-$6.153E$+$01 & $+$3.377E$+$00 & $+$7.536E$-$01 & $-$1.433E$-$01 & $+$3.057E$-$03 & $-$2.938E$-$05 & $+$1.052E$-$07 & $+$2.709E$+$00 \\%
46 & $+$3.146E$-$01 & $-$1.956E$+$00 & $-$5.650E$+$01 & $-$1.164E$+$00 & $+$2.172E$+$00 & $-$2.057E$-$01 & $+$4.785E$-$03 & $-$4.771E$-$05 & $+$1.745E$-$07 & $+$6.449E$-$01 \\%
\midrule%
\bottomrule%
\end{tabular}
\end{table*}

\end{landscape}



\end{document}